\documentclass[pdflatex,sn-basic]{sn-jnl}

\usepackage{graphicx}%
\usepackage{multirow}%
\usepackage{amsmath,amssymb,amsfonts}%
\usepackage{amsthm}%

\usepackage{mathrsfs}%
\usepackage[title]{appendix}%
\usepackage{xcolor}%
\usepackage{textcomp}%
\usepackage{manyfoot}%
\usepackage{booktabs}%
\usepackage{algorithm}%
\usepackage{algorithmicx}%
\usepackage{algpseudocode}%
\usepackage{listings}%
\usepackage{natbib}
\usepackage{braket}
\usepackage{dirtytalk}
\usepackage{array}
\usepackage{bm}
\usepackage{caption} 
\usepackage{subcaption}
\usepackage{tablefootnote}

\theoremstyle{thmstyleone}%
\theoremstyle{thmstyletwo}%

\theoremstyle{thmstylethree}%

\usepackage{tikz}
\usetikzlibrary{quantikz2}
\usepackage{mathtools}
\usepackage{bbm}
\usepackage{xspace}
\usepackage{placeins}

\newcommand\norm[1]{\left\lVert#1\right\rVert}

\newcommand{\eg}{e.g.,\xspace}
\newcommand{\idest}{i.e.,\xspace}

\begin{document}

\title[Hamiltonian Expressibility for Ansatz Selection in Variational Quantum Algorithms]{Hamiltonian Expressibility for Ansatz Selection in Variational Quantum Algorithms}


\author*[1]{\fnm{Filippo} \sur{Brozzi}}\email{filippo.brozzi@unifi.it}
\author*[2]{\fnm{Gloria} \sur{Turati}}\email{gloria.turati@polimi.it}

\author*[2]{\fnm{Maurizio} \sur{Ferrari Dacrema}}\email{maurizio.ferrari@polimi.it}

\author[1]{\fnm{Filippo} \sur{Caruso}}\email{filippo.caruso@unifi.it}

\author[2]{\fnm{Paolo} \sur{Cremonesi}}\email{paolo.cremonesi@polimi.it}

\affil[1]{\orgname{Università degli Studi di Firenze}, \city{Firenze}, \country{Italy}}

\affil[2]{\orgname{Politecnico di Milano}, \city{Milano}, \country{Italy}}

\abstract{
In the context of Variational Quantum Algorithms (VQAs), selecting an appropriate ansatz is crucial for efficient problem-solving. Hamiltonian expressibility has been introduced as a metric to quantify a circuit's ability to uniformly explore the energy landscape associated with a Hamiltonian ground state search problem. However, its influence on solution quality remains largely unexplored.
In this work, we estimate the Hamiltonian expressibility of a well-defined set of circuits applied to various Hamiltonians using a Monte Carlo-based approach.
We analyse how ansatz depth influences expressibility and identify the most and least expressive circuits across different problem types.
We then train each ansatz using the Variational Quantum Eigensolver (VQE) and analyse the correlation between solution quality and expressibility.
Our results indicate that, under ideal or low-noise conditions and particularly for small-scale problems, ansätze with high Hamiltonian expressibility yield better solution quality for problems with non-diagonal Hamiltonians and superposition state solutions. Conversely, circuits with low expressibility are more effective for problems whose solutions are basis states, including those defined by diagonal Hamiltonians. Under noisy conditions, low-expressibility circuits remain preferable for problems with solutions in a computational basis state, while intermediate expressibility yields better results for some problems involving superposition state solutions.
}

\keywords{Variational Quantum Algorithms, Expressibility, Ansatz, Correlation Analysis}



\maketitle

\section{Introduction}
\label{sec:introduction}

In the field of quantum computing, hybrid classical-quantum algorithms have emerged as a practical solution to the limitations of current Noisy Intermediate-Scale Quantum (NISQ) devices~\citep{NISQ-Preskill_2018}. Among these, Variational Quantum Algorithms (VQAs)~\citep{cerezo_2021} represent a prominent class that leverages classical optimization to tune the parameters of a quantum circuit, referred to as an \textit{ansatz}, with the goal of minimizing a problem-specific cost function. One notable example is the Variational Quantum Eigensolver (VQE)~\citep{peruzzo_2014}, which optimizes the ansatz parameters to prepare a quantum state that minimizes the expectation value of a given Hamiltonian.

One of the major challenges within these algorithms is the choice of the ansatz, since it significantly affects both trainability and solution quality.
Problem-inspired designs~\citep{QAOA_review-Blekos2024, ChemInspandHEAReview-Fedorov2022} can achieve high accuracy but are often difficult to implement on quantum hardware. 
In contrast, hardware-efficient ansätze~\citep{kandala_2017} are easier to implement due to their reduced depth and lower resource requirements, although they typically do not exploit problem-specific features, such as symmetries~\citep{meyer_2023}, which could help guide the optimization process more effectively.
Various strategies have been proposed to construct ansätze that incorporate problem knowledge while remaining compatible with hardware constraints. These include adaptive algorithms~\citep{turati_2023, AdaptiveVQEReview-Claudino2020} and reinforcement learning approaches~\citep{fodera_2024}. However, such methods often rely on heuristics and require a large number of executions and optimizations of candidate circuits.

Given the limitations of these techniques, part of the research has focused on analyzing circuit-based metrics~\citep{sim_2019, Marrero2021, Fontana_2021} as a means to support the selection of suitable ansätze. Among these, the notion of expressibility has received considerable attention in the literature~\citep{sim_2019, funcke_2021, du_2022, wu_2021, holmes_2022}. This metric quantifies a circuit’s ability to explore either the full space of quantum states or the energy landscape defined by a specific problem. The underlying idea is that greater exploration capability should increase the likelihood that the optimal solution lies within the region explored by the circuit, but it can also make the identification of high-quality solutions challenging. As a result, determining the optimal level of expressibility and whether it can meaningfully inform circuit selection remains an open research question.

In this work, we analyse the definition of expressibility introduced by~\citep{holmes_2022}, which characterises a circuit's ability to explore the energy landscape induced by a target Hamiltonian. We refer to this metric as \textit{Hamiltonian expressibility}. The utility of this metric has not been thoroughly investigated in the literature, and we argue that analyzing its properties and its potential contribution to ansatz selection is of particular interest due to its problem-specific nature. Its direct relationship with the structure of the energy landscape distinguishes it from other commonly used metrics, which are typically problem-agnostic, and suggests that it may serve as a more tailored indicator for the solution of specific target problems. Using a set of well-established ansatz templates and Hamiltonian problem instances, we develop and apply a Monte Carlo method to estimate this metric on systems of 4 and 8 qubits.

Our analysis first investigates the relationship between Hamiltonian expressibility and ansatz depth, revealing that expressibility generally increases with depth, until it reaches a saturation point. We then highlight the problem-dependent character of Hamiltonian expressibility by identifying, for each problem class, the most and least Hamiltonian-expressive circuits, and determining, for each circuit, the problem classes in which it is most and least expressive. Furthermore, we observe that all considered ansätze achieve higher levels of Hamiltonian expressibility on problems defined by diagonal Hamiltonians.

To evaluate the practical usefulness of these metrics for ansatz selection, we train each circuit on all problem instances using the VQE algorithm, under both ideal (noiseless) and noisy simulation settings, and analyse the correlation between solution quality and expressibility.

Our results suggest that Hamiltonian expressibility can provide valuable guidance for ansatz selection in small-scale problems. Specifically, under ideal or low-noise conditions, for problems defined by non-diagonal Hamiltonians with superposition state solutions, selecting circuits with high Hamiltonian expressibility leads to higher-quality solutions. Conversely, for problems whose solutions are basis states, such as those associated with diagonal Hamiltonians, including instances derived from Quadratic Unconstrained Binary Optimization (QUBO) problems, ansätze with low expressibility tend to yield superior results.

However, as the number of qubits increases, the correlation between expressibility and solution quality tends to weaken. This observation is consistent with theoretical findings that associate high expressibility with the emergence of barren plateaus, a phenomenon in which the gradient magnitude vanishes as the system size increases, leading to trainability issues in variational algorithms, regardless of whether gradient-based or gradient-free optimization methods are used.

Under noisy conditions, the relationship between Hamiltonian expressibility and solution accuracy changes significantly. Highly expressive circuits are typically the most affected by noise due to their increased complexity and circuit depth. As a result, for problems with basis state solutions and some problems with superposition state solutions, low-expressibility is advantageous, while for some problems with superposition state solutions, such as those involving the Heisenberg XXZ model, circuits with intermediate levels of expressibility tend to yield the best solution quality.
This behaviour likely arises from the need to balance exploration capability with resilience to noise.

Overall, our findings highlight the potential of Hamiltonian expressibility as a practical metric for evaluating ansatz suitability in VQE, under both ideal and noisy conditions, especially in small-scale scenarios.
This work lays the groundwork for further integration of expressibility-based metrics into circuit design strategies, including adaptive algorithms and machine learning-based techniques.

The remainder of this paper is structured as follows. In Section~\ref{sec:background}, we provide the necessary theoretical background, including an overview of Variational Quantum Algorithms, the barren plateau phenomenon, and different notions of ansatz expressibility.
Section~\ref{sec:methods} details the experimental methodology, describing the circuits and Hamiltonians used in the experiments, the procedure for estimating Hamiltonian expressibility, the VQE implementation setup, and the approach adopted for the correlation analysis.
In Section~\ref{sec:results}, we present and discuss the main results, focusing on the characteristics of Hamiltonian expressibility and its correlation with solution quality under both ideal and noisy conditions.
Finally, Section~\ref{sec:conclusions} concludes the paper and outlines potential directions for future work.

\section{Background}
\label{sec:background}

In this section, we briefly present the theoretical background underlying our work. We begin by introducing the class of Variational Quantum Algorithms (VQAs), focusing on two fundamental challenges: the barren plateau phenomenon and the ansatz selection problem. We then introduce the concept of circuit expressibility, which constitutes the main subject of this paper.

\subsection{Variational Quantum Algorithms}
\label{subsec:vqas}

Variational Quantum Algorithms (VQAs)~\citep{cerezo_2021} are a class of hybrid quantum-classical algorithms designed to mitigate the limitations of currently available noisy quantum hardware.
These algorithms rely on a parametrized quantum circuit whose parameters are iteratively optimized by a classical routine to minimize a problem-specific cost function.
Once the optimal parameters are found, the quantum circuit can be executed to generate a solution to the target problem.
Although VQAs do not necessarily guarantee a quantum advantage, they provide a practical framework for exploring the potential of near-term quantum devices and represent a viable approach for quantum computation in the NISQ era.

\subsubsection{Variational Quantum Eigensolver}

The Variational Quantum Eigensolver (VQE)~\citep{peruzzo_2014, tilly_2021} is one of the most widely adopted VQAs. It is designed to compute the ground state of a Hamiltonian operator $H$, \idest the eigenvector corresponding to its minimum eigenvalue.

VQE operates by using an ansatz circuit for preparing a parametrized quantum state $\ket{\psi(\theta)}$, where $\theta$ is a vector of tunable parameters.
The objective is to minimize the expectation value $\bra{\psi(\theta)} H \ket{\psi(\theta)}$, \idest the energy of the final quantum state, using a variational optimization procedure.
For brevity, we denote this expectation value as $\langle H \rangle$.

On real quantum hardware, $\langle H \rangle$ is not directly accessible.
Instead, an estimate $\langle H \rangle^*$ can be obtained by executing the circuit $n_{\text{shots}}$ times and averaging the resulting measurement outcomes $o_k$:
\begin{equation}
 \label{eq:expectation_estimate}
 \langle H \rangle^* = \frac{1}{n_{\text{shots}}} \sum_{k=1}^{n_{\text{shots}}} o_k.
\end{equation}

VQE has been primarily applied in quantum chemistry~\citep{cao_2019, mcclean_2016}, where it is used to estimate the ground-state energy of molecular systems. However, the algorithm is sufficiently general to be employed in a broader range of domains, including combinatorial optimization, where problems can be reformulated as ground-state searches of suitably constructed Hamiltonians.

\subsubsection{Barren Plateaus}

A key challenge in the optimization of variational quantum circuits is the phenomenon of \textit{barren plateaus}~\citep{mcclean_2018, arrasmith_2021, arrasmith_2022, cerezo_2021_bp, holmes_2022, larocca_2022, volkoff_2021}.
This effect occurs when the gradients of the cost function vanish exponentially with the number of qubits, resulting in flat regions in the optimization landscape.
In such cases, classical optimizers struggle to identify meaningful parameter updates, thereby hindering the convergence of the algorithm.

Formally, a cost function $C(\theta)$, with $\theta = (\theta_1, \ldots, \theta_p) \in \Theta$, is said to exhibit a barren plateau if, for all $i = 1, \ldots, p$, the following conditions hold:
\begin{align*}
 &\bullet\quad \mathbb{E}_{\theta}[\partial_i C(\theta)] = 0, \\
 &\bullet\quad \text{Var}_\theta[\partial_i C(\theta)] \leq F(n), \quad \text{with} \quad F(n) \in \mathcal{O}\left(\frac{1}{b^n}\right) \quad \text{for some} \quad b > 1,
\end{align*}
where $\partial_i C(\theta)$ denotes the partial derivative of the cost function with respect to parameter $\theta_i$, and $n$ is the number of qubits.

To understand the practical implications of barren plateaus, we recall Chebyshev’s inequality:
\[
P\left(|\partial_i C(\theta) - \mathbb{E}_{\theta}[\partial_i C(\theta)]| \geq \delta\right) \leq \frac{\text{Var}_\theta[\partial_i C(\theta)]}{\delta^2}.
\]

When the variance decays exponentially with $n$, the probability that any partial derivative $\partial_i C(\theta)$ deviates significantly from zero becomes vanishingly small.
As a consequence, optimization algorithms receive little to no useful information about the parameter update direction, significantly reducing their effectiveness, particularly as the system size increases.

\subsubsection{Ansatz Choice Problem}

Another central challenge in the design of VQAs lies in selecting a suitable parametrized quantum circuit, or \textit{ansatz}, for a given problem~\citep{sim_2019, qin_2023, wurtz_2021, du_2020}.

An effective ansatz should satisfy several criteria: it should minimize circuit depth to reduce the impact of noise and decoherence, leverage the native gate set of the hardware to simplify implementation, and exhibit robustness against the barren plateau phenomenon.

Crucially, the ansatz should also possess sufficient expressive power to explore the relevant portion of the solution space and identify high-quality solutions.
This expressive capacity, referred to as \textit{expressibility}, has been the subject of extensive research, which has introduced various metrics to quantify it.
A detailed investigation of expressibility constitutes the main focus of this work.

Commonly adopted ansätze can be broadly categorized into two families:
\begin{itemize}
 \item Hardware-efficient ansätze~\citep{kandala_2017}, which prioritize compatibility with current quantum devices by minimizing resource requirements.
 \item Problem-inspired ansätze~\citep{peruzzo_2014, farhi_2014}, which incorporate domain knowledge to tailor the circuit structure to the problem at hand.
\end{itemize}

While hardware-efficient ansätze offer practical feasibility, they may lack sufficient structure to yield high-quality solutions in complex problems.
Conversely, problem-inspired ansätze tend to achieve better results due to their alignment with the problem structure, but are more difficult to implement and may still suffer from barren plateaus if not carefully designed.

To address this trade-off, several adaptive approaches have been proposed. These include adaptive variational algorithms, which dynamically construct the ansatz during training by iteratively adding or removing gates~\citep{turati_2023, grimsley_2019, tang_2021, yordanov_2021, feniou_2023, zhu_2022, rattew_2020, chivilikhin_2020, las_heras_2016, cincio_2018, du_2022_adapt, bilkis_2023, ostaszewski_adapt_2021}, as well as reinforcement learning-based techniques~\citep{kuo_2021, zhu_2023, giordano_2022, pirhooshyaran_2021, ostaszewski_2021, fodera_2024}.

Despite the progress achieved through hardware-efficient, problem-inspired, and adaptive strategies, the ansatz selection problem remains an open challenge. Designing ansätze that simultaneously optimize expressibility, trainability, and hardware compatibility is still an active area of investigation.
In this work, we contribute to this ongoing research by investigating the role of Hamiltonian expressibility in guiding the selection of suitable ansätze for solving specific problems using variational quantum algorithms.

\subsection{Expressibility}
\label{subsec:expressibility}

Ansatz expressibility is a fundamental concept in quantum computing that describes the ability of a parametrized quantum circuit to explore the space of quantum states or the associated energy landscape.

The underlying intuition is that a highly expressive circuit is more likely to capture the optimal solution to a given problem, as it can explore a broader portion of the solution space.
However, circuits with high expressibility have been shown to be more susceptible to the barren plateau phenomenon~\citep{holmes_2022}, which leads to vanishing gradients and training difficulties, especially as the number of qubits increases.
Thus, understanding expressibility is crucial for guiding ansatz design, allowing to balance expressive power with trainability.

Several notions of expressibility have been proposed in the literature~\citep{wu_2021, funcke_2021, du_2022, ghosh_2023}.
In this work, we adopt a definition based on the ability of an ansatz to generate quantum states or associated energies uniformly over the entire solution space~\citep{sim_2019, holmes_2022}.

\subsubsection{Haar Measure and Expressibility}

Following prior studies, we quantify the uniformity in exploring the solution space using the Haar measure~\citep{meckes_2019, collins_2006, simon_1996}, which induces a uniform probability distribution over the unitary group $\mathcal{U}(d)$, consisting of all $d \times d$ unitary matrices~\citep{mele_2024}.
Sampling $d$-dimensional unitary operators from the Haar measure ensures that each element of $\mathcal{U}(d)$ is selected with equal probability.

In quantum computing, applying a unitary operator sampled from the Haar measure to a fixed reference state (typically $\ket{0}$) produces a quantum state that is itself Haar-distributed.
Consequently, the Haar measure induces a uniform distribution over quantum states in the Hilbert space. The expressibility of a parametrized quantum circuit can therefore be assessed by comparing the distribution of its output states to this Haar-induced distribution. The closer the match, the higher the expressibility of the circuit.

This approach was originally introduced in a problem-independent setting by~\citet{sim_2019}, and later extended to a problem-dependent formulation by~\citet{holmes_2022}.
To clarify the connection, we begin by presenting the general formulation in~\citet{holmes_2022}, from which two specific cases are derived.
The first corresponds to the original, problem-independent definition in~\citet{sim_2019}, while the second introduces a problem-dependent metric that constitutes the primary focus of this work.

Given a parametrized ansatz with associated unitary matrix $U$, we denote by $\mathcal{C}(U)$ the set of unitary operators generated by varying its parameters, \idest,
\[
\mathcal{C}(U) \coloneq \{ U(\theta) \mid \theta \in \Theta \}.
\]
To quantify the deviation of the distribution induced by $\mathcal{C}(U)$ from Haar uniformity, \citet{holmes_2022} introduced the following superoperator:
\begin{equation}
\label{eq:expressibility_superoperator}
 A^{(t)}_U(\cdot) = \int_{\mathcal{U}(d)} d\mu_H(V) V^{\otimes t} (\cdot) (V^\dagger)^{\otimes t} 
 - \int_{\mathcal{C}(U)} dW W^{\otimes t} (\cdot) (W^\dagger)^{\otimes t},
\end{equation}

where $V$ is a generic element of the unitary group $\mathcal{U}(d)$, with $d = 2^n$ for a system of $n$ qubits. The symbol $\mu_H$ denotes the Haar measure over $\mathcal{U}(d)$, while $W$ is a generic circuit sampled from the ensemble $\mathcal{C}(U)$, and $dW$ represents the measure induced by uniform sampling from the ansatz-generated ensemble $\mathcal{C}(U)$. If $A^{(t)}_U(\cdot) = 0$, then averaging over elements of $\mathcal{C}(U)$ agrees with averaging over elements of the Haar distribution over $\mathcal{U}(d)$ up to the t-th moment. For our purposes it is sufficient to consider the behaviour of $A^{(t)}_U(\cdot)$, for $t=2$. Henceforth, we drop the t-superscript, i.e., $A_U(\cdot) := A^{(2)}_U$.

By evaluating the superoperator $A_U(\cdot)$ on specific input operators and computing its 2-norm, we obtain different notions of expressibility.
Two commonly used choices are $\rho^{\otimes 2}$, where $\rho \coloneq \ket{0}\bra{0}$, and $H^{\otimes 2}$, where $H$ is the problem Hamiltonian.
These choices lead to two distinct metrics, referred to as \textit{state expressibility} and \textit{Hamiltonian expressibility}, defined as follows:
\begin{itemize}
 \item \textbf{State expressibility:} $\varepsilon^{\mathcal{S}}(U) \coloneq \norm{A_U(\rho^{\otimes 2})}_2$. 
 This metric is obtained by applying (\ref{eq:expressibility_superoperator}) to $\rho^{\otimes 2} = \ket{0}\bra{0}^{\otimes 2}$ and measures how uniformly the ansatz explores the Hilbert space of quantum states, which corresponds to the original formulation in~\citep{sim_2019}.
 
 \item \textbf{Hamiltonian expressibility:} $\varepsilon^{\mathcal{H}}(U, H) \coloneq \norm{A_U(H^{\otimes 2})}_2$. 
 This metric is obtained by applying (\ref{eq:expressibility_superoperator}) to $H^{\otimes 2}$ and quantifies how uniformly the ansatz explores the energy landscape defined by the problem Hamiltonian $H$. This metric is the primary focus of the present work.
\end{itemize}

\subsubsection{Frame Potential Formulation}

Hamiltonian expressibility can also be formulated in terms of frame potentials \citep{FramePots-Roberts2017}, which provide a convenient way to quantify how closely a distribution of unitaries approximates the Haar measure.

To this purpose, we define the \textit{ansatz-Hamiltonian frame potential} as:
\begin{equation}
 \label{eq:ansatz_ham_frame_pot}
 \mathcal{F}(U,H) \coloneq \int_{\Theta} \int_{\Theta} \mathrm{Tr}\left[ H U^{\dagger}(\theta') U(\theta'') H U^{\dagger}(\theta'') U(\theta') \right]^2 d\theta' d\theta'',
\end{equation}
where $\mathrm{Tr}(\cdot)$ denotes the trace operator.

The corresponding \textit{Haar-Hamiltonian frame potential} is given by:
\begin{equation}
 \label{eq:haar_ham_frame_pot}
 \mathcal{F}_{\text{Haar}}(H) \coloneq \int_{\mathcal{U}(d)} \int_{\mathcal{U}(d)} \mathrm{Tr}\left[ H W^{\dagger} V H V^{\dagger} W \right]^2 d\mu_H(W) d\mu_H(V),
\end{equation}
where the integrals are taken over the Haar measure $\mu_H$ on the unitary group $\mathcal{U}(d)$.

Using these definitions, Hamiltonian expressibility can be expressed as:
\begin{equation}
 \label{eq:ham_expr_frame_pots}
 \varepsilon^{\mathcal{H}}(U,H) = \sqrt{ \mathcal{F}(U,H) - \mathcal{F}_{\text{Haar}}(H)}.
\end{equation}

In particular, for an $n$-qubit system, the Haar-Hamiltonian frame potential admits a closed-form expression~\citep{holmes_2022}:
\begin{equation}
 \label{eq:haar_ham_frame_pot_solution}
 \mathcal{F}_{\text{Haar}}(H) = \frac{ \mathrm{Tr}[H]^4 + \mathrm{Tr}[H^2]^2 }{ 2^{2n} - 1 } - \frac{ 2 \mathrm{Tr}[H^2] \mathrm{Tr}[H]^2 }{ 2^n (2^{2n} - 1) }.
\end{equation}

By definition, $\mathcal{F}(U,H) \geq \mathcal{F}_{\text{Haar}}(H)$, with equality if and only if the ansatz is maximally Hamiltonian-expressive for $H$. Accordingly, a smaller value of $\varepsilon^{\mathcal{H}}(U,H)$ indicates higher Hamiltonian expressibility, with $\varepsilon^{\mathcal{H}}(U,H) = 0$ corresponding to the case of an ansatz with maximal expressibility.

\paragraph{Hamiltonian Expressibility Ratio}
Frame potentials can also be used to define an alternative expressibility-related metric that reduces sensitivity to the magnitude of the Hamiltonian norm. Following~\citep{holmes_2022}, we adopt the \textit{Hamiltonian expressibility ratio}, defined as:
\begin{equation}
 \label{eq:ham_ratio}
 \gamma^{\mathcal{H}}(U,H) \coloneq \frac{\mathcal{F}(U,H)}{\mathcal{F}_{\text{Haar}}(H)}.
\end{equation}

This ratio satisfies $\gamma^{\mathcal{H}}(U,H) \geq 1$, with equality if and only if the ansatz ensemble exactly matches the Haar distribution in terms of its frame potential. Lower values of $\gamma^{\mathcal{H}}$ thus indicate higher Hamiltonian expressibility.

\subsubsection{Expressibility and Barren Plateaus}

Theoretical studies~\citep{holmes_2022} have established a direct connection between circuit expressibility and the variance of the cost function gradient in variational quantum algorithms.
Specifically, under certain conditions, this variance is upper-bounded by both state and Hamiltonian expressibility, up to a constant factor.
As a result, circuits with low expressibility values (\idest highly expressive ansätze) tend to exhibit vanishing gradients, thereby hindering their trainability.

This observation highlights a critical trade-off: high expressibility enhances the circuit’s ability to explore the solution space effectively, but simultaneously increases the risk of barren plateaus.
Consequently, achieving optimal results requires a careful balance between exploration capability and trainability.

Determining when high expressibility is beneficial and when it hinders training remains an open research question.
While existing studies suggest that limiting expressibility can help mitigate barren plateaus and improve trainability, the optimal trade-off between exploration capability and result quality appears to be highly problem-dependent and is not yet fully understood.
In this work, we aim to contribute to this understanding by empirically investigating the relationship between Hamiltonian expressibility and the effectiveness of variational quantum algorithms.

\section{Methods}
\label{sec:methods}

In this section, we present the experimental methodology employed in our study.
We begin by describing the quantum circuits and problem instances under consideration.
Subsequently, we detail the Monte Carlo procedure used to estimate the Hamiltonian expressibility and the Hamiltonian expressibility ratio of a given ansatz with respect to each problem Hamiltonian.
Finally, we describe the specific choices for the implementation of the VQE algorithm and the approach used for correlation analysis.

\subsection{Circuits and Hamiltonians}
\label{subsec:circuits_and_hamiltonians}

In this section, we describe the circuits utilized in our analysis and offer a comprehensive overview of the Hamiltonian operators selected to ensure a diversified dataset for our study.

\subsubsection{Circuits}

We adopt the set of parametrized circuits originally introduced in~\citep{sim_2019}, which are inspired by earlier works, including~\citep{sousa_2006, romero_2017, johnson_2017, wilson_2019, kandala_2017, geller_2018, schuld_2020}. In particular, some of these circuits were originally proposed for applications such as quantum autoencoders, quantum error correction, and quantum classification algorithms, indicating that they cover a wide variety of applications. Each circuit in the set operates on $4$ qubits, with the number of layers ranging from $1$ to $5$. Given that the set comprises $19$ distinct circuit architectures, this yields a total of $95$ circuit instances evaluated and trained at the $4$-qubit level. All circuits are built with one- and two-qubit gates, both parametric and non-parametric. The parametric gates consist of single-qubit rotations ($R_X,R_Y,R_Z)$ and controlled rotations. A detailed description of the circuit structures is provided in Appendix~\ref{app:circuit_specifics}.

To investigate expressibility at larger system sizes, we extend these circuits to $8$ qubits by appropriately modifying their structure while preserving their original design principles. Specifically, the same structural patterns are maintained: for example, if a 4-qubit circuit includes a layer of single-qubit $X$ gates (one per qubit), the corresponding 8-qubit circuit features the same pattern applied to all qubits. Similarly, if the original circuit contains controlled rotations arranged in a ring topology, this structure is consistently mapped onto the 8-qubit system.
For this setting, we retain only a single layer per circuit in order to limit circuit depth and computational cost. This extension results in an additional $19$ circuits.

\subsubsection{Problem Hamiltonians}

Our dataset of Hamiltonian matrices includes approximately equal numbers of diagonal and non-diagonal instances, enabling us to explore potential differences between the two categories.

Diagonal Hamiltonians are characterized by having optimal solutions that are always basis states.
In contrast, non-diagonal Hamiltonians generally admit ground states that are superpositions, although this is not guaranteed; some non-diagonal Hamiltonians may still have basis states as their minimum-energy eigenvectors.

This distinction allows us to further classify Hamiltonians based on the nature of their ground states: we refer to those with basis state solutions as \textit{basis state} Hamiltonians, and those whose ground states are superpositions as \textit{superposition state} Hamiltonians.
Although this classification is determined a posteriori, since the nature of the ground state for non-diagonal Hamiltonians is not known in advance, it nonetheless offers a valuable perspective for interpreting the results of our analysis.

Within the diagonal category, we include well-known QUBO problems such as Maximum Cut, Minimum Vertex Cover, and Maximum Clique. The non-diagonal category features Hamiltonians derived from the Heisenberg XXZ model, the Transverse Field Ising model, and the Adiabatic problem. The detailed descriptions and theoretical formulations of these problems are provided in Appendix~\ref{app:notable_problems}.

In addition to these well-established Hamiltonians, we also generate random Hamiltonians, both diagonal and non-diagonal. This inclusion allows us to analyse a wider class of problem instances and to investigate expressibility-related phenomena in greater generality. Construction details and design rationale for these Hamiltonians are also provided in Appendix~\ref{app:notable_problems}.

The complete set of 4-qubit Hamiltonians used in our experiments is listed in Table~\ref{tab:4qb_problem_instances}, while Table~\ref{tab:8qb_problem_instances} presents the corresponding 8-qubit instances.

\begin{table}[ht]
\centering
\caption{Composition of the set of Hamiltonian matrices used for the experiments in the 4-qubit case}
\label{tab:4qb_problem_instances}
\begin{tabular}{|>{\raggedright\arraybackslash}m{2.85cm}|
 >{\centering\arraybackslash}m{1.85cm}|
 >{\raggedright\arraybackslash}m{7cm}|} 
\hline
\textbf{Problem Class} & \textbf{\# Instances} & \textbf{Implementation Details} \\
\hline
Maximum Cut & 10 & \vspace{0.1cm} Underlying graphs covering all possible 4-node topologies. \vspace{0.1cm} \\
\hline
Minimum Vertex Cover & 10 & \vspace{0.1cm}Penalty $p = 8$, covering all possible 4-node topologies \vspace{0.1cm}\\
\hline
Maximum Clique & 10 & \vspace{0.1cm}Underlying graphs covering all possible 4-node topologies. \vspace{0.1cm}\\
\hline
Random Diagonal & 60 &\vspace{0.1cm} 20 with elements sampled uniformly from the interval $[-10, 10]$, 20 with elements sampled from a normal distribution ($\mu = 0$, $\sigma = 4$), truncated to the interval $[-10, 10]$, 20 with elements sampled from a log-uniform distribution over $[a, b]$, where $a$ and $b$ are chosen such that the final values lie within $[-10, 10]$.\vspace{0.1cm} \\
\hline
Heisenberg XXZ& 23 & \vspace{0.1cm}Anisotropy parameter $\Delta \in \{ -1.5, \allowbreak -1, \allowbreak -0.5, 0.5, 1, 1.5 \}$ and coupling strengths $g$ determined by $g_c = 2(1 + \Delta)$: specifically, $g \in \{ -g_c, \allowbreak -g_c/2, \allowbreak g_c/2, \allowbreak g_c \}$ when $g_c \neq 0$, and $g \in \{-1, 0, 1\}$ when $g_c = 0$ \vspace{0.1cm}\\
\hline
Transverse Field Ising& 24 & \vspace{0.1cm} Interaction strengths $J \in \{ -1, -0.5, 0.5, 1 \}$ and transverse fields $g \in \{ -1.5, -1, -0.5, 0.5, 1, 1.5 \}$ \vspace{0.1cm} \\
\hline
Adiabatic& 36 & \vspace{0.1cm} Derived from the three previously mentioned QUBO problems, following the D-Wave schedule \tablefootnote{See the D-Wave schedule at \url{https://docs.dwavequantum.com/en/latest/index.html}} and adopting 4 graph topologies (4 nodes with 3, 4, 5, and 6 edges). \vspace{0.1cm}\\
\hline
Random Non-Diagonal& 60 & \vspace{0.1cm} Formed as linear combinations of Pauli string matrices with real coefficients. 20 with coefficients sampled uniformly from the interval $[-10, 10]$, 20 with coefficients sampled from a normal distribution ($\mu = 0$, $\sigma = 4$), truncated to the interval $[-10, 10]$, 20 with coefficients sampled from a log-uniform distribution over $[a, b]$, where $a$ and $b$ are chosen such that the final values lie within $[-10, 10]$. \vspace{0.1cm}\\
\hline
\end{tabular}
\end{table}

\begin{table}[ht]
\centering
\caption{Composition of the set of Hamiltonian matrices used for the experiments in the 8-qubit case}
\label{tab:8qb_problem_instances}
\begin{tabular}{|>{\raggedright\arraybackslash}m{2.85cm}|
 >{\centering\arraybackslash}m{1.85cm}| 
 >{\raggedright\arraybackslash}m{7cm}|} 
\hline
\textbf{Problem Class} & \textbf{\# Instances} & \textbf{Implementation Details} \\
\hline
Maximum Cut & 20 & \vspace{0.1cm} Built from four distinct 8-node topologies: 10 Erdős-Rényi graphs with edge probability $p \in \{0.3, 0.7\}$, 5 3-regular graphs, and 5 Barabási–Albert graphs with $m=2$ \vspace{0.1cm} \\
\hline
Minimum Vertex Cover & 20 & \vspace{0.1cm}Penalty $p = 8$, built from four distinct 8-node topologies: 10 Erdős-Rényi graphs with edge probability $p \in \{0.3, 0.7\}$, 5 3-regular graphs, and 5 Barabási–Albert graphs with $m=2$ \vspace{0.1cm}\\
\hline
Maximum Clique & 20 & \vspace{0.1cm}Built from four distinct 8-node topologies: 10 Erdős-Rényi graphs with edge probability $p \in \{0.3, 0.7\}$, 5 3-regular graphs, and 5 Barabási–Albert graphs with $m=2$ \vspace{0.1cm}\\
\hline
Random Diagonal & 60 &\vspace{0.1cm} 20 with elements sampled uniformly from the interval $[-10, 10]$, 20 with elements sampled from a normal distribution ($\mu = 0$, $\sigma = 4$), truncated to the interval $[-10, 10]$, 20 with elements sampled from a log-uniform distribution over $[a, b]$, where $a$ and $b$ are chosen such that the final values lie within $[-10, 10]$.\vspace{0.1cm} \\
\hline
Heisenberg XXZ & 23 & \vspace{0.1cm} Anisotropy parameter $\Delta \in \{ -1.5, \allowbreak -1, \allowbreak -0.5, 0.5, 1, 1.5 \}$ and coupling strengths $g$ determined by $g_c = 2(1 + \Delta)$: specifically, $g \in \{ -g_c, \allowbreak -g_c/2, \allowbreak g_c/2, \allowbreak g_c \}$ when $g_c \neq 0$, and $g \in \{-1, 0, 1\}$ when $g_c = 0$ \vspace{0.1cm}\\
\hline
Transverse Field Ising & 24 & \vspace{0.1cm} Interaction strengths $J \in \{ -1, -0.5, 0.5, 1 \}$ and transverse fields $g \in \{ -1.5, -1, -0.5, 0.5, 1, 1.5 \}$ \vspace{0.1cm} \\
\hline
Adiabatic & 36 & \vspace{0.1cm} Derived from the three previously mentioned QUBO problem, following the D-Wave schedule and adopting 4 graph topologies (4 nodes with 3, 4, 5, and 6 edges). \vspace{0.1cm}\\
\hline
Random Non-Diagonal & 60 & \vspace{0.1cm} Formed as linear combinations of Pauli string matrices with real coefficients. 20 with coefficients sampled uniformly from the interval $[-10, 10]$, 20 with coefficients sampled from a normal distribution ($\mu = 0$, $\sigma = 4$), truncated to the interval $[-10, 10]$, 20 with coefficients sampled from a log-uniform distribution over $[a, b]$, where $a$ and $b$ are chosen such that the final values lie within $[-10, 10]$. \vspace{0.1cm}\\
\hline
\end{tabular}
\end{table}

\subsection{Expressibility Estimation}
\label{subsec:expressibility_estimation}

To estimate the Hamiltonian expressibility and the Hamiltonian expressibility ratio, we adopt a Monte Carlo method inspired by~\citep{sim_2019}, originally developed for computing state expressibility.

Both quantities rely on the ansatz-Hamiltonian frame potential $\mathcal{F}^{\mathcal{H}}(U, H)$ and the Haar frame potential $\mathcal{F}^{\mathcal{H}}_{\text{Haar}}(H)$, defined in~(\ref{eq:ham_expr_frame_pots}) and~(\ref{eq:ham_ratio}), respectively.
Since a closed-form expression exists for $\mathcal{F}^{\mathcal{H}}_{\text{Haar}}(H)$ (see~(\ref{eq:haar_ham_frame_pot_solution})), it is sufficient to estimate $\mathcal{F}^{\mathcal{H}}(U, H)$ in order to compute both metrics.

We employ a Monte Carlo sampling strategy with sample size $k$. To account for the inherent statistical uncertainty of the method, we also report confidence intervals for the estimated quantities.

Given a parametrized circuit $U(\theta)$ and a fixed Hamiltonian $H$, the procedure is as follows:

\begin{enumerate}

 \item \textbf{Sampling and Circuit Generation:} Generate $2k$ random parameter sets:
 \[
 \left\{ \theta^1, \theta^2, \ldots, \theta^{2k} \right\},
 \]
 and construct the corresponding unitaries:
 \[
 \left\{ U(\theta^i) \mid i = 1, \ldots, 2k \right\}.
 \]

 \item \textbf{Trace Evaluation:} Construct $k$ independent pairs $(\theta^i, \theta^{k+i})$ and evaluate:
 \[
 \text{Trace}_i = \left( \operatorname{Tr}\left[ H U^\dag(\theta^i) U(\theta^{k+i}) H U^\dag(\theta^{k+i}) U(\theta^i) \right] \right)^2.
 \]

 \item \textbf{Frame Potential and Uncertainty Estimation:} Estimate the ansatz-Hamiltonian frame potential by averaging the trace values, compute the corresponding standard deviation, and compute an estimate of the width of the confidence interval for the frame potential, necessary for a control on the quality of the estimate\footnote{We choose a $l = 0.99$ confidence level for $\tilde{\mathcal{F}}^{\mathcal{H}}(U, H)$, using the $t$-distribution with $k-1$ degrees of freedom. $t^{*}(\alpha)$ denotes the critical $t$-value at significance level $\alpha = \frac{1}{2}(1 + \frac{l}{100})$ for $k-1$ degrees of freedom.}:
 
 \begin{align*}
 & \quad \tilde{\mathcal{F}}(U, H) \coloneq \frac{1}{k} \sum_{i=1}^{k} \text{Trace}_i, \\
 & \quad \tilde{\sigma} \coloneq \sqrt{ \frac{1}{k} \sum_{i=1}^{k} \left( \text{Trace}_i - \tilde{\mathcal{F}}(U,H) \right)^2 }, \\
 & \quad \text{err} \coloneq t^*(0.995) \cdot \tilde{\sigma}.
 \end{align*}

 \item \textbf{Metric Estimation:} Estimate Hamiltonian expressibility and Hamiltonian expressibility ratio, and provide the respective confidence intervals.
 \begin{itemize}
 \item \textbf{Hamiltonian Expressibility\footnote{The use of the $\max$ function in the confidence interval computation ensures the lower bound is non-negative when subtracting the error term.}:}
 \begin{align}
 \label{eq:ham_expr_final_estimate}
 & \quad \tilde{\varepsilon}^{\mathcal{H}}(U, H) \coloneq \sqrt{\tilde{\mathcal{F}}(U,H) - \mathcal{F}_{\text{Haar}}(H) },\\
 & \quad \Bigg[
 \sqrt{ \max\left\{ \tilde{\mathcal{F}}(U,H)  - \text{err}, \mathcal{F}_{\text{Haar}}(H)  \right\} - \mathcal{F}_{\text{Haar}}(H)}, \nonumber \\
 & \quad\sqrt{ \tilde{\mathcal{F}}(U,H)  + \text{err} - \mathcal{F}_{\text{Haar}}(H) } \Bigg]. 
 \end{align}

 \item \textbf{Hamiltonian Expressibility Ratio\footnote{The use of the $\max$ function in the confidence interval computation ensures the lower bound does not fall below the theoretical minimum value of $1$.}:}
 \begin{align}
 \label{eq:ham_expr_ratio_final_estimate}
 & \quad\tilde{\gamma}^{\mathcal{H}}(U, H) \coloneq \frac{ \tilde{\mathcal{F}}(U,H) }{ \mathcal{F}_{\text{Haar}}(H)  },\\
 & \quad\left[
 \frac{ \max\left\{ \tilde{\mathcal{F}}(U,H)  - \text{err}, \mathcal{F}_{\text{Haar}}(H)  \right\} }{ \mathcal{F}_{\text{Haar}}(H) }, 
 \frac{ \tilde{\mathcal{F}}(U,H)  + \text{err} }{ \mathcal{F}_{\text{Haar}}(H)  }
 \right].
 \end{align}
 \end{itemize}
\end{enumerate}

The choice of the sample size $k$ is critical and discussed in Appendix~\ref{app:sample_size}.

\subsubsection{Thresholds for Maximal Expressibility}
Due to the finite nature of sampling, a statistical bias is inherently introduced and must be accounted for.
In particular, even when unitary operators are sampled from the Haar distribution, the empirical estimates $\tilde{\varepsilon}^{\mathcal{H}}_{\text{Haar}}(H)$ and $\tilde{\gamma}^{\mathcal{H}}_{\text{Haar}}(H)$ may deviate from their theoretical values of $0$ and $1$, respectively.

To mitigate this issue, we adopt a thresholding approach: any circuit for which $\tilde{\varepsilon}^{\mathcal{H}}(U, H) \leq \tilde{\varepsilon}^{\mathcal{H}}_{\text{Haar}}(H)$ or, equivalently, $\tilde{\gamma}^{\mathcal{H}}(U, H) \leq \tilde{\gamma}^{\mathcal{H}}_{\text{Haar}}(H)$ is considered maximally expressive. Further justification for this procedure is provided in Appendix~\ref{app:thresholds}.

In rare cases, sampling fluctuations may lead to $\tilde{\mathcal{F}}^{\mathcal{H}}(U, H) < \mathcal{F}^{\mathcal{H}}_{\text{Haar}}(H)$, resulting in an undefined value for $\tilde{\varepsilon}^{\mathcal{H}}(U, H)$ and in $\tilde{\gamma}^{\mathcal{H}}(U, H) < 1$, which is theoretically inadmissible.
To address these anomalies, we introduce minor adjustments to both metrics, as detailed in Appendix~\ref{app:thresholds}.

\subsection{VQE Execution}
\label{subsec:vqe_execution}

This section presents the implementation details and methodological choices adopted for executing the VQE algorithm to solve the optimization problems associated with the Hamiltonians described in Section~\ref{subsec:circuits_and_hamiltonians}.

To assess the practical relevance of Hamiltonian expressibility in realistic scenarios, we implement the VQE algorithm using a hypothetical quantum hardware model that reflects constraints typically encountered in commercially available quantum devices. Specifically, we limit the set of available quantum gates and adopt a non–fully connected qubit topology inspired by existing hardware architectures.

Our analysis comprises two distinct scenarios. The first assumes an ideal, noise-free environment, while the second incorporates realistic quantum noise sources, including decoherence effects due to limited qubit coherence times, single- and two-qubit gate error rates, and measurement errors.

The hardware topology and supported gate set, along with specifications common to both scenarios, are described in Section~\ref{subsubsec:execution_specifics}. Additional considerations specific to the noisy setting are provided in Section~\ref{subsubsec:noisy_setting_specifics}.

\subsubsection{Execution Specifics}
\label{subsubsec:execution_specifics}

The VQE algorithm is executed in both noiseless and noisy simulation environments. In both cases, $n_{\text{shots}}=1000$ are used to estimate the expectation value of each final circuit state\footnote{The VQE implementation is based on the \texttt{qiskit\_algorithms} library: \url{https://github.com/qiskit-community/qiskit-algorithms}}. 
Each ansatz from the dataset introduced in Section~\ref{subsec:circuits_and_hamiltonians} is evaluated on all the problem instances described in the same section. For each ansatz–problem pair, we perform 50 independent optimization runs to account for stochastic variability in the training process. 

In order to rule out potential dependencies or biases arising from the choice of a specific optimizer and to obtain more robust results, we employed five different optimizers. In particular, we used three gradient-free optimizers: COBYLA~\citep{powell_1994}, Powell \citep{Powell_1964}, and Nelder–Mead \citep{Nelder_1965}; and two gradient-based optimizers: SLSQP \citep{Kraft_1988}, and L-BFGS-B \citep{Byrd_1995}. All optimizers are used to minimize the expectation value of the target Hamiltonian. The results obtained from different optimizers are then aggregated following a procedure described later in this section.
\begin{figure}[!ht]
 \centering
 \includegraphics[width=0.4\textwidth]{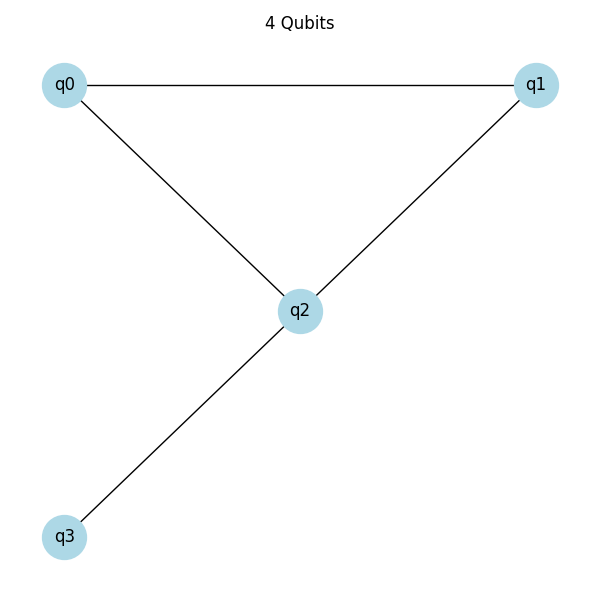}\hfill
 \includegraphics[width=.4\textwidth]{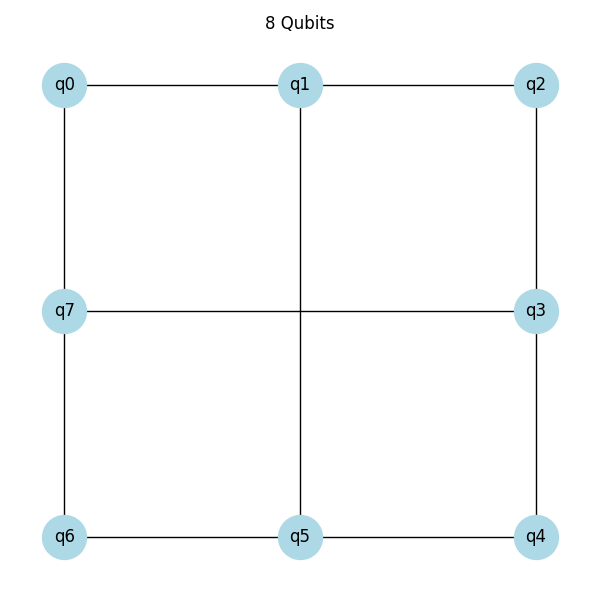}\hfill
 \caption{Topologies selected for the VQE simulations in the 4-qubit case (left) and 8-qubit case (right).}
 \label{fig:topologies}
\end{figure}

To better reflect the impact of realistic hardware constraints, we adopt non–fully connected qubit topologies for both the 4- and 8-qubit configurations, as shown in Figure~\ref{fig:topologies}. These topologies are inspired by the architecture of commercially available quantum processors and account for the limited qubit connectivity that characterizes current NISQ devices. In addition, we restrict the available gate set to the default basis gates defined by Qiskit, namely $Id$, $R_z$, $\sqrt{X}$, $X$, and $CX$. This choice ensures compatibility with typical hardware-native gate sets and facilitates comparisons under realistic constraints.

\paragraph{Approximation Ratio}

To assess solution quality, we adopt the normalized approximation ratio A.R., defined by:
\begin{equation}
 \label{eq:normalized_approx_ratio}
 \text{A.R.} \coloneq \frac{\langle H \rangle^* - \langle H \rangle_{\text{max}}}{\langle H \rangle_{\text{min}} - \langle H \rangle_{\text{max}}},
\end{equation}
where $\langle H \rangle_{\text{min}}$ and $\langle H \rangle_{\text{max}}$ are the minimum and maximum achievable expectation values of the Hamiltonian, respectively, and $\langle H \rangle^*$ is the estimated expectation value defined in~(\ref{eq:expectation_estimate}), obtained from circuit sampling.

This metric approaches $1$ when $\langle H \rangle^*$ is close to $\langle H \rangle_{\text{min}}$, indicating that the circuit is effectively sampling states with near-optimal energies.
To ensure that the ratio is well-defined and bounded within the interval $[0, 1]$, the quantity $\langle H \rangle_{\text{max}}$ is subtracted from both the numerator and the denominator.
Notably, in the specific case of the Maximum Cut problem, we have $\langle H \rangle_{\text{max}} = 0$.

We emphasize that this metric is used exclusively for benchmarking purposes, as it requires prior knowledge of both the exact minimum and maximum values of the objective function, which are generally unavailable in practical applications.

For each circuit–problem pair $(U, H)$, each of the 5 optimizers is used to perform 10 independent optimization runs, corresponding to a total of 50 optimizations per circuit–problem pair. Out of these 50 runs, we first discard the 10 with the worst normalized approximation ratio, in order to discard unsuccessful optimization runs due to bad initializations, and then report the average normalized approximation ratio across the remaining 40 optimization runs, denoted $\text{A.R.}(U, H)$, together with the corresponding standard deviation.

\subsubsection{Noisy Setting Specifics}
\label{subsubsec:noisy_setting_specifics}

In the noisy setting, the analysis is restricted to 4-qubit systems.
We employ a Qiskit noise model designed to emulate a generic quantum backend, incorporating realistic noise sources such as qubit decoherence, gate infidelities, and readout errors.
Specifically, the noise level is controlled by four parameters: the relaxation time $T_1$, the dephasing time $T_2$, the error rate of single-qubit gates and measurements $err_1$, and the error rate of two-qubit gates $err_2$. For simplicity, we assume $T_1 = T_2$ in all simulations and impose $err_2 = 25 \times err_1$.
Both assumptions are realistic and commonly observed in current quantum hardware \citep{löschnauer2024, Shi_2022, AbuGhanem2025, Evered2023}. In our main analysis, we set the noise parameters as follows: $T_1 = 200 \mu s$, $T_2 = 200 \mu s$, $err_1 = 1.6 \times 10^{-4}$, $err_2 = 4 \times 10^{-3}$.

\paragraph{Error Rate Metric}

Each of the parameters mentioned above affects the level of noise present during the execution of the VQE algorithm and, consequently, introduces a certain amount of error into the final results. By adjusting these four parameters together, it is possible to simulate different noise levels, each of which impacts the simulation outcomes in distinct ways. To effectively characterize and control the overall noise in the system, it is useful to define a single, unified metric that reflects the backend's error rate and depends on these parameters. By calculating the value of this metric, we can then estimate the amount of noise introduced during execution, and tune the parameters accordingly in order to gradually add in noise.

In our case, the error rate is determined by the default noise model generated by the GenericBackendV2 class in Qiskit. When creating the backend, we can specify the noise parameters, $T_1$, $T_2$, $err_1$, and $err_2$, which are used to construct a noise model based on their values. In the following, we describe how Qiskit builds this noise model in order to derive a single, representative metric for the backend’s error rate.

First, for each type of gate $g$ supported by the backend, the noise model specifies the corresponding type of noise to be applied during circuit execution. Two scenarios are possible:
\begin{itemize}
 \item The gate is deterministically affected by a quantum noise channel, represented by a set of Kraus operators. In this case, the effective operation is a fixed noisy version of the gate.
 \item Alternatively, the gate is randomly sampled from a finite set of noisy variants according to a probability distribution. This set includes the ideal gate $g$, and some perturbed versions.
\end{itemize}

This choice, deterministic or probabilistic, is fixed for each gate $g$ and remains the same every time the gate is applied within a given execution. However, the specific Kraus operators and distribution of noisy gates depend on the qubits the gate acts upon, meaning that the same logical gate may exhibit different noise models if applied to different target qubits. It follows that, if $g$ is a single-qubit gate, a different noise profile, whether represented by a set of Kraus operators or by a probability distribution over noisy variants, is defined for every possible target qubit $q_i$. Likewise, if $g$ is a two-qubit gate, the noise model, either Kraus-based or probabilistic, varies depending on the specific qubit pair (i, j) $g$ acts upon. We note that, since many of the gates used in our circuits are controlled two-qubit gates, the noise model also depends on which qubit has the role of control and which has the role of target. This means that a two-qubit gate $g$ will have different noise models when applied as $g(q_i,q_j)$ compared to when applied as $g(q_j,q_i)$.

From now on we will denote with ${g}_k(g, q_i)$ the Kraus version of the theoretical single-qubit gate $g$ acting on qubit $q_i$ and with $P(g, q_i)$ the probability of sampling and implementing the ideal gate $g$, out of the set of noisy versions. Similarly, we will denote with ${g}_k(g,q_i,q_j)$ the Kraus version of the theoretical two-qubit gate $g$ acting on qubits $q_i$ and $q_j$, and with $P(g, q_i,q_j)$ the probability of sampling and implementing the ideal gate $g$, out of the set of noisy versions.

Based on this information, we can then define a quantity that describes the accuracy of the generic single-qubit gate $g$ when applied to qubit $q_i$.

\begin{equation}
\eta_1(g, q_i) \coloneq
\begin{cases}
F(g, g_k(g, q_i)) & \text{if Kraus-induced noise}, \\
P(g, q_i) & \text{otherwise},
\end{cases}
\label{eq:etagate_1qb}
\end{equation}

And similarly, the accuracy of the generic two-qubit gate $g$ when applied to qubits $q_i,q_j$.

\begin{equation}
\eta_{2}(g,(q_i,q_j)) \coloneq
\begin{cases}
F(g, g_k(g,q_i,q_j)) & \text{if Kraus-induced noise}, \\
P(g,q_i,q_j) & \text{otherwise},
\end{cases}
\label{eq:etagate_2qb}
\end{equation}

where $F(\cdot ,\cdot)$ denotes the \textit{average gate fidelity} \citep{Bowdrey2002}, in our case between the ideal gates $g^1$ or $g^2$ and their Kraus noisy implementation.

At this point, we can define the average gate accuracy. Denoting by $N$ the total number of gates applicable on the backend, which depends on the set of available gates and the topology, with $G_1$ and $G_2$ the sets of available single and two-qubit gates respectively, with $n$ the number of qubits and $E$ the set of edges in the topology, it is defined as:

\begin{equation}
\eta \coloneq \frac{1}{N} \left( \sum_{g \in G_1} \sum_{i=0}^{n-1} \eta_1(g,q_i) + \sum_{g \in G_2} \sum_{(i,j) \in E} \eta_2(g,(q_i, q_j)) \right),
\label{eq:eta_new}
\end{equation}

In addition to gate noise, each qubit’s measurement is subject to readout error. This is modelled by a set of conditional probabilities describing the likelihood of observing each possible measurement outcome, given the actual state of each qubit $q_i$
 \[
 \begin{aligned}
 &P(M(q_i) = 0 \mid q_i = |0\rangle), \quad P(M(q_i) = 0 \mid q_i = |1\rangle), \\
 &P(M(q_i) = 1 \mid q_i = |0\rangle), \quad P(M(q_i) = 1 \mid q_i = |1\rangle).
 \end{aligned}
 \]

Through these values, the definition of conditional probability, and assuming for simplicity that over all measurements in each execution we have $P(q_i=\ket{0}) = P(q_i=\ket{1})=\frac{1}{2}$ for each qubit $q_i$, we can define $\delta$, which denotes the average measurement accuracy across the $n$ qubits:

\begin{equation}
\delta \coloneq \prod_{i=1}^n \frac{1}{2}\sum_{k \in \{0,1\}} P(M(q_i) = k \mid q_i = |k\rangle).
\label{eq:delta}
\end{equation}

At this point, we combine the information on gate fidelity and measurement accuracy to define a metric for the average error rate $\text{Err}$ of a simulated backend configured with parameters $T_1$, $T_2$, $err_1$, and $err_2$ as:

\begin{equation}
\text{Err} \coloneq 1 - (\eta \times \delta),
\label{eq:err}
\end{equation}

This metric enables us, during the simulation phase, to analyse how the results of the correlation analysis discussed in Section~\ref{subsec:corr_analysis} evolve as the average error rate $\text{Err}$ increases.
In particular, it allows us to investigate whether and how the correlation coefficients change as noise is gradually introduced, transitioning from the ideal case to the noisy setting described at the beginning of this section.

\subsection{Correlation Analysis}
\label{subsec:corr_analysis}

To investigate potential relationships between Hamiltonian expressibility (or Hamiltonian expressibility ratio) and the quality of the solutions obtained using the VQE algorithm described in Section~\ref{subsec:vqe_execution}, we compute a set of correlation coefficients and mutual information scores.

\subsubsection{Measures of Correlation and Mutual Information}

We briefly review the statistical measures employed, each defined for a pair of general metrics evaluated over a dataset. The sets of values corresponding to each metric across the dataset are denoted by the vectors $x$ and $y$.

\begin{itemize}

\item \textbf{Pearson correlation coefficient}~\citep{Pearson-Pearson1896}, denoted $r_p(x, y)$, quantifies the strength and direction of linear relationships. Values range from $1$ (perfect positive linear correlation) to $-1$ (perfect negative linear correlation), with $0$ indicating no linear dependence.

\item \textbf{Spearman correlation coefficient}~\citep{spearman_1904}, denoted $r_s(x, y)$, captures monotonic relationships by computing the Pearson coefficient on ranked variables. A value of $1$ indicates a strictly increasing monotonic trend, $-1$ indicates a strictly decreasing trend, and $0$ suggests the absence of any monotonic relationship.

\item \textbf{Kendall’s Tau}~\citep{kendall_1938}, denoted $\tau(x, y)$, also evaluates monotonic associations but is based on the count of concordant and discordant pairs\footnote{A pair of observations $(x_i, y_i)$ and $(x_j, y_j)$ is said to be \textit{concordant} if the orderings of both $x$ and $y$ agree, \idest if $(x_j - x_i)(y_j - y_i) > 0$. 
Conversely, the pair is \textit{discordant} if the orderings disagree, \idest if $(x_j - x_i)(y_j - y_i) < 0$.}
Kendall’s Tau is generally considered more robust to ties and outliers compared to Spearman’s coefficient.

\item \textbf{Mutual information}~\citep{cover_2006}, denoted $I(x, y)$, measures the amount of shared information between two variables. It is always non-negative and unbounded above. A mutual information value close to zero implies near independence, while larger values indicate stronger dependence.

\end{itemize}

\subsubsection{Correlation Analysis Specifics}

For each ansatz $U(\theta)$ and each problem Hamiltonian $H$ from the datasets described in Section~\ref{subsec:circuits_and_hamiltonians}, we compute the Pearson, Spearman, and Kendall Tau correlation coefficients between:
\begin{itemize}
 \item the Hamiltonian expressibility $\varepsilon^{\mathcal{H}}(U, H)$ and the average normalized approximation ratio $\text{A.R.}(U, H)$,
 \item the Hamiltonian expressibility ratio $\gamma^{\mathcal{H}}(U, H)$ and $\text{A.R.}(U, H)$.
\end{itemize}

The results are aggregated across problem instances belonging to the same problem class by reporting the mean and standard deviation of each correlation coefficient. The problem classes, as introduced in Section~\ref{subsec:circuits_and_hamiltonians}, include Maximum Cut, Minimum Vertex Cover, Maximum Clique, Random Diagonal, Heisenberg XXZ, Transverse Ising Field, Adiabatic, and Random Non-Diagonal.

To facilitate a higher-level interpretation of the results, we further group the Hamiltonians according to two broad criteria:
\begin{itemize}
 \item \textbf{Diagonal vs. non-diagonal}: this criterion distinguishes between Hamiltonians that are diagonal in the computational basis (\eg Maximum Cut, Minimum Vertex Cover, Maximum Clique, Random Diagonal) and those that are not diagonal (\eg Heisenberg XXZ, Transverse Ising Field, Adiabatic, Random Non-Diagonal).
 \item \textbf{Basis state vs. superposition ground states}: this classification depends on whether the ground state is expected to be a basis state (as is always the case for diagonal Hamiltonians and some instances of Heisenberg XXZ) or a superposition state (non-diagonal Hamiltonians, excluding specific Heisenberg instances).
\end{itemize}

Lastly, we compute mutual information scores between $\varepsilon^{\mathcal{H}}(U, H)$ (or $\gamma^{\mathcal{H}}(U, H)$) and $\text{A.R.}(U, H)$ to quantify the amount of information each expressibility metric provides about VQE results.
This complementary analysis captures non-linear and non-monotonic dependencies and offers an additional perspective on the predictive relevance of each expressibility measure.

\section{Results and Discussion}
\label{sec:results}
In this section, we present our findings. The experiments were split into two stages. In the first stage, the Hamiltonian expressibility and the Hamiltonian expressibility ratio were estimated for each 4- and 8-qubit circuit across all Hamiltonians in our dataset. 

In the second stage, the VQE algorithm was executed on the same set of circuits and Hamiltonians under different settings. In the ideal setting, 95 circuits with 4 qubits and 19 circuits with 8 qubits were optimized over 233 and 263 different Hamiltonians, respectively, using the VQE algorithm. As described in Section~\ref{subsec:vqe_execution}, each optimization involved 10 independent runs with 5 different optimizers. Overall, the ideal setting required more than 1.3 million VQE simulations. In the noisy setting, 95 circuits with 4 qubits were optimized over 233 different Hamiltonians using the VQE algorithm. In total, this setting required more than 1.1 million VQE noisy simulations.

Section~\ref{subsec:hamiltonian_expressibility_analysis} discusses the estimates of Hamiltonian expressibility and Hamiltonian expressibility ratio, while Section~\ref{subsec:relationship_with_approximation_ratio} examines the correlations between these metrics, introduced in Section~\ref{subsec:expressibility}, and the average normalized approximation ratio defined in Section~\ref{subsec:vqe_execution}.

\subsection{Hamiltonian Expressibility Analysis}
\label{subsec:hamiltonian_expressibility_analysis}
This section presents our findings on the estimation of the expressibility metrics for all circuits and problem classes from Section~\ref{subsec:circuits_and_hamiltonians}, focusing on layer dependence and circuit expressiveness per problem class.

\subsubsection{Relationship with Depth}
\label{subsubsec:relationship_with_depth}
\begin{figure}[!ht]
 \includegraphics[width=.5\textwidth]{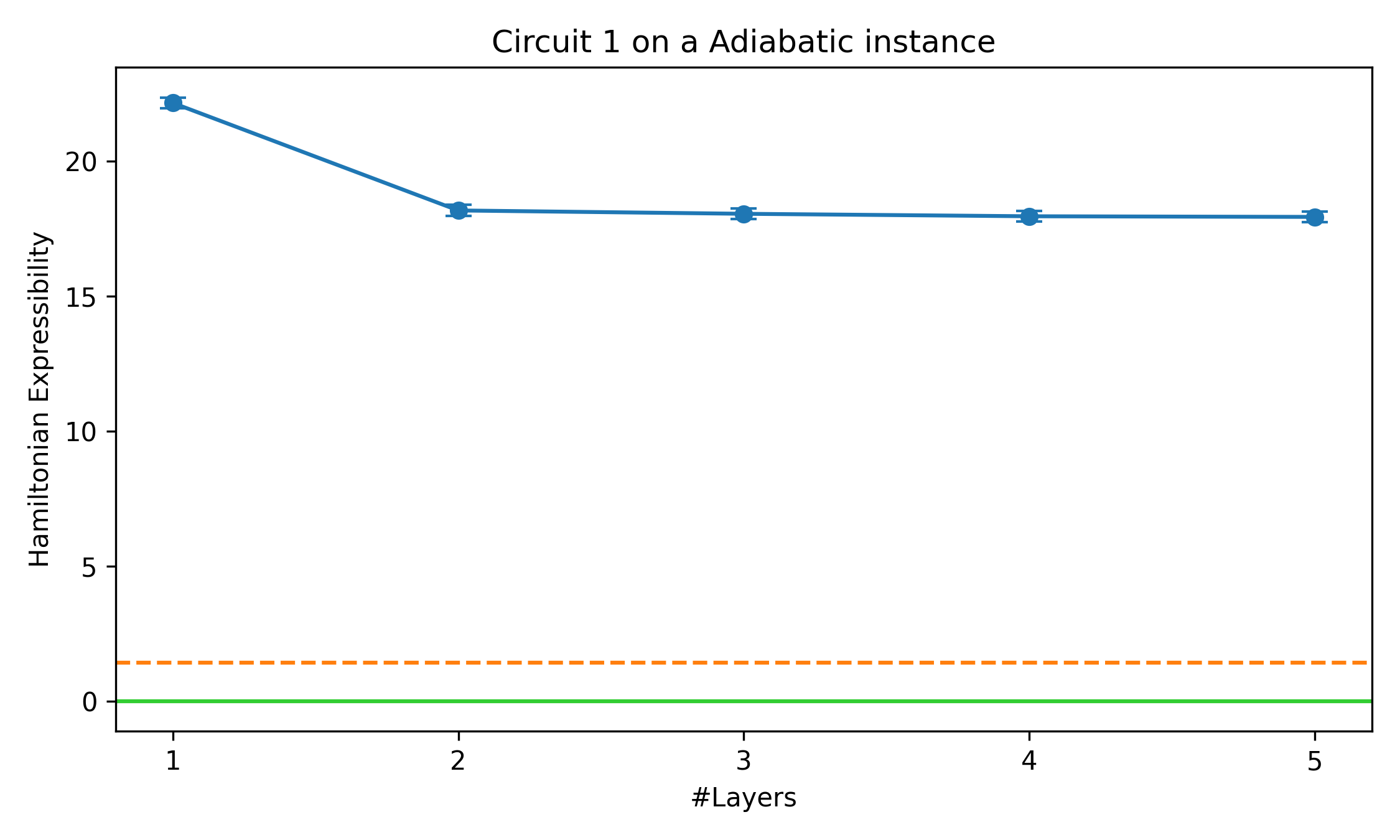}\hfill
 \includegraphics[width=.5\textwidth]{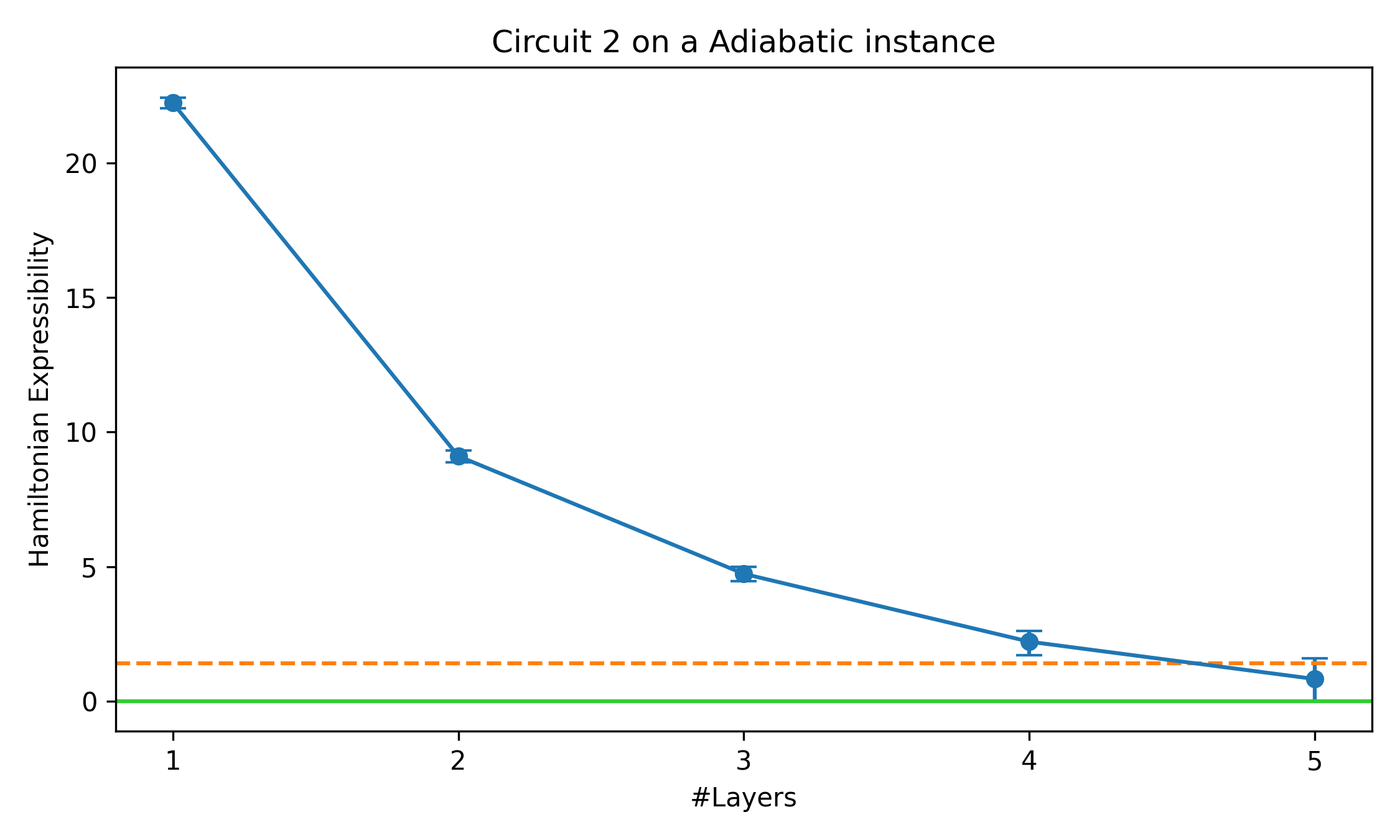}\hfill
 \\[\smallskipamount]
 \includegraphics[width=.5\textwidth]{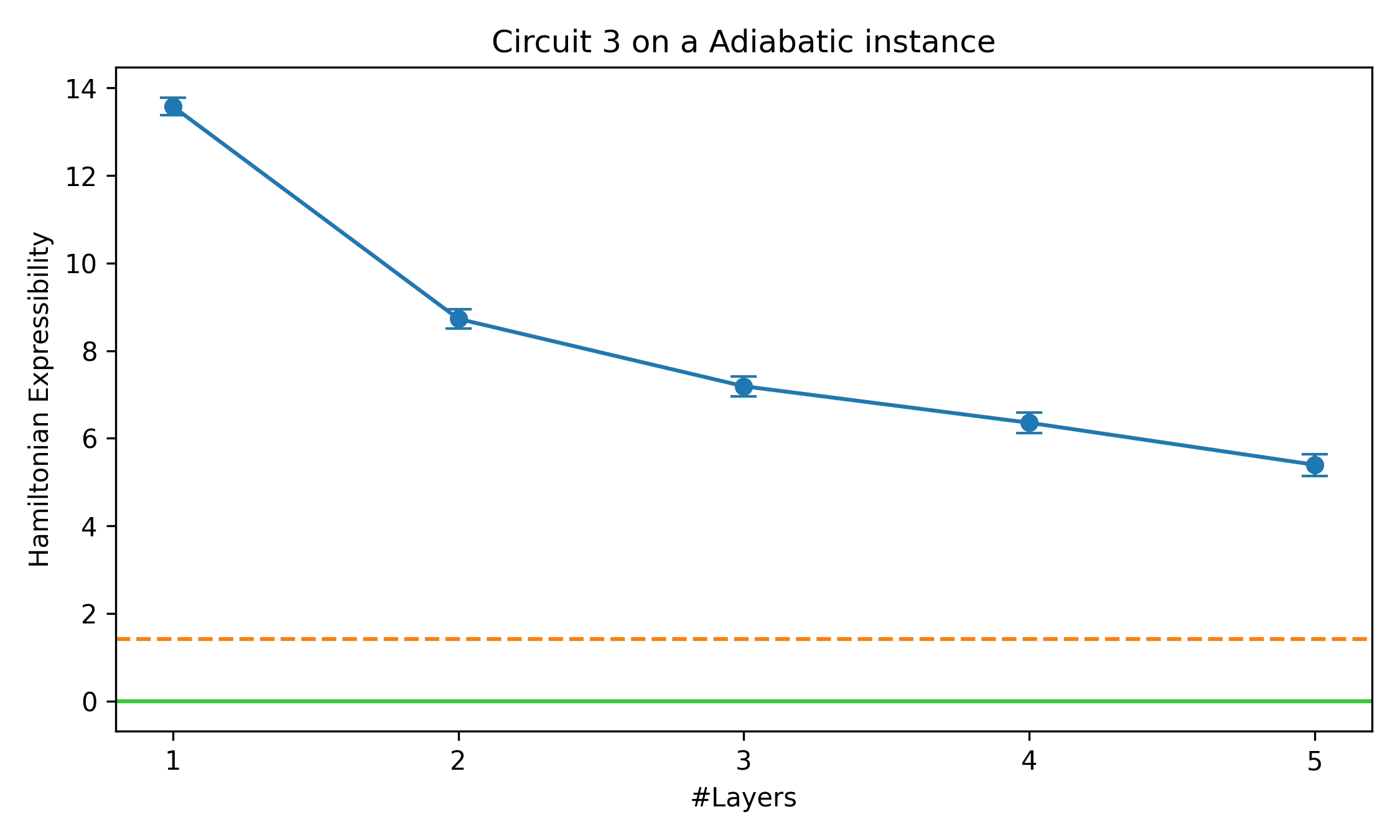}\hfill
 \includegraphics[width=.5\textwidth]{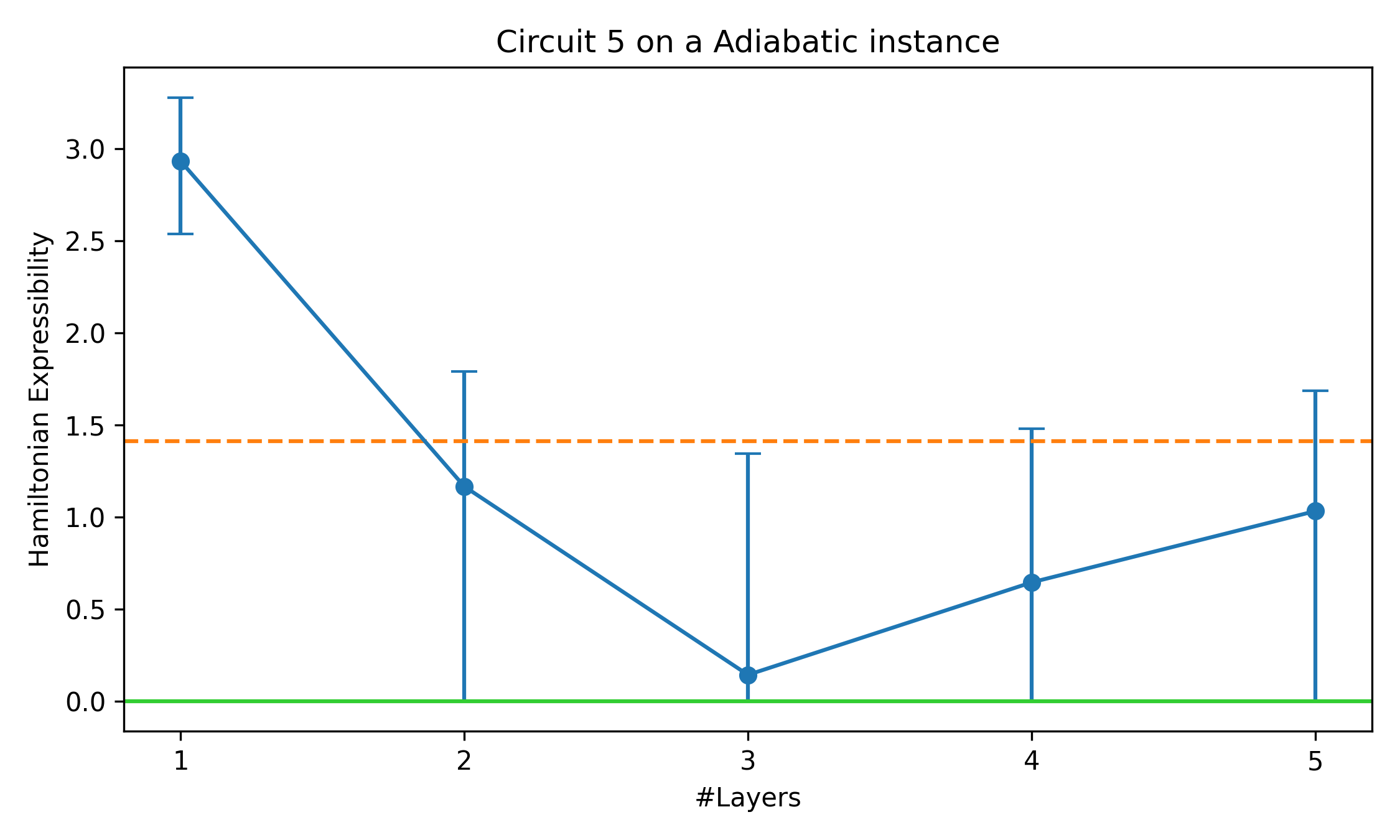}\hfill
 \\
 \includegraphics[width=\textwidth]{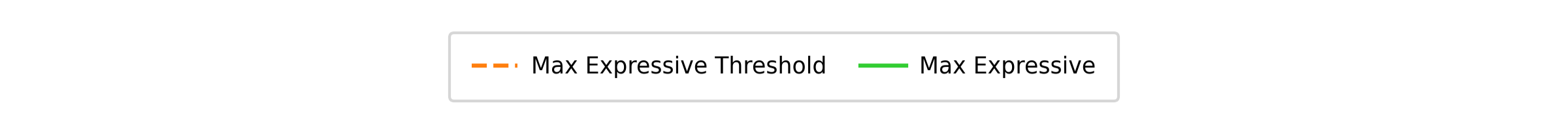}
 \caption{Hamiltonian expressibility values for some of the circuits with 4 qubits on a specific instance of the Adiabatic problem, with respect to the number of layers. Each value is plotted along with its corresponding confidence interval. The orange dashed line indicates the value of Hamiltonian expressibility below which a circuit can be considered maximally expressive, while the green line represents the theoretical minimum of $0$.}
 \label{fig:RatovsLayers}
\end{figure}
Firstly, we observe that the Hamiltonian expressibility of a circuit improves as the number of layers increases. Specifically, as partially shown in Figure~\ref{fig:RatovsLayers}, we note that the Hamiltonian expressibility values for all circuits of $4$ qubits decrease with the increase of the number of layers. However, this decrease may plateau before reaching the minimum (as seen in Circuit 1 in Figure~\ref{fig:RatovsLayers}), around the threshold (Circuit 2), or immediately after exceeding the maximum expressibility threshold (Circuit 5) which is derived from sampling Haar uniform unitaries, as described in Section~\ref{subsec:expressibility_estimation} and Appendix~\ref{app:thresholds}.
Circuit 3 may require additional layers, but it is expected to eventually reach saturation. Although the graph presents results for a specific instance of the Adiabatic problem, similar trends of decreasing Hamiltonian expressibility are observed for all Hamiltonians in the dataset presented in Section~\ref{subsec:circuits_and_hamiltonians}. 

\begin{figure}[!ht]
 \includegraphics[width=.5\textwidth]{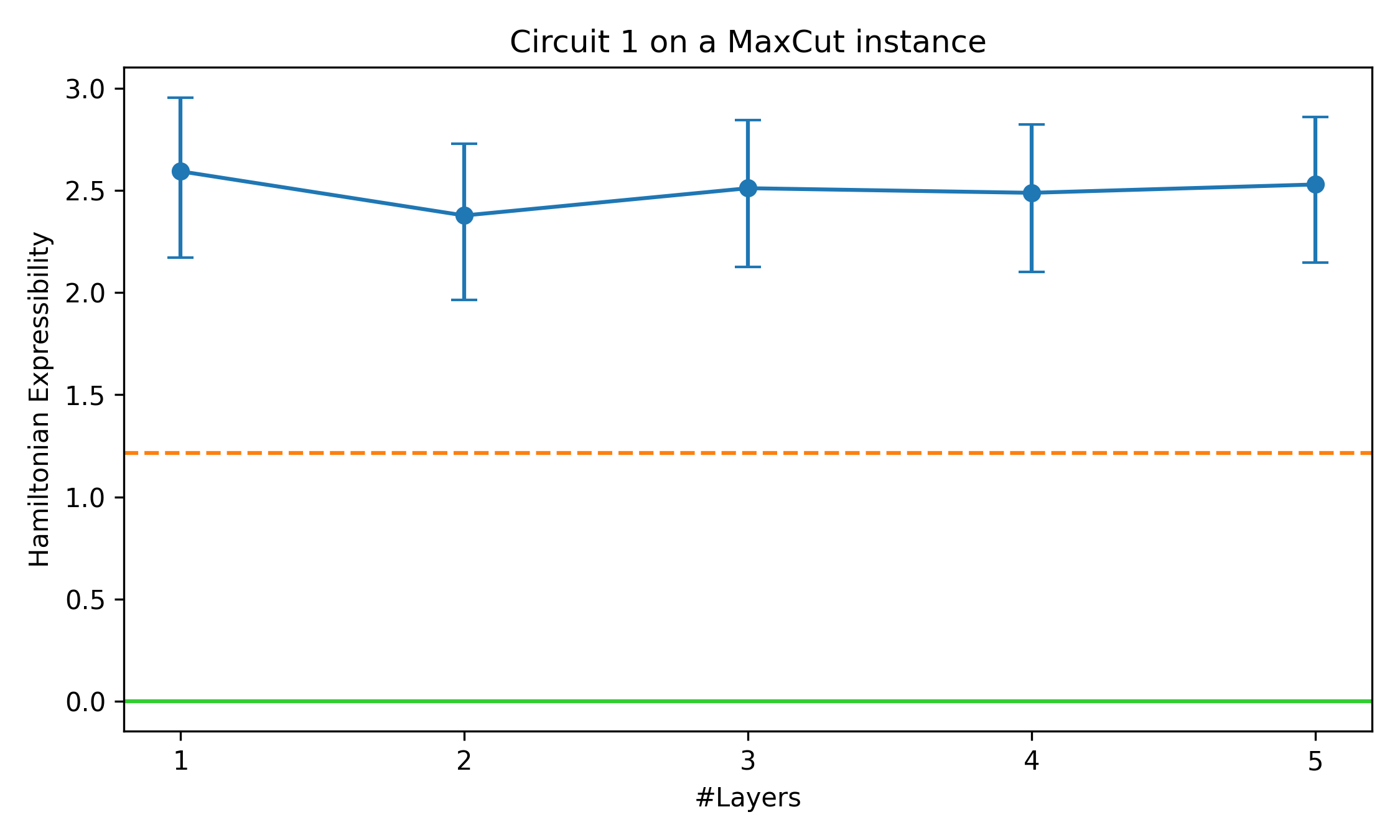}\hfill
 \includegraphics[width=.5\textwidth]{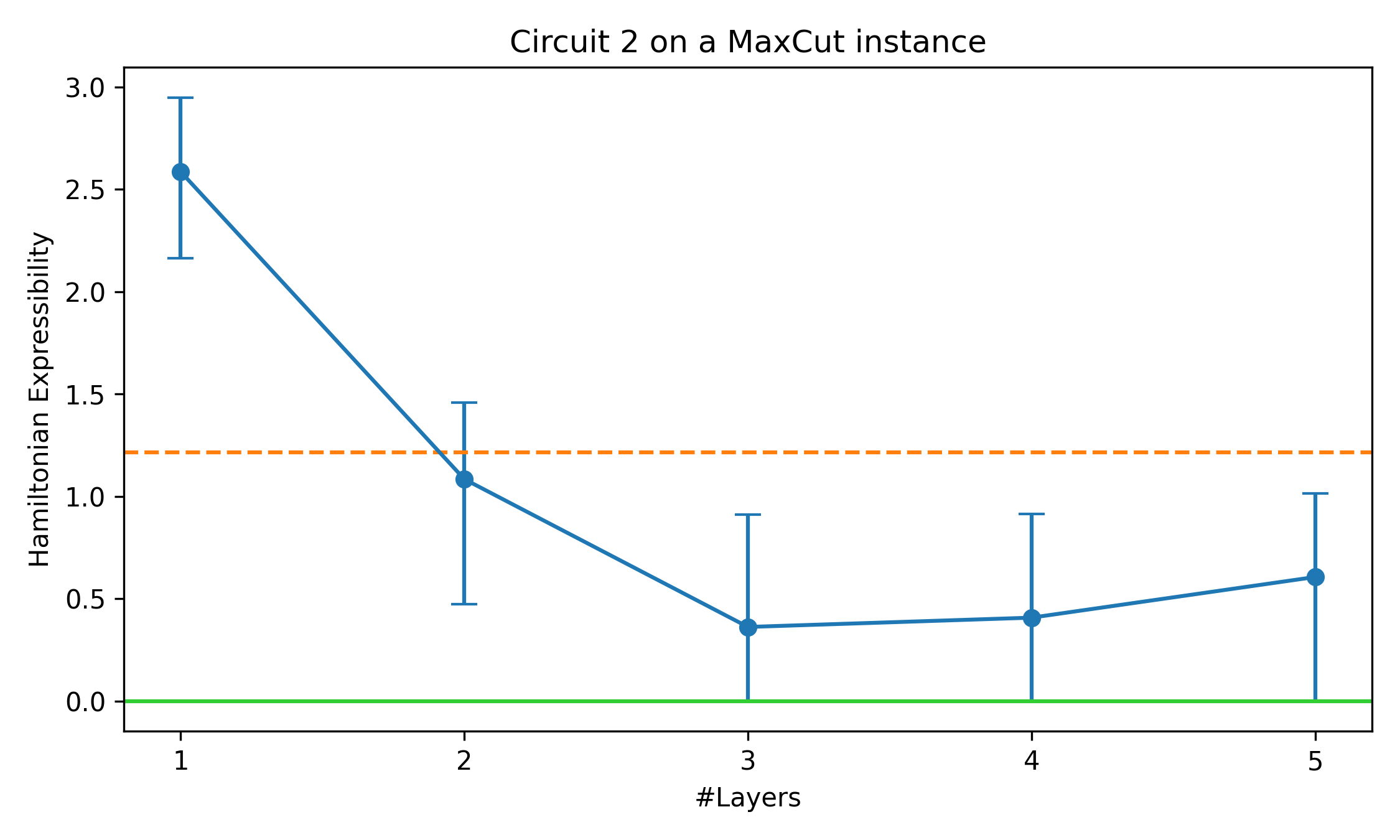}\hfill
 \\[\smallskipamount]
 \includegraphics[width=.5\textwidth]{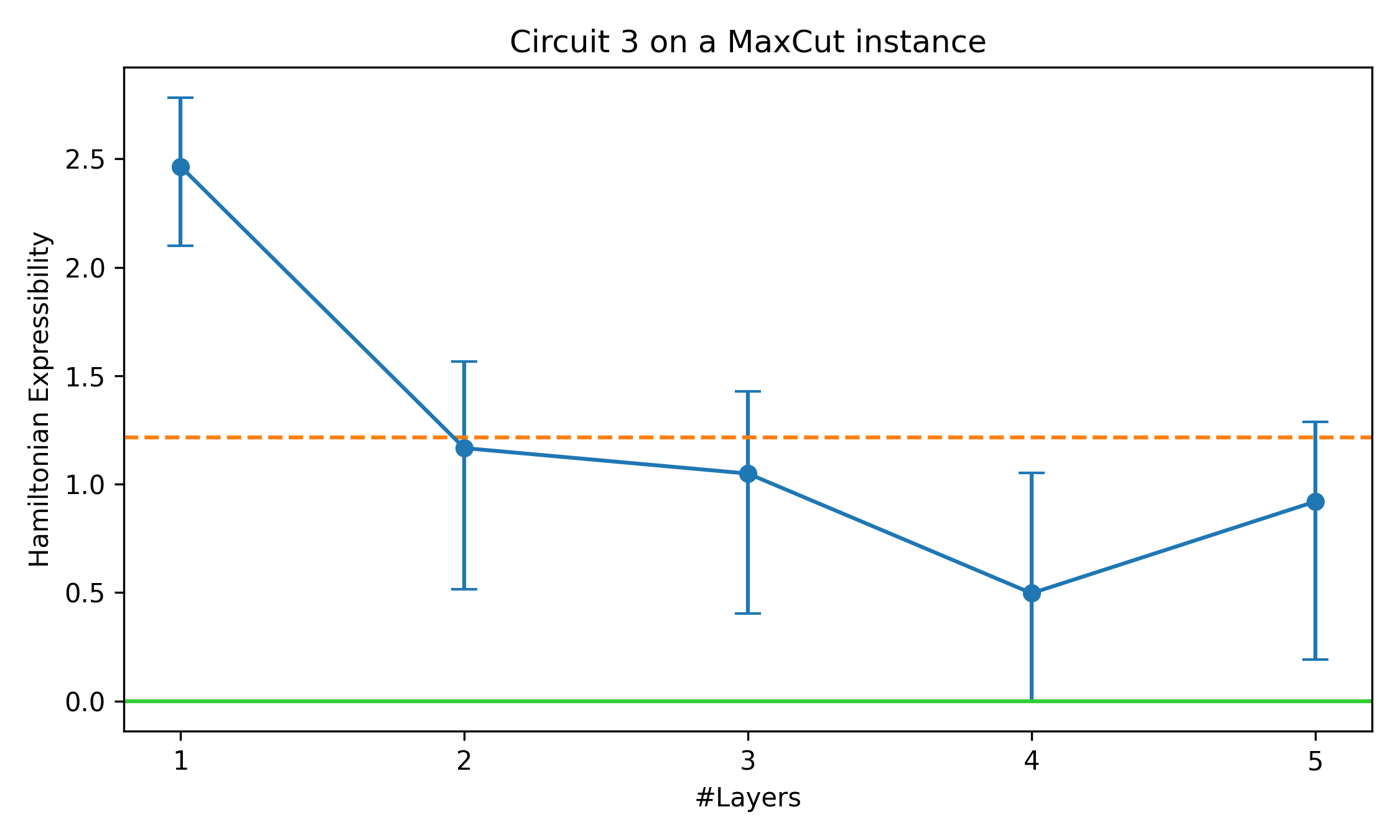}\hfill
 \includegraphics[width=.5\textwidth]{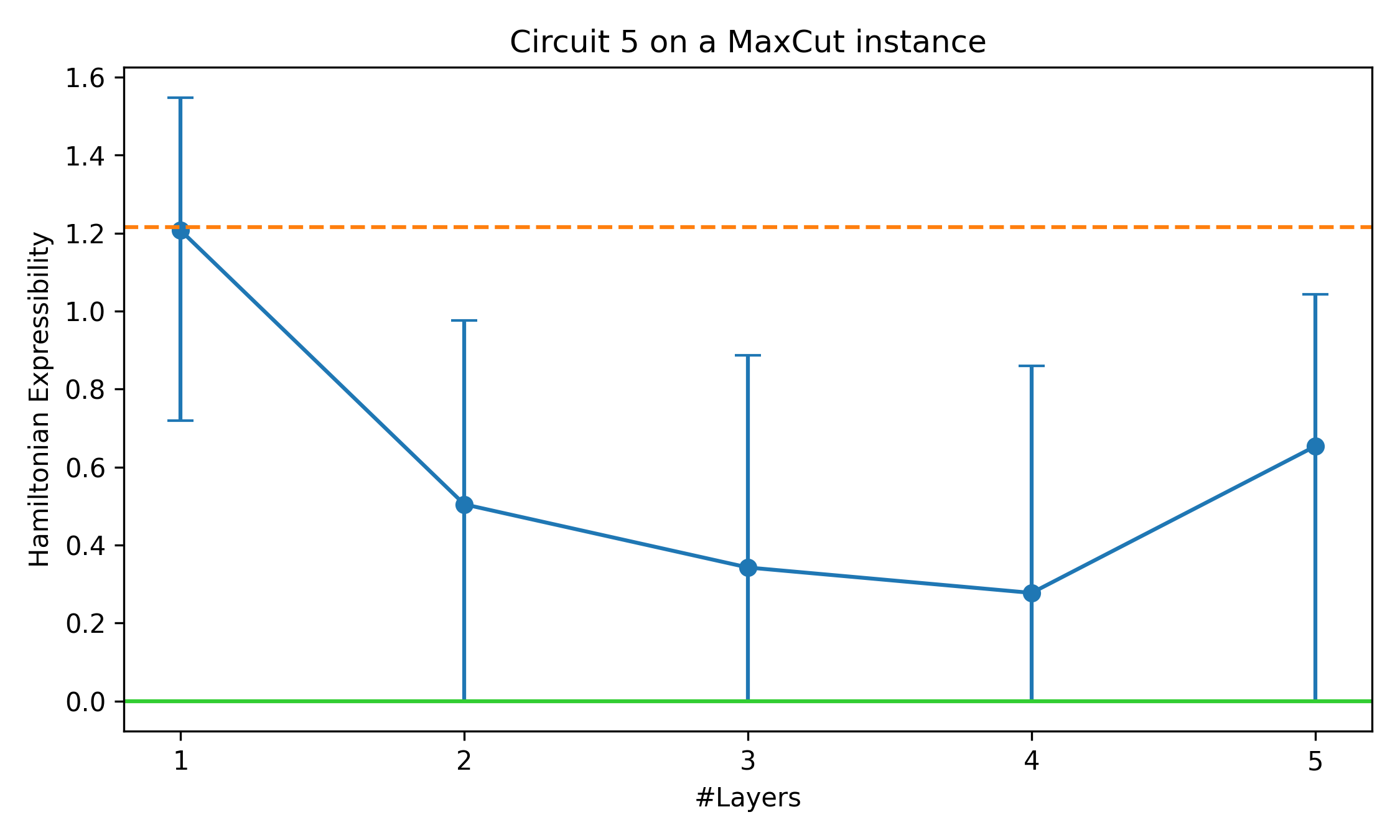}\hfill
 \\
 \includegraphics[width=\textwidth]{DepthLegend.png}
 \caption{Hamiltonian expressibility values of some of the circuits of 4 qubits on a specific instance of the MaxCut class, with respect to the number of layers. Each value is plotted along with its corresponding confidence interval. The orange dashed line indicates the value of Hamiltonian expressibility below which a circuit can be considered maximally expressive, while the green line represents the theoretical minimum of $0$.}
 \label{fig:ExprvsLayersMAXCUT}
\end{figure}

In some cases, increasing the number of layers does not enhance Hamiltonian expressibility. For example, in Circuit 5, the expressibility value appears to increase and not decrease, after 3 layers. However, by examining the confidence interval plotted in the graphs, it can be inferred that this behaviour is likely due to precision errors in the expressibility, which stem from the limited sample size. This phenomenon typically occurs once the Hamiltonian expressibility has passed the aforementioned threshold, indicating that the corresponding ansatz can be considered maximally expressive. In such cases, adding more layers becomes unnecessary, and the Hamiltonian expressibility value starts to oscillate within the \say{maximally expressive zone}, which lies between $0$ and the threshold.

Similar observations apply to the Hamiltonian expressibility ratio. As the number of layers increases, the metric value decreases (indicating improved expressibility), approaches the theoretical minimum of $1$, and saturates before or right after the maximally expressive thresholds (where it can potentially exhibit an oscillating behaviour).

Finally, it is noteworthy that, when the problem is either an instance of a QUBO or Random Diagonal problem, and therefore when the Hamiltonian is diagonal, all circuits achieve a high level of Hamiltonian expressibility and Hamiltonian expressibility ratio more quickly. In particular, the thresholds are easily reached or even passed with fewer layers compared to scenarios with a non-diagonal Hamiltonian, as shown in Figure~\ref{fig:ExprvsLayersMAXCUT} for a specific instance of the MaxCut class. 
\subsubsection{Circuits with Highest Hamiltonian Expressibility For a Given Problem Class}
\label{subsec:expr_class_relationship}

\begin{figure}[!ht]
 \includegraphics[width=0.5\textwidth{}]{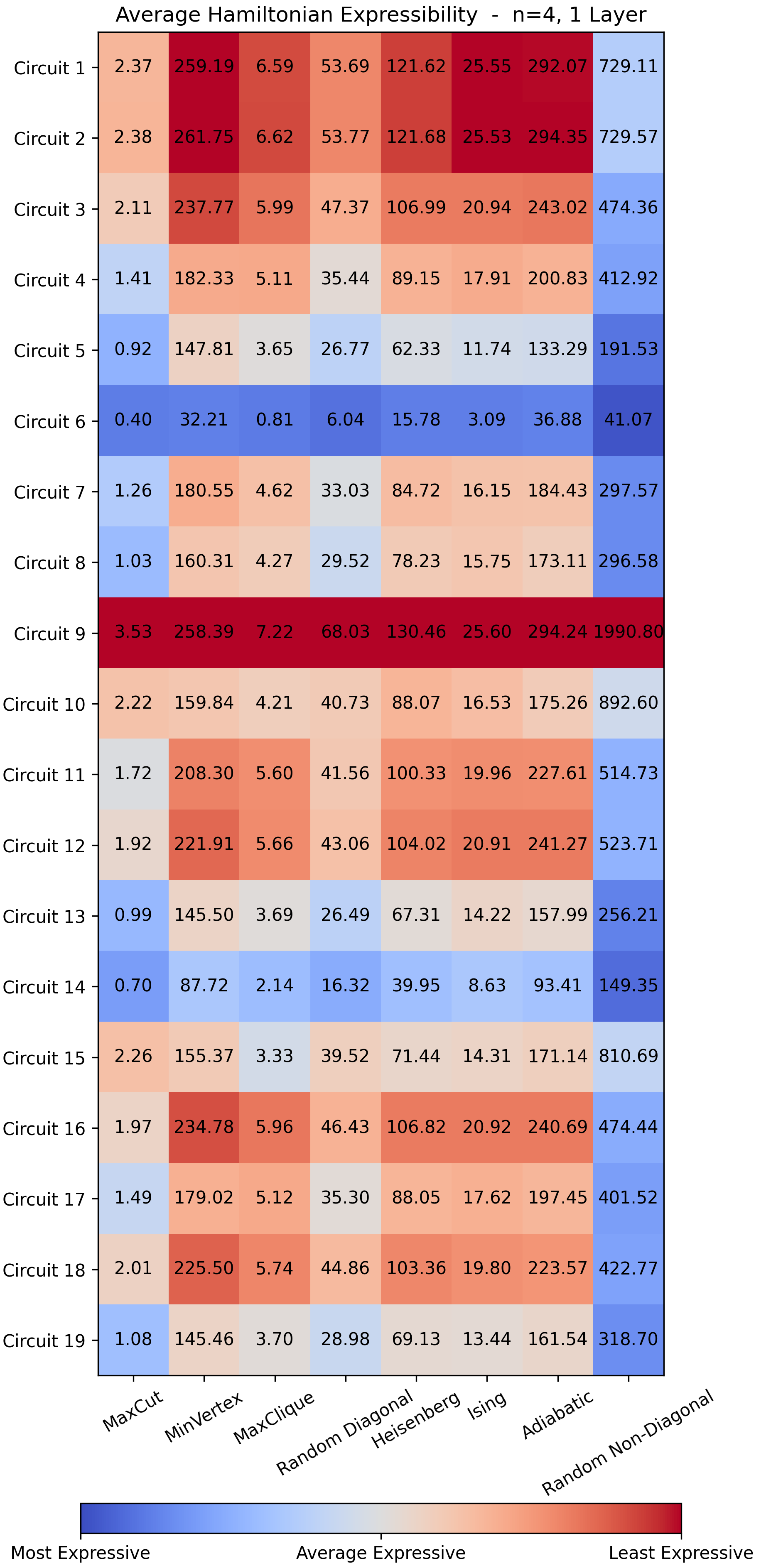}
 \includegraphics[width=0.5\textwidth{}]{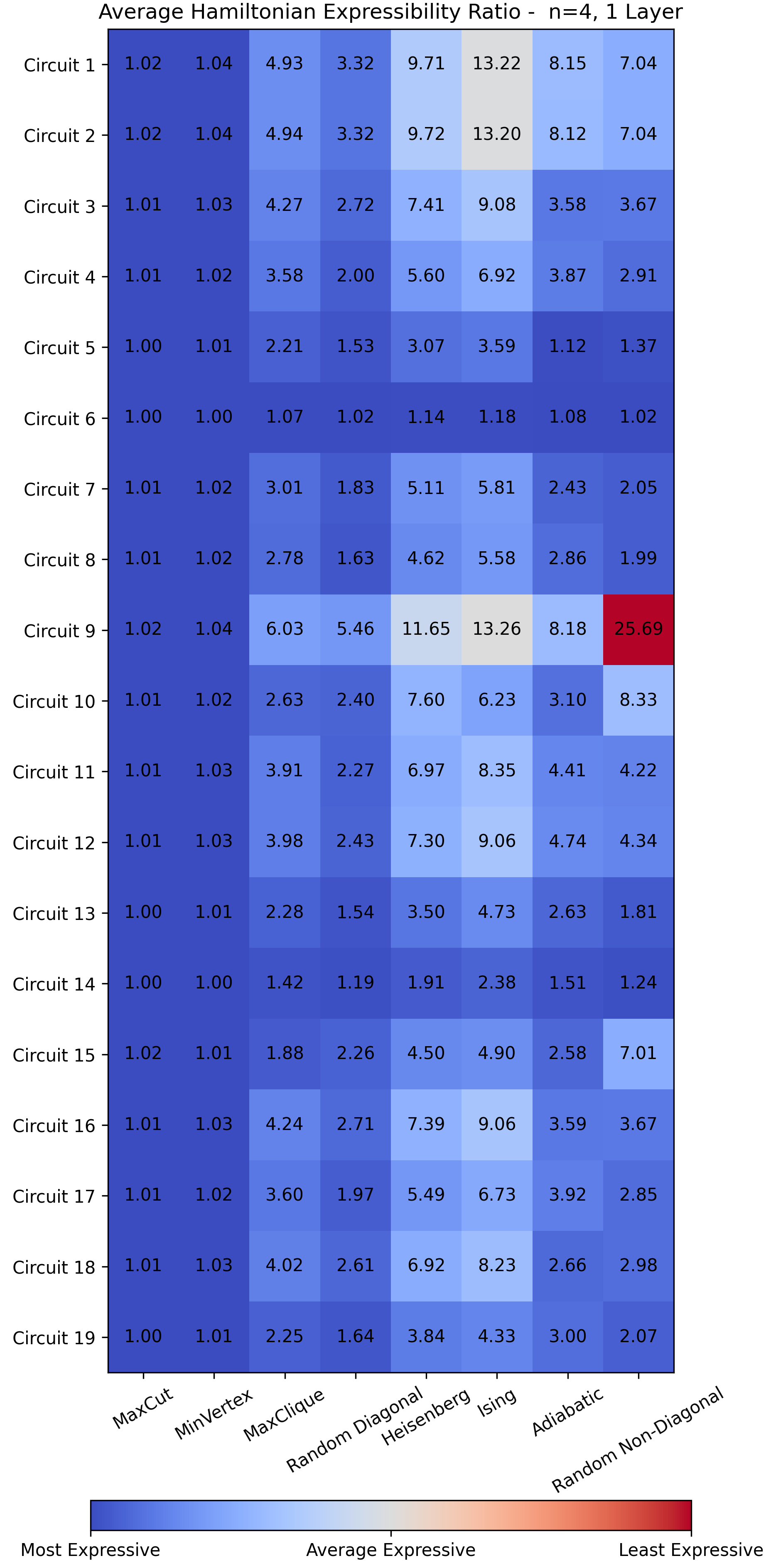}
 \caption{Average Hamiltonian expressibility and Hamiltonian expressibility ratio of all circuits of 4 qubits and with 1 layer, for all the problem classes.
 Blue indicates high expressibility (\idest Hamiltonian expressibility near $0$ or Hamiltonian expressibility ratio near $1$), and red indicates low expressibility. In the Hamiltonian expressibility heatmap (left panel), the colour scale is normalized from $0$ to the maximum value in each column, enabling relative comparisons within each Hamiltonian class. For Hamiltonian expressibility ratio (right panel), the colour scale is normalized globally, from $1$ to the overall maximum, allowing absolute comparisons across all classes. The first four classes have diagonal Hamiltonians, while the others have non-diagonal Hamiltonians. }
 \label{fig:Average_expr_and_ratio}
\end{figure}

Figure~\ref{fig:Average_expr_and_ratio} shows the average Hamiltonian expressibility and expressibility ratio for all 19 circuits of 4 qubits with a single layer across all problem classes.

It is noteworthy that, while the most expressive circuits are often the same across all problem classes, there are some instances where an ansatz may be more expressive than another circuit for a particular problem class, and less expressive than the same circuit for a different problem category. For instance, for the MaxCut class, Circuit 11 is more Hamiltonian-expressive than Circuit 10 (i.e., it has a lower Hamiltonian expressibility value). However, for the MinVertex class, Circuit 10 is the most Hamiltonian-expressive circuit, as Circuit 11 has a higher Hamiltonian expressibility value.

Additionally, our results reveal that all 19 circuits achieve better expressibility, both in terms of Hamiltonian expressibility and Hamiltonian expressibility ratio, when applied to problems with a diagonal Hamiltonian. In contrast, circuits are less Hamiltonian-expressive on problems with a non-diagonal Hamiltonian. From this, we can conclude that diagonal problems are more easily \say{explorable}, meaning that most circuits reach a level of expressibility sufficient to explore the system's energy space exhaustively. This is likely due to the fact that diagonal problems are guaranteed to have the solution in one of the basis states, making the energy space inherently narrower and easier to navigate. This distinction will be crucial for interpreting relationships between expressibility and solution quality, as shown in the next sections.
\subsection{Relationship with Approximation Ratio}
\label{subsec:relationship_with_approximation_ratio}

In this section, we present the results of our correlation analysis, which aims to draw conclusions about the relationship between a circuit's Hamiltonian expressibility and its final accuracy in the VQE algorithm. Section~\ref{subsec:noiseless_results} analyses the results obtained in the ideal setting, while Section~\ref{subsec:noisy_results} deals with the noisy setting. In both scenarios, we decided to focus on the Spearman correlation coefficient between Hamiltonian expressibility and the average normalized approximation ratio for all problem classes, because it was the most effective correlation metric. The results of the Kendall-Tau correlation are very similar to Spearman, and Pearson proved to be very sensitive to outliers, hence the results of these two coefficients are left in Appendix \ref{app:Pearson_KT}. Similarly, we focus on Hamiltonian expressibility and leave the results of Hamiltonian expressibility ratio in Appendix~\ref{app:Ham_Ratio} because it led to similar results. Next, we focus on the differences between problems with diagonal versus non-diagonal Hamiltonians, as well as between problems whose ground states are basis states versus those requiring superposition states. Finally, we examine the mutual information between Hamiltonian expressibility and solution accuracy, while that of Hamiltonian expressibility ratio can be found in Appendix \ref{app:Ham_Ratio}. The results obtained using Hamiltonian expressibility are also compared with those obtained using two simpler, strictly problem-agnostic metrics: the number of parameters of a circuit and its Parameter Dimension \citep{Haug_2021}. The results of this comparison are reported in Appendix \ref{app:N_Params_Param_Dim}.

\subsubsection{Results in the Noiseless Setting}
\label{subsec:noiseless_results}
In this section, we present the results of the correlation analysis performed when the VQE algorithm was executed under ideal conditions, as described in Section~\ref{subsubsec:execution_specifics}. The results are reported for the 4- and 8-qubit circuits and problems introduced in Section~\ref{subsec:circuits_and_hamiltonians}.

\paragraph{Correlation Coefficients in the 4-Qubit Case}
\begin{figure}[!ht]
\centering
 \includegraphics[width=0.75\textwidth]{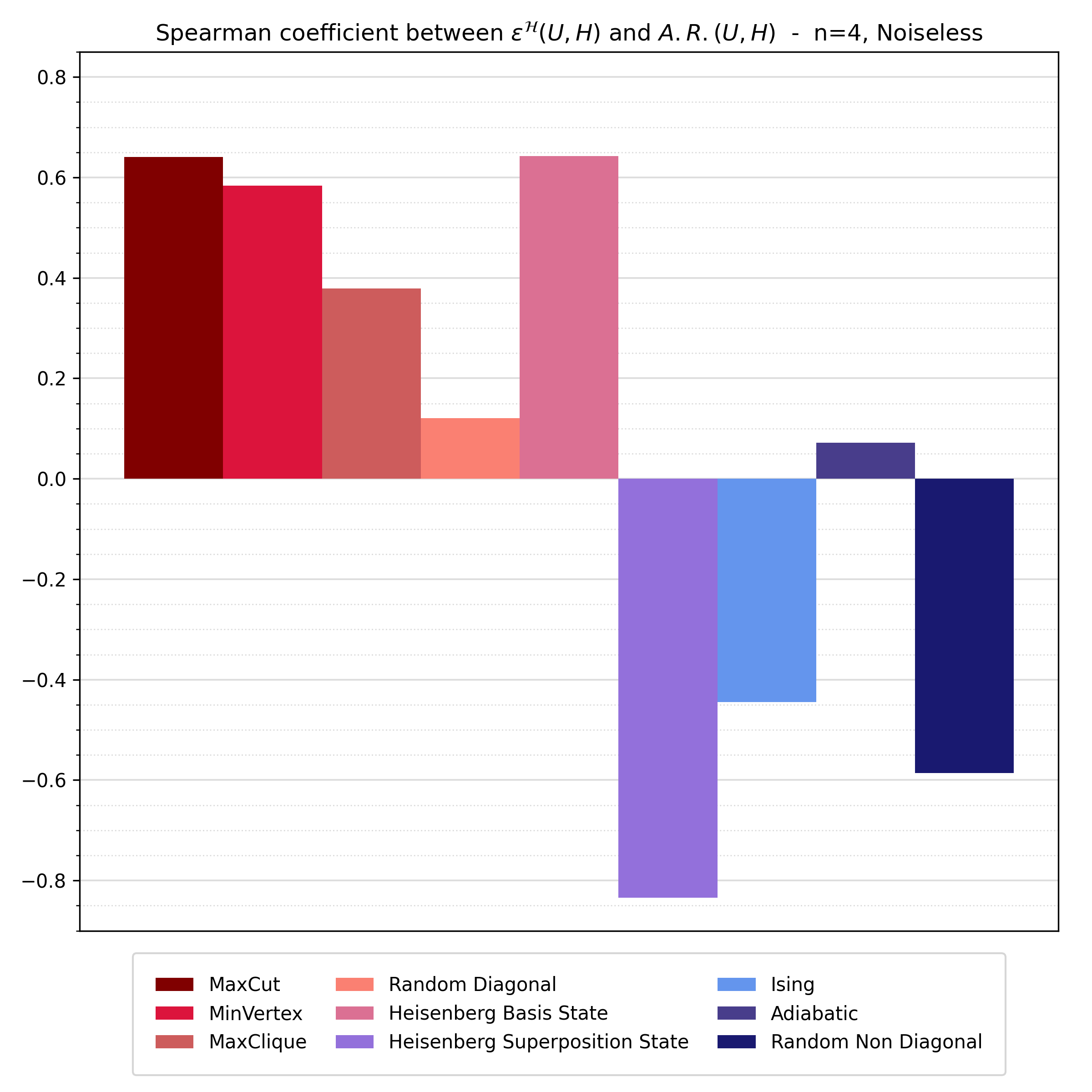}
 \caption{Average Spearman correlation coefficient between Hamiltonian expressibility and average normalized approximation ratio, when adopting a noiseless setting for the VQE. The values are obtained for circuits and problems of 4 qubits. Colouring in shades of red indicates problem classes with diagonal Hamiltonians or with a basis state solution, while blue colours represent problem classes based on non-diagonal Hamiltonians with a superposition state solution.}
 \label{fig:Correlations_4_Classes}
\end{figure}
Figure~\ref{fig:Correlations_4_Classes} presents, for all problem classes considered and presented in Section~\ref{subsec:circuits_and_hamiltonians}, the results of the computation of the Spearman correlation coefficient as detailed in Section~\ref{subsec:corr_analysis}, for circuits of 4 qubits. Note that the Heisenberg XXZ class was further split into two subclasses, one for instances where the solution is a basis state, and the other for instances where the solution is a superposition state.

It is important to recall that all coefficients were computed between the expressibility metrics, where low values indicate good expressibility, and average normalized approximation ratio, where high values indicate a high accuracy. Therefore, any negative value of the coefficient suggests a monotonically decreasing (Spearman, Kendall Tau) or linearly decreasing (Pearson) relationship between the expressibility metric considered and accuracy in the VQE algorithm. In this case, lower values of the metric (better expressibility) correlate with a higher normalized approximation ratio (better accuracy). Conversely, positive correlation coefficients suggest the opposite: less expressive circuits will result in higher solution accuracy.

As shown in Figure~\ref{fig:Correlations_4_Classes}, different problem classes exhibit varying levels of correlation between expressibility and accuracy. Furthermore, we observe a substantial difference in the values of the Spearman coefficient between the problem classes based on a diagonal Hamiltonian, or more generally those whose solutions lie in a computational basis state (the first five coloured bars in shades of red), and the problem classes described by a non-diagonal Hamiltonian with solutions in a superposition state (the last four, in shades of blue and violet).

Specifically, we observe how non-diagonal problem classes such as Ising, Heisenberg Superposition State and Random Non-Diagonal, each of which has a superposition state solution, show moderately or highly negative Spearman coefficients, reaching $-0.85$ in some cases. Recall that the Spearman coefficient is designed to capture monotonic relationships, these results imply the presence of a decreasing monotonic correlation for these classes. Based on our previous considerations, this suggests that selecting highly Hamiltonian-expressive circuits could be beneficial for solving problems whose solutions lie in a superposition state. The only exception is the Adiabatic class, which shows a Spearman coefficient close to $0$.

In contrast, for classes with a diagonal Hamiltonian or a basis state solution, we observe low or moderately positive Spearman values. In particular, there is little to no monotonic correlation for the Random Diagonal class, and weak to moderate increasing monotonic correlations for the MaxCut, MinVertex, MaxClique and Heisenberg Basis State classes. This suggests that highly Hamiltonian-expressive circuits are not necessary for solving problems whose solutions lie in a basis state; instead, they may even be counterproductive. In some of these cases, in fact, it could be preferable to select circuits with low Hamiltonian expressibility.

\paragraph{Trend Analysis of Approximation Ratio vs Hamiltonian Expressibility in the 4 Qubits Case}

\begin{figure}[!ht]
\centering
 \includegraphics[width=0.85\textwidth]{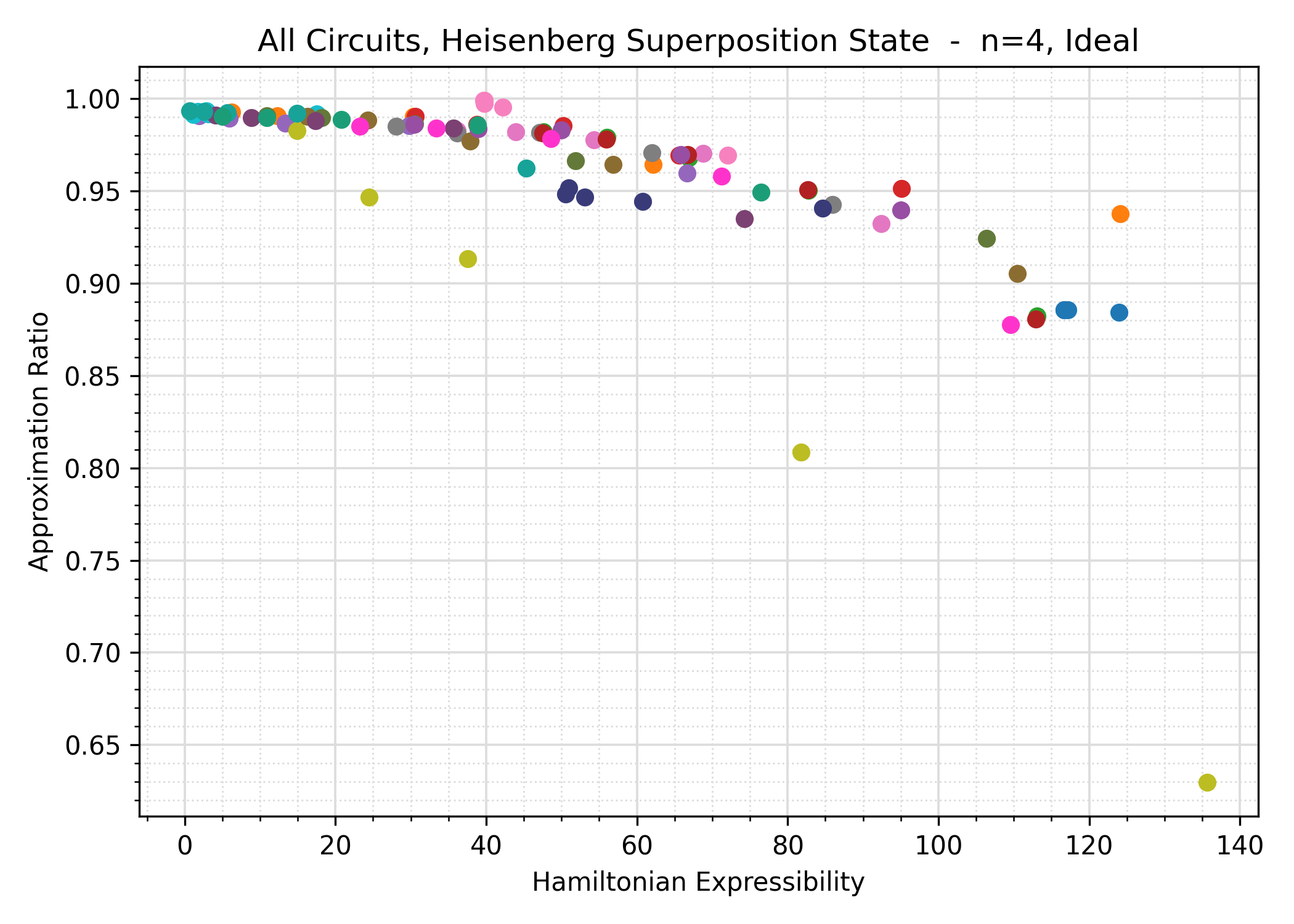}
 \caption{Average normalized approximation ratio as a function of Hamiltonian expressibility of all 4-qubit circuits over a specific instance of the Heisenberg Superposition State class, in the ideal case. Colours indicate the base circuit pattern from which each ansatz is derived. Ansätze built by varying the number of layers from the same pattern share the same colour.}
 \label{fig:HamExpr_vs_ar_HEIS_4}
\end{figure}

\begin{figure}[!ht]

 \includegraphics[width=0.565\textwidth]{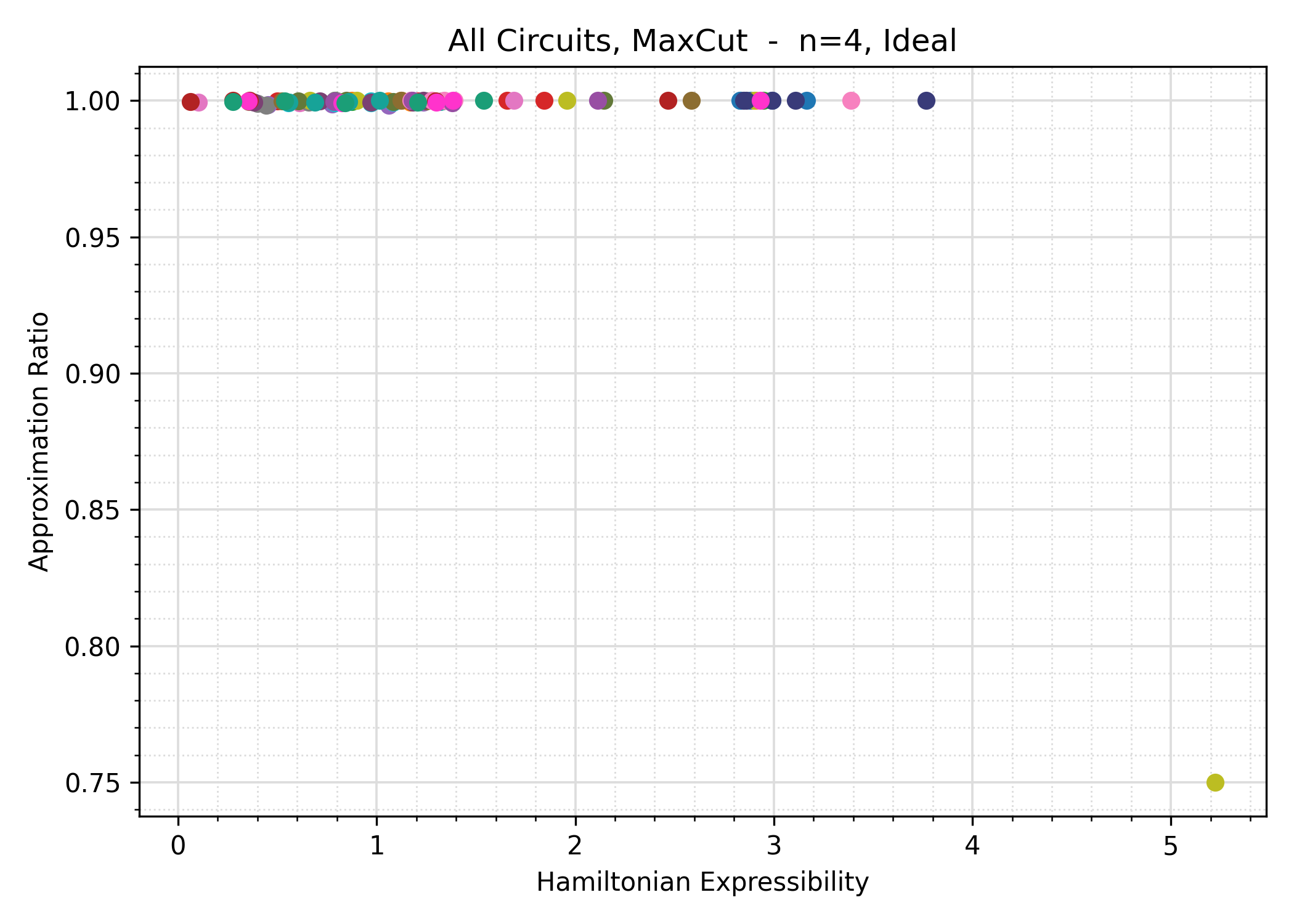}\hfill
 \includegraphics[width=0.435\textwidth]{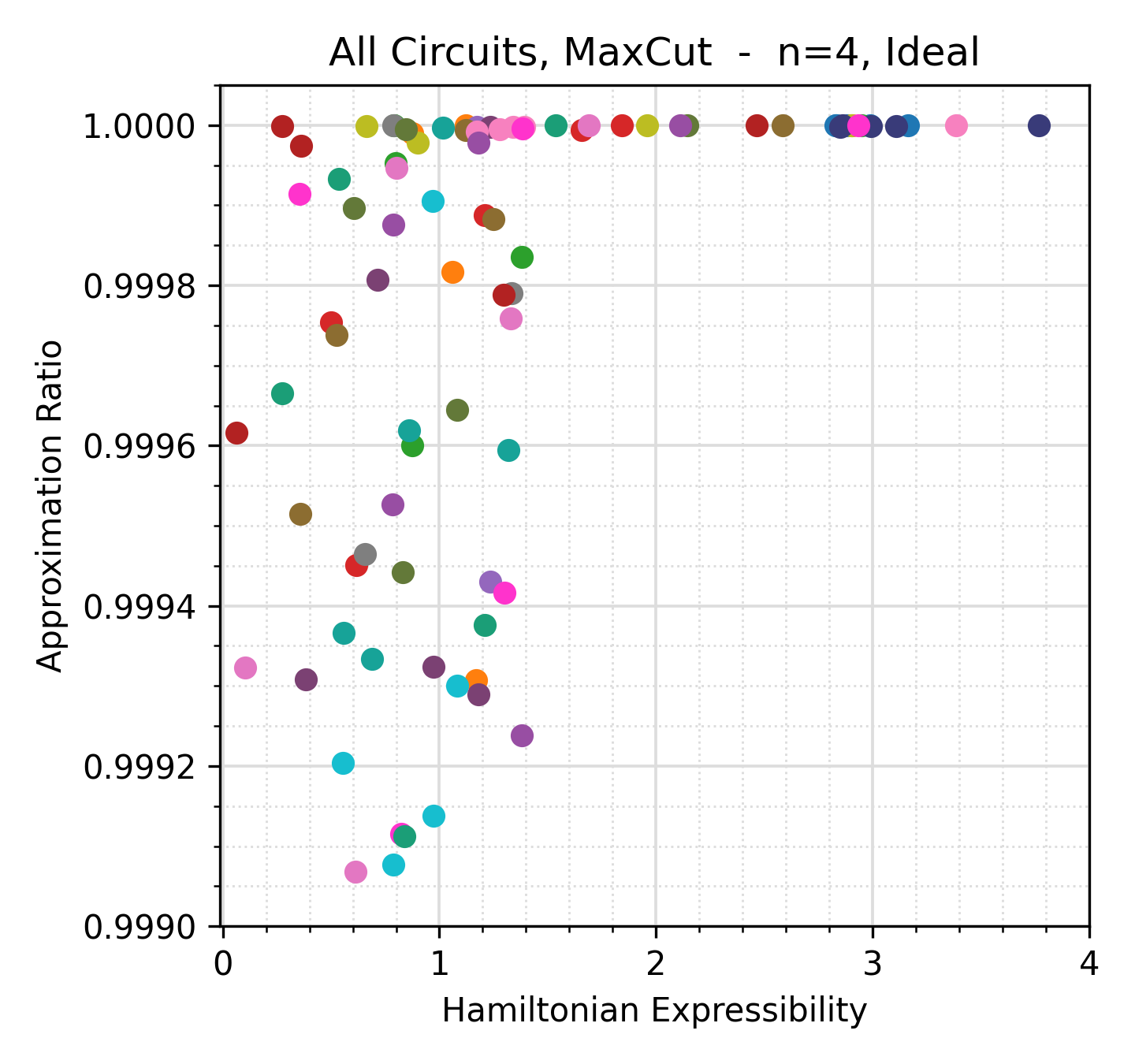}\hfill

 \caption{Average normalized approximation ratio as a function of Hamiltonian expressibility of all 4-qubit circuits over a specific instance of the MaxCut class, in the ideal case. The graph on the right is a zoomed section of the graph on the left. Colours indicate the base circuit pattern from which each ansatz is derived. Ansätze built by varying the number of layers from the same pattern share the same colour.}
 \label{fig:HamExpr_vs_ar_MaxCut_4}
\end{figure}

To better illustrate the relationship between Hamiltonian expressibility and the average normalized approximation ratio, we report these two quantities for all 4-qubit circuits under ideal (noiseless) conditions in Figure~\ref{fig:HamExpr_vs_ar_HEIS_4} and Figure~\ref{fig:HamExpr_vs_ar_MaxCut_4}, for a specific instance of the Heisenberg Superposition State class and of the MaxCut class, respectively. Based on these sets of data, one per instance, the statistics of the correlation coefficients presented in Figure~\ref{fig:Correlations_4_Classes} are derived for each class. 

In the Heisenberg Superposition State case (Figure~\ref{fig:HamExpr_vs_ar_HEIS_4}), a clear decreasing trend can be observed, as captured by the Spearman correlation coefficient. It is evident that the more Hamiltonian-expressive circuits, those on the left side of the x-axis, achieve higher normalized approximation ratios compared to the less expressive ones on the right.

Figure~\ref{fig:HamExpr_vs_ar_MaxCut_4} shows the results for an instance of the MaxCut problem and is useful to explain a discrepancy observed in basis state or diagonal problems between the Spearman and Kendall Tau coefficients (which are close to zero or moderately positive) and the Pearson coefficient (which is weakly or moderately negative, see Figure~\ref{fig:Correlations_HamExpr_Pears_KT} in Appendix~\ref{app:Pearson_KT}).

Indeed, in the full plot on the left, we clearly see a strong outlier in the lower-right corner: a circuit with relatively low Hamiltonian expressibility and very low approximation ratio. This outlier heavily influences the Pearson coefficient, which is sensitive to such points. 
However, the zoomed-in plot on the right shows that most data points follow a mild increasing trend, suggesting a positive monotonic relationship. This supports the interpretation given by the Spearman coefficient and shows that the Pearson coefficient is misleading in this case.
We observed similar behaviour in other diagonal or basis state problem instances, reinforcing the conclusion that the Pearson coefficient is less useful in such settings.


\paragraph{Correlation Coefficients in the 8-Qubit Case}

\begin{figure}[!ht]
\centering
 \includegraphics[width=0.75\textwidth]{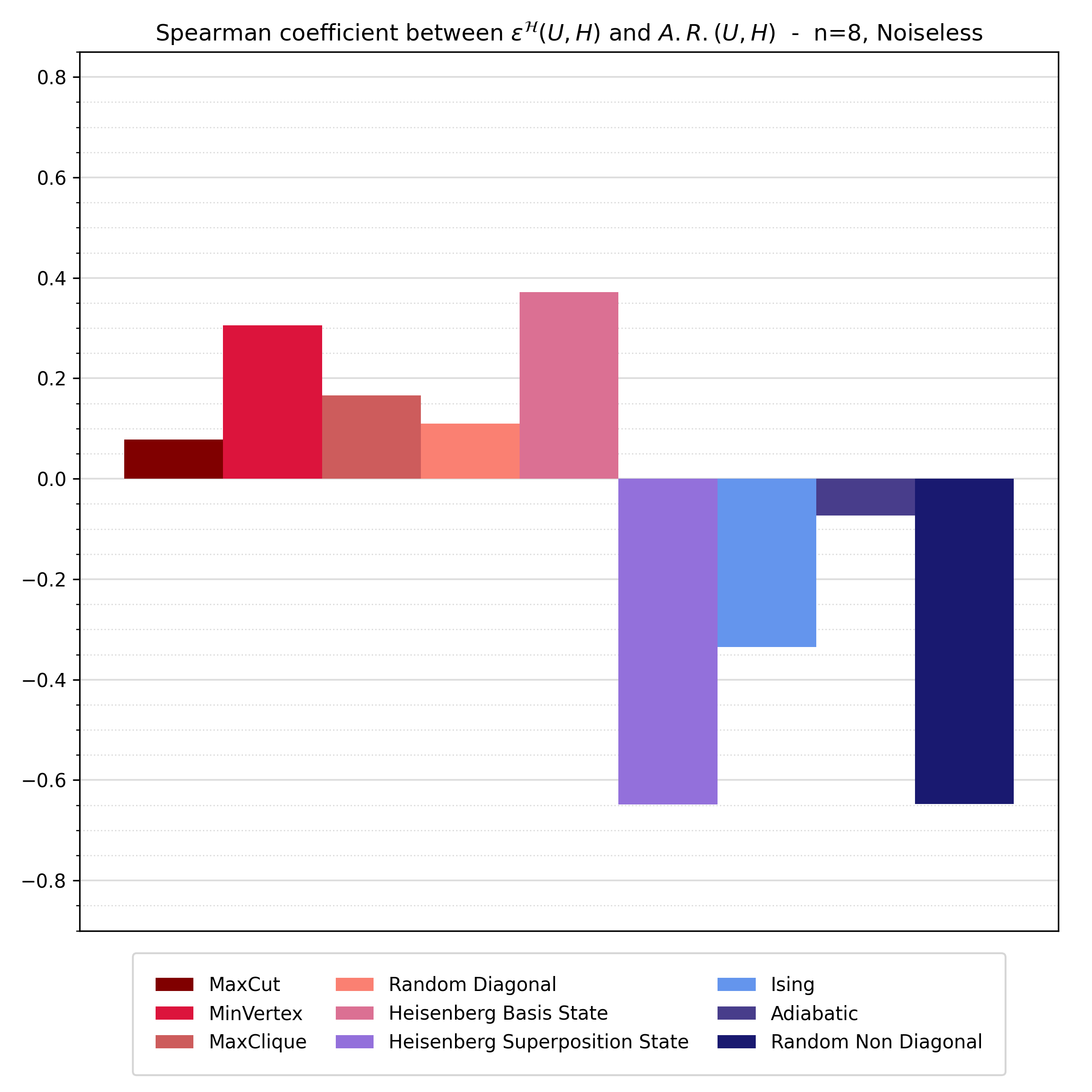}
 \caption{Average Spearman correlation coefficient between Hamiltonian expressibility and average normalized approximation ratio, when adopting a noiseless setting for the VQE. The values are obtained for circuits and problems of 8 qubits. Colouring in shades of red indicates problem classes with diagonal Hamiltonians or with a basis state solution, while blue colours represent problem classes based on non-diagonal Hamiltonians with a superposition state solution.}
 \label{fig:Correlations_8_Classes}
\end{figure}

Figure~\ref{fig:Correlations_8_Classes} shows the results of the Spearman correlation coefficient obtained under ideal conditions, considering 8-qubit circuits and problems with only one layer.

Firstly, we observe that the contrast between problem classes with basis state solutions and those with superposition state solutions, previously evident in the Spearman and Kendall Tau coefficients, appears to diminish. Specifically, diagonal and basis state problems now show Spearman values closer to $0$, while problems with superposition state solutions, which had strongly negative coefficients in the 4-qubit case, now exhibit mostly moderately negative values.
For example, the Heisenberg Superposition State and Ising classes, which previously had Spearman values around $-0.85$ and $-0.45$, now decrease to approximately $-0.65$ and $-0.3$, respectively. An exception is the Random Non-Diagonal class, which retains correlation values similar to the 4-qubit case.

These results, which show a slight weakening of the relationship between high expressibility and good results in most superposition state problem classes, lead us to hypothesize the emergence of barren plateaus (see Section~\ref{subsec:expressibility}), which are well-known to be associated with high expressibility and can severely hinder trainability.

Consequently, for problems where low-expressibility ansätze perform better (such as those with diagonal Hamiltonians or basis state solutions), it is reasonable to expect low-expressive ansätze to remain the most useful, as the number of qubits increases. In contrast, for problems where high-expressibility circuits are beneficial (problems with non-diagonal Hamiltonians or superposition state solutions), the ability of such ansätze to produce better results will likely diminish as the system size grows due to trainability issues induced by the possible presence of barren plateaus.

\begin{figure}[!ht]
\begin{center}

 \includegraphics[width=\textwidth]{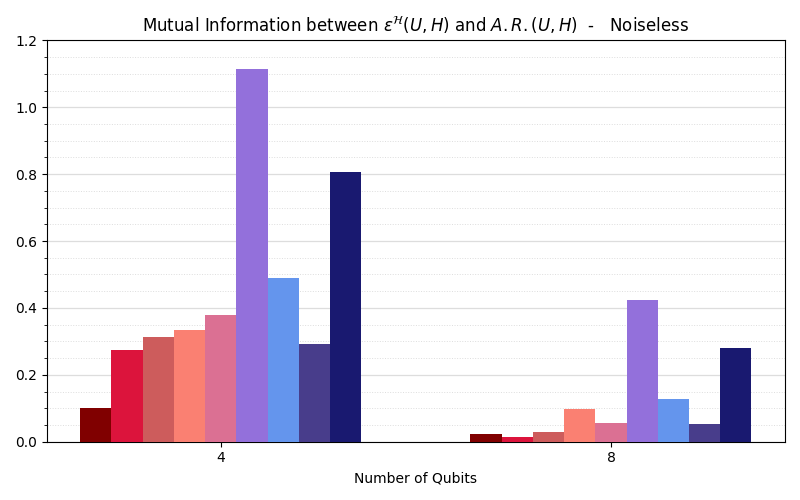}\hfill
 \includegraphics[width=\textwidth]{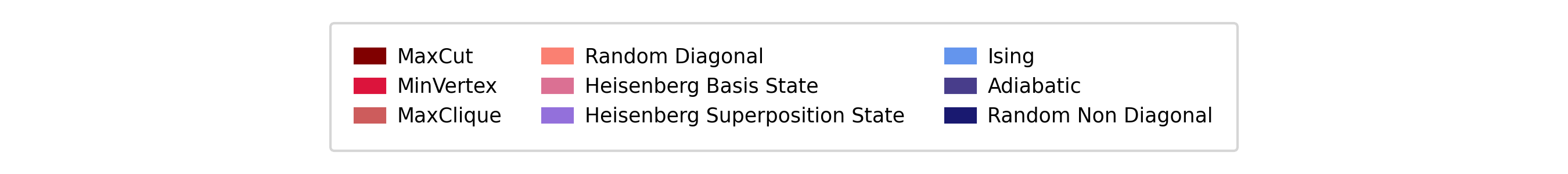}
 \caption{Average mutual information between Hamiltonian expressibility and average normalized approximation ratio, when adopting a noiseless setting for the VQE. The values are obtained for circuits and problems of 4 and 8 qubits. Colouring in shades of red indicates problem classes with diagonal Hamiltonians or with a basis state solution, while blue colours represent problem classes based on non-diagonal Hamiltonians with a superposition state solution.}
  \label{fig:Mutual_Info}

\end{center}
\end{figure}

\paragraph{Mutual Information}
\label{sec:mutual_info}
We now present the results, for all problem classes considered and presented in Section~\ref{subsec:circuits_and_hamiltonians}, of the mutual information between Hamiltonian expressibility and average normalized approximation ratio in Figure~\ref{fig:Mutual_Info}. 

Firstly, we observe a clear distinction between the values for 4-qubit and 8-qubit circuits, with the former showing much higher mutual information values. This aligns with our previous findings on the correlation coefficients, where we noted that Hamiltonian expressibility and Hamiltonian expressibility ratio are more informative and useful to select the most suitable ansätze primarily in smaller problems. 

In the 4-qubit case, the distinction between diagonal classes (or those with solutions in basis states) and non-diagonal classes with superposition state solutions remains clear. 
Specifically, we observe low mutual information values for the former across both expressibility metrics, suggesting that Hamiltonian expressibility carries limited predictive power for such problems. 
In contrast, significantly higher mutual information values are found for classes with superposition state solutions. 
Notably, the Ising class shows mutual information around $0.5$, the Random Non-Diagonal class reaches approximately $0.8$ and the Heisenberg Superposition State class achieves a value of $1.1$. 
These results further support the relevance of expressibility-based metrics for this class of problems and highlight the potential utility of such metrics in particular for the Heisenberg Superposition State class.

In any case, the relatively low Pearson, Spearman, and Kendall Tau correlation coefficients observed for our metrics may indicate the presence of a stronger underlying relationship between expressibility and accuracy, which is neither linear nor monotonic. 
This suggests that future research could be conducted to identify possible non-linear dependencies, which may better capture the connection between expressibility and results.

\paragraph{Basis State Solution vs Superposition State Solution}
A clear difference emerges from the results in the relationship between Hamiltonian expressibility (and the Hamiltonian expressibility ratio) and the quality of the final solution, measured by the normalized approximation ratio, when comparing problems with solutions in computational basis states to those with solutions in superposition states. In particular, when a problem is described by a diagonal Hamiltonian, or by a non-diagonal one whose ground state is still a computational basis state, circuits with Hamiltonian expressibility close to zero, i.e., nearly maximally expressive, tend to perform worse than less expressive circuits. Conversely, when the Hamiltonian is non-diagonal and the ground state is a more complex superposition, circuits that are closer to being maximally expressive tend to achieve better solution quality.

Although our results are based on a purely empirical study, and establishing a formal theoretical link between Hamiltonian expressibility and the average normalized approximation ratio is a complex task which goes beyond the scope of this work, we believe that this behaviour could be intuitively justified by the structural differences in the solution spaces of the two types of problems. In practice, the exploration capability provided by a significant Hamiltonian expressibility may be beneficial only for complex and large solution spaces, as for problems described by non-diagonal Hamiltonians. More specifically, Hamiltonian expressibility appears to be related to the complexity of the final solution. When the solution is a computational basis state, a higher exploration capability is not only unnecessary but can even be harmful.

Several further considerations can be made on this point. First, as shown in Section~\ref{subsec:expr_class_relationship}, for problems described by diagonal Hamiltonians, all circuits exhibit very similar numerical values of expressibility, especially in terms of Hamiltonian expressibility ratio and these values are typically close to $1$, the theoretical minimum of such metric. This indicates that all circuits explore the energy landscape of the problem in a similar manner, closely replicating the Haar-induced uniformity, and that leads to the conclusion that such problems are more easily \say{explorable}. Second, as observed in Section~\ref{subsubsec:relationship_with_depth}, circuits with a better Hamiltonian expressibility are generally deeper, and therefore have a higher number of gates and parameters.

This explains the observed results: for problems with simple solution spaces, such as those with diagonal Hamiltonians or solutions in computational basis states, highly expressive circuits perform worse because their advantage in exploration is small, while their higher complexity makes optimization harder. Conversely, in problems with superposition state solutions, circuits show a wider range of expressibility (see Section~\ref{subsec:expr_class_relationship}), and often the deeper ones are significantly more expressive. Consequently, when these circuits are employed to explore such a larger solution space in search of a more complex solution, the exploration advantage of more expressive (and thus deeper) circuits is likely to outweigh the potential disadvantage associated with a hard trainability.

In this regard, the unusual behaviour of the Adiabatic class could still be justified. Despite involving a non-diagonal Hamiltonian with a superposition ground state, it shows a weaker Spearman correlation compared to similar cases. However, this can still be consistent with our interpretation. As detailed in Appendix~\ref{app:notable_problems} its Hamiltonian can be seen as a weighted combination of a diagonal Hamiltonian (from the underlying QUBO problem) and a non-diagonal one with a superposition solution (a Pauli-X string). This makes it an intermediate case between \say{simple} and \say{complex} problems. By varying the coefficients, the problem shifts between these two regimes, which explains why its Spearman coefficient lies roughly between the ones of the two categories.

\subsubsection{Results in the Noisy Setting}
\label{subsec:noisy_results}

In this section, we present the results of the correlation analysis performed when the VQE algorithm was executed under noisy conditions, as described in Section~\ref{subsubsec:noisy_setting_specifics}. 
The analysis is conducted on the 4-qubit circuits and problem instances introduced in Section~\ref{subsec:circuits_and_hamiltonians}.
Specifically, we first compare the noisy-setting correlation coefficients with those obtained in the ideal case, focusing on the Spearman coefficient between Hamiltonian expressibility and the average normalized approximation ratio. Details on the other coefficients and on the correlations involving Hamiltonian expressibility ratio are presented in Appendix~\ref{app:Pearson_KT} and Appendix~\ref{app:Ham_Ratio}.
Then, we illustrate the observed relationship between Hamiltonian expressibility and the average normalized approximation ratio through selected examples.
Finally, using the average error rate metric defined in~\eqref{eq:err}, we investigate how the correlation coefficients evolve as the level of noise increases.

\paragraph{Correlation Coefficients}

\begin{figure}[!ht]
\centering
 \includegraphics[width=0.85\textwidth]{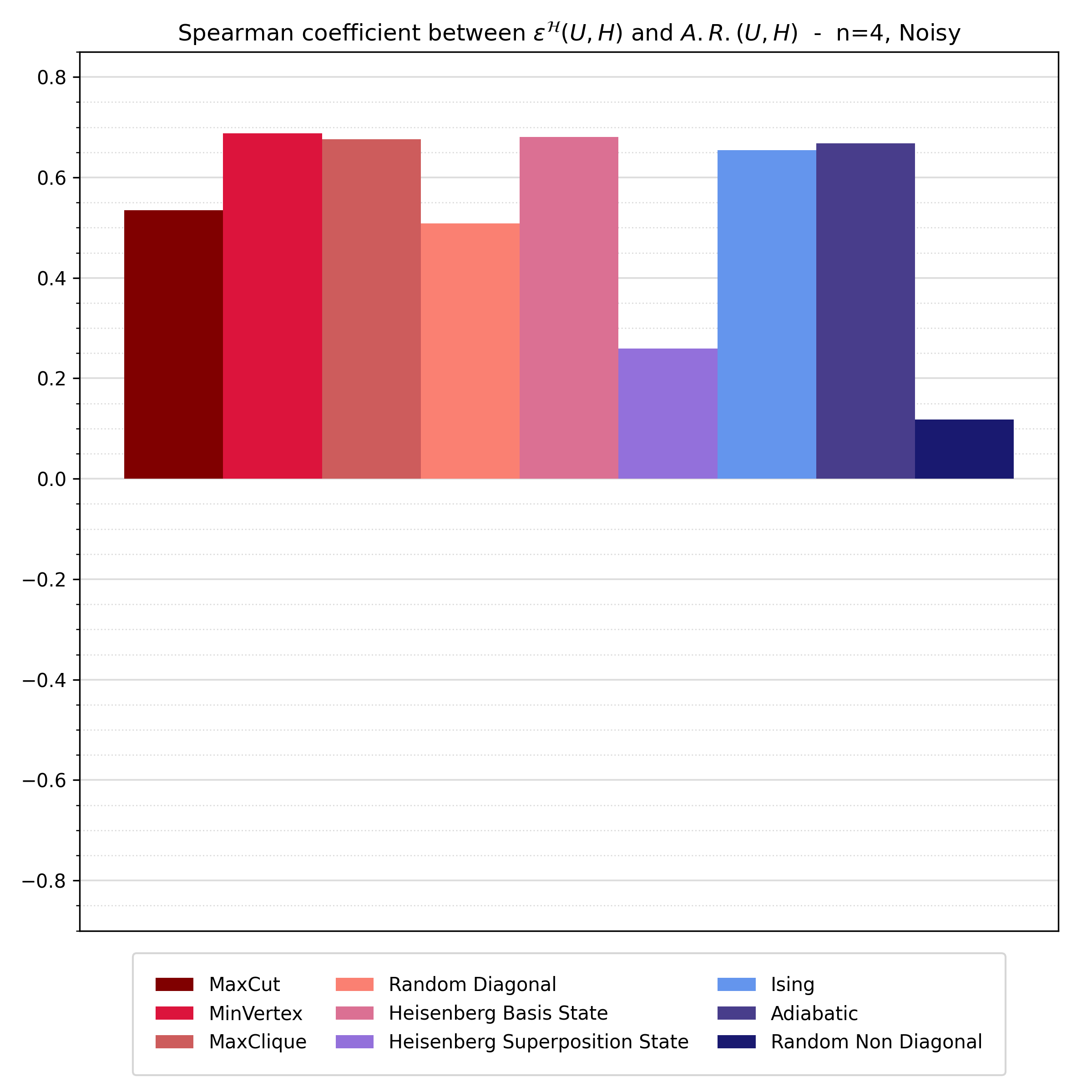}
 \caption{Average Spearman correlation coefficient between Hamiltonian expressibility and average normalized approximation ratio, when adopting a noisy setting for the VQE. The values are obtained for circuits and problems of 4 qubits. Colouring in shades of red indicates problem classes with diagonal Hamiltonians or with a basis state solution, while blue colours represent problem classes based on non-diagonal Hamiltonians with a superposition state solution.}
 \label{fig:Noisy_Correlations_4_Classes}
\end{figure}

Figure~\ref{fig:Noisy_Correlations_4_Classes} reports the Spearman correlation coefficient computed as described in Section~\ref{subsec:corr_analysis} for the 4-qubit circuits under the noisy conditions defined in Section~\ref{subsubsec:noisy_setting_specifics}. The graph shows a significant deviation from the ideal case, clearly illustrating how the introduction of noise radically changes the situation.

Most notably, the clear distinction observed in the ideal case between diagonal classes (or those with ground-state solutions, shown in shades of red) and classes with superposition state solutions (shown in shades of blue and violet) seems to disappear under noisy conditions. Compared to the ideal setting, we observe a strengthening of the positive monotonic correlation in the diagonal problem classes. The Spearman coefficients are significantly higher, indicating that in these cases, the use of highly Hamiltonian-expressive circuits should be avoided. In contrast, for non-diagonal classes, the Spearman coefficients, previously moderately or strongly negative under ideal conditions, now exhibit low or even moderately positive values. The decreasing monotonic correlation observed in the ideal case, which led us to recommend the use of highly expressive circuits for such problems, now appears significantly weaker or even reversed.

\begin{figure}[!ht]
\centering
 \includegraphics[width=0.85\textwidth]{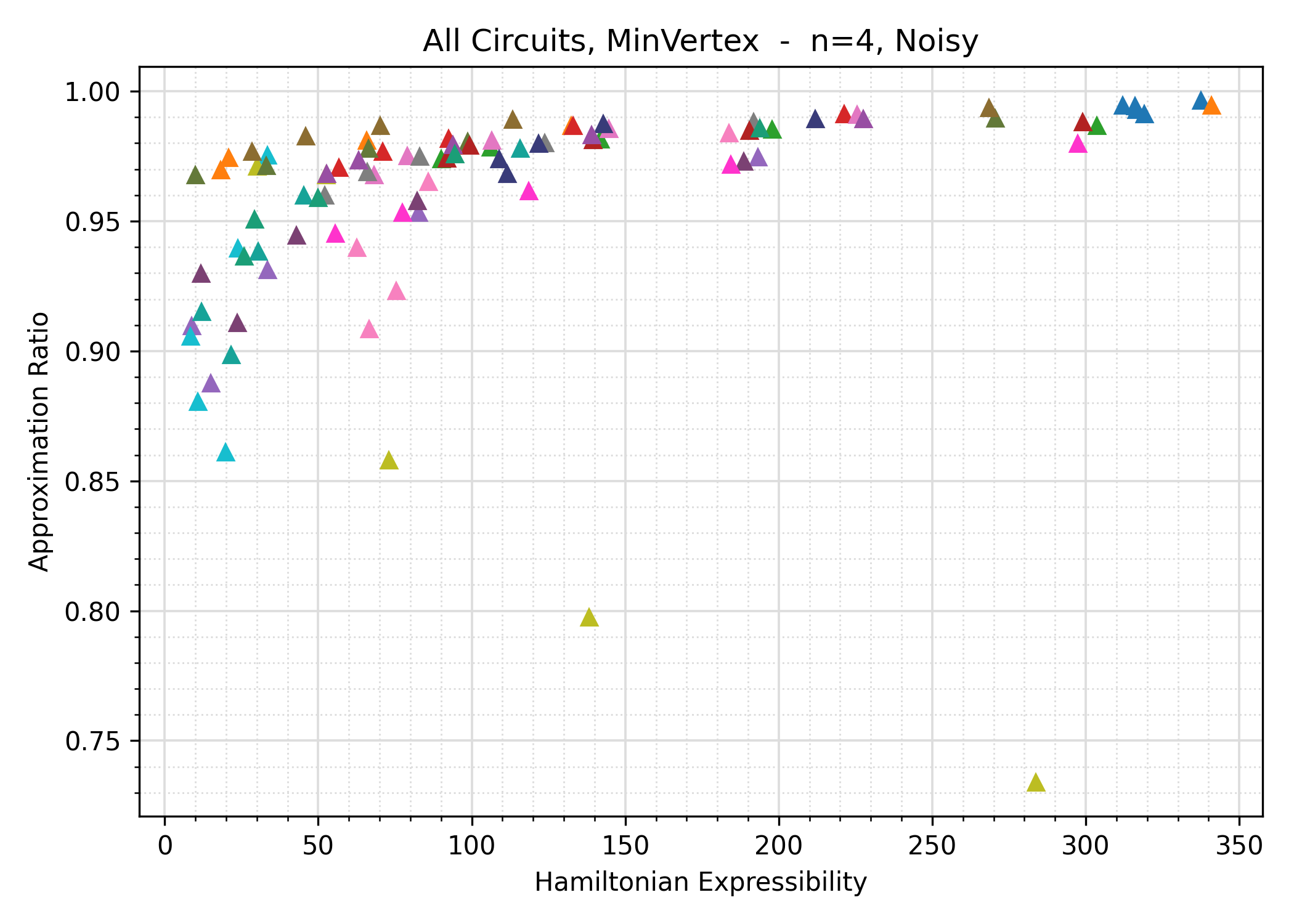}
 \caption{Average normalized approximation ratio as a function of Hamiltonian expressibility of all 4-qubit circuits over a specific instance of the MinVertex class, in the noisy case. Colours indicate the base circuit pattern from which each ansatz is derived. Ansätze built by varying the number of layers from the same pattern share the same colour.}
 \label{fig:MinVertexNoisy}
\end{figure}

\begin{figure}[!ht]
 \includegraphics[width=0.5\textwidth]{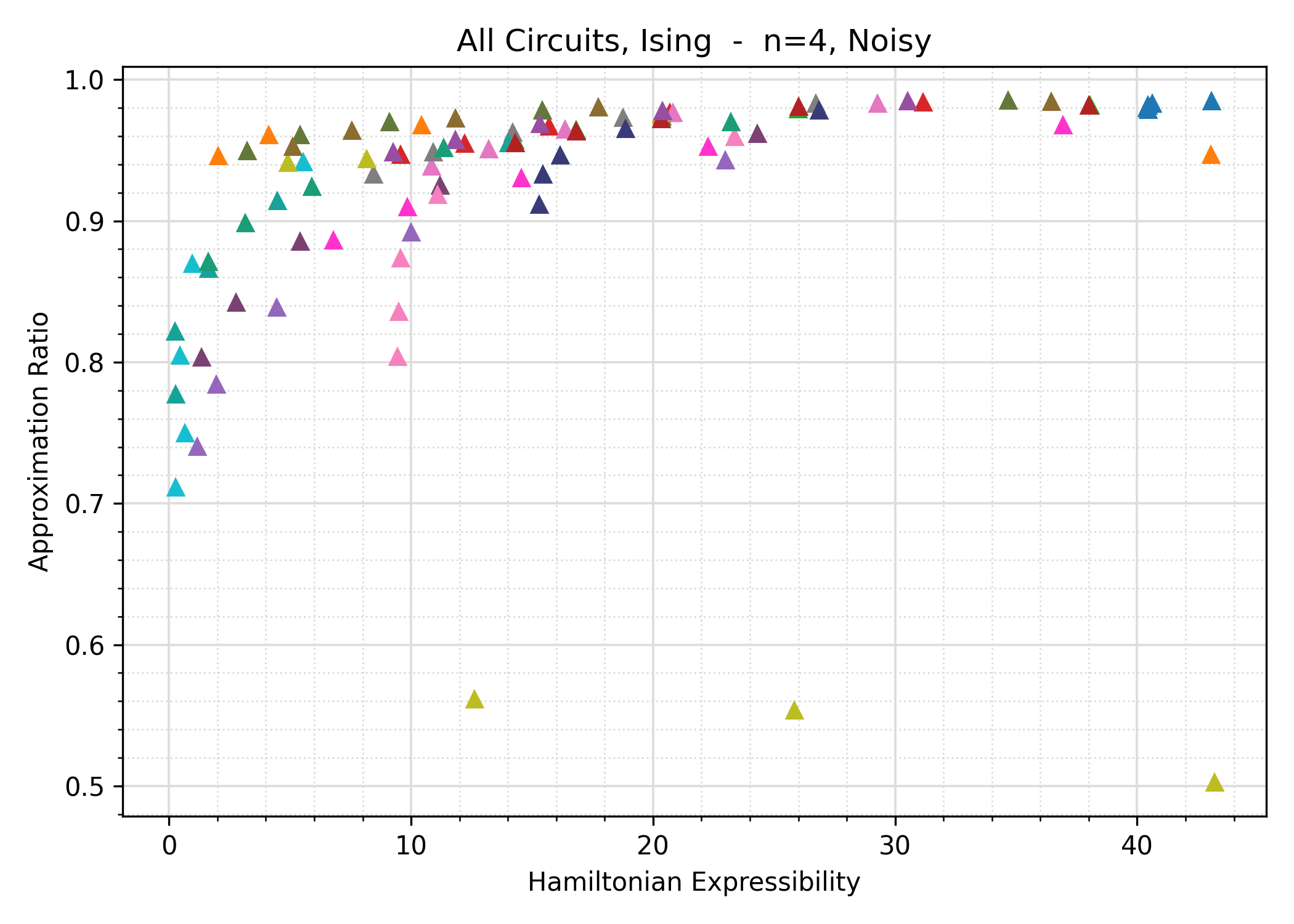}\hfill
 \includegraphics[width=0.5\textwidth]{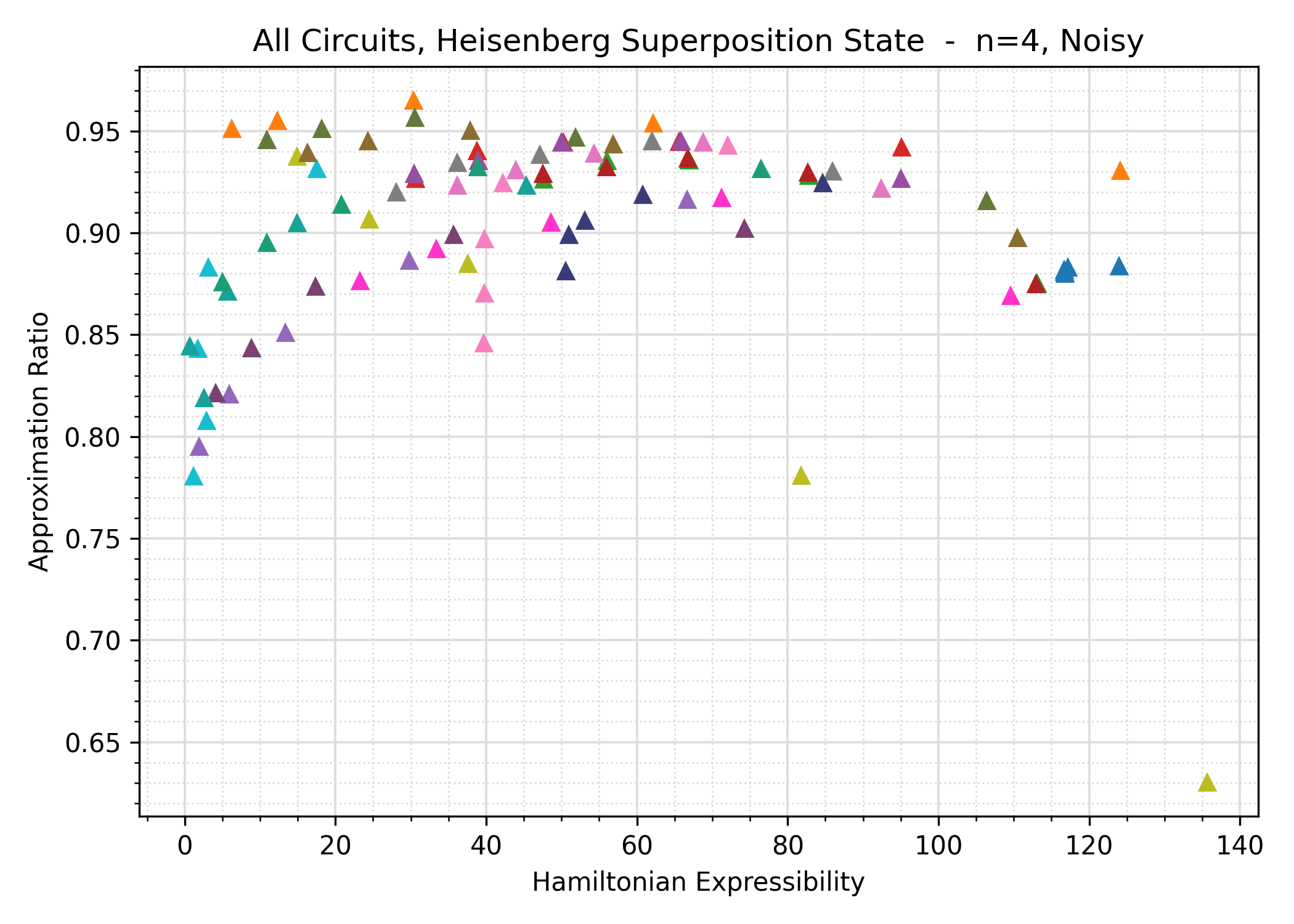}\hfill
 \caption{Average normalized approximation ratio as a function of Hamiltonian expressibility of all 4-qubit circuits over a specific instance of the Ising class (left panel) and over a specific instance of the Heisenberg Superposition State class (right panel), in the noisy case. Colours indicate the base circuit pattern from which each ansatz is derived. Ansätze built by varying the number of layers from the same pattern share the same colour.}
 \label{fig:NoisyHeis_Is}
\end{figure}

\paragraph{Trend Analysis of Approximation Ratio vs Hamiltonian Expressibility}
To better understand the effects of noise on the relationship between Hamiltonian expressibility and the average normalized approximation ratio, Figure~\ref{fig:MinVertexNoisy} reports these quantities for a representative instance of the MinVertex class, while Figure~\ref{fig:NoisyHeis_Is} reports them for representative instances of the Ising and Heisenberg Superposition State classes.

For the MinVertex class, which is representative of all the diagonal problem classes, Figure~\ref{fig:MinVertexNoisy} shows a generally increasing trend, apart from a few outliers, with a steep initial rise that gradually levels off.

For the non-diagonal classes, we observe two distinct trends: one shared by the Ising and Adiabatic classes, and another characteristic of the Heisenberg Superposition State and Random Non-Diagonal classes. Figure~\ref{fig:NoisyHeis_Is} illustrates examples of both of these patterns.

For the Ising class, the trend closely resembles that observed in diagonal problems: a steep initial increase in A.R. as expressibility improves, followed by a plateau.
In contrast, the Heisenberg class exhibits a different pattern. Here, the A.R. increases rapidly with expressibility, reaches a peak at an intermediate expressibility level, and then declines. Since lower numerical values of Hamiltonian expressibility correspond to higher expressibility, this suggests that increasing expressibility is beneficial initially, but then becomes less so when circuits are too expressive. A similar bell-shaped behaviour, with varying sharpness, is observed across all instances of the Heisenberg Superposition State and Random Non-Diagonal classes.

The origin of this bell-shaped trend is likely due to the greater A.R. degradation experienced by highly expressive circuits under noisy conditions. These circuits are generally more complex, involving deeper structures and a larger number of two-qubit gates, making them more susceptible to noise. In contrast, weakly expressive circuits tend to be simpler and shallower, and are therefore less affected by noise.

In any case, it is evident that there exists an optimal range of Hamiltonian expressibility which allows one to achieve good A.R. on these types of problems. The ideal circuit should be sufficiently complex to be expressive enough and able to explore the relevant solution space, but at the same time not too complex as it would suffer significant degradation due to noise.
Determining where this optimal range lies, whether it depends on individual problem instances or more broadly on the problem class, and, more generally, understanding the precise relationship between Hamiltonian expressibility and the average normalized approximation ratio underlying these bell-shaped trends, are important directions for future research.

Overall, these findings highlight the relevance of the Hamiltonian expressibility metric for the Heisenberg Superposition State and Random Non-Diagonal classes even in noisy scenarios. However, unlike in the ideal setting, its practical application here is not as straightforward as in the ideal case.

\paragraph{Evolution of Correlation Coefficients with Increasing Noise}
To investigate how the transition from ideal to noisy conditions affects the relationship between expressibility and A.R. (as expressed by the correlation coefficients), we employ the error rate metric defined in~\eqref{eq:err} to analyse how these coefficients evolve as the level of noise increases.

\begin{figure}[!ht]
\centering
 \includegraphics[width=0.85\textwidth]{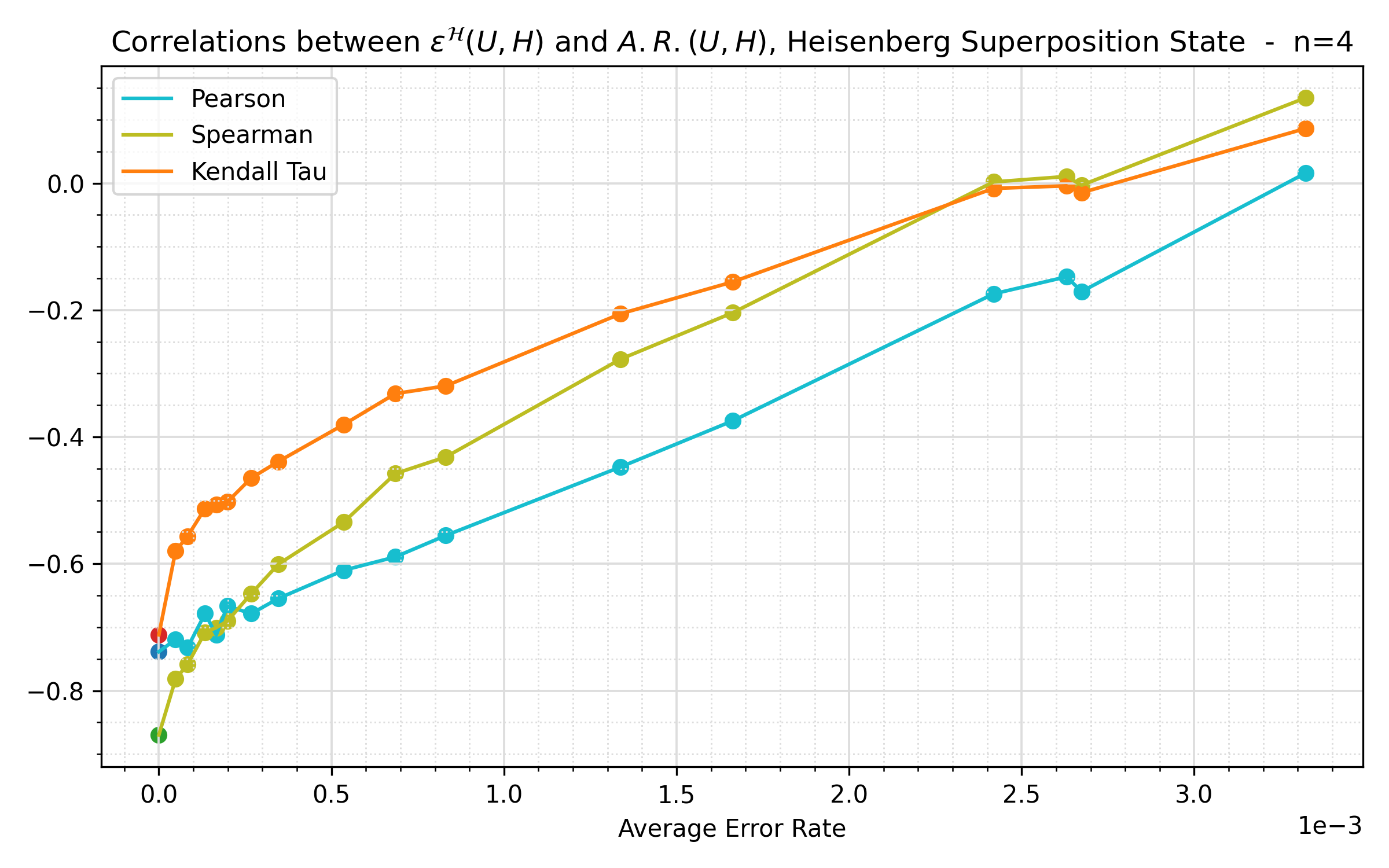}
 \caption{Trend of all correlation coefficients between average normalized approximation ratio and Hamiltonian expressibility, as a function of the average error rate of the backend adopted. All coefficients are evaluated on a specific instance of the Heisenberg Superposition State class, with all circuits of 4 qubits. Darker coloured points represent the values of the ideal case.}
 \label{fig:Error_Tuning}
\end{figure}

Figure~\ref{fig:Error_Tuning} shows the evolution of Pearson, Spearman, and Kendall Tau coefficients as the average hardware-induced error rate increases, on a specific instance of the Heisenberg Superposition State class. Note that the rightmost data point corresponds to the previously analysed noisy case with noise parameters $T_1 = T_2 = 200 \mu s$, $ err_1 = 1.6 \times 10 ^{-4} $, and $err_2 = 4 \times 10 ^{-3}$, yielding an average error rate of approximately $3.3 \times 10^{-3}$.

From this graph, we observe that as the error rate increases, the correlation coefficients transition from moderately negative values to approximately $+0.1$, following a nearly linear and gradual trend.
In the ideal case, for the Heisenberg Superposition State class (as for most problems whose solutions lie in superposition states), we previously observed a decreasing trend between the average normalized approximation ratio and Hamiltonian expressibility (Figure~\ref{fig:HamExpr_vs_ar_HEIS_4}). In contrast, the noisy case revealed a bell-shaped relationship (Figure~\ref{fig:NoisyHeis_Is}). These results suggest that the emergence of the bell-shaped trend occurs gradually as noise increases.

\begin{figure}[!ht]
\centering
 \includegraphics[width=0.85\textwidth]{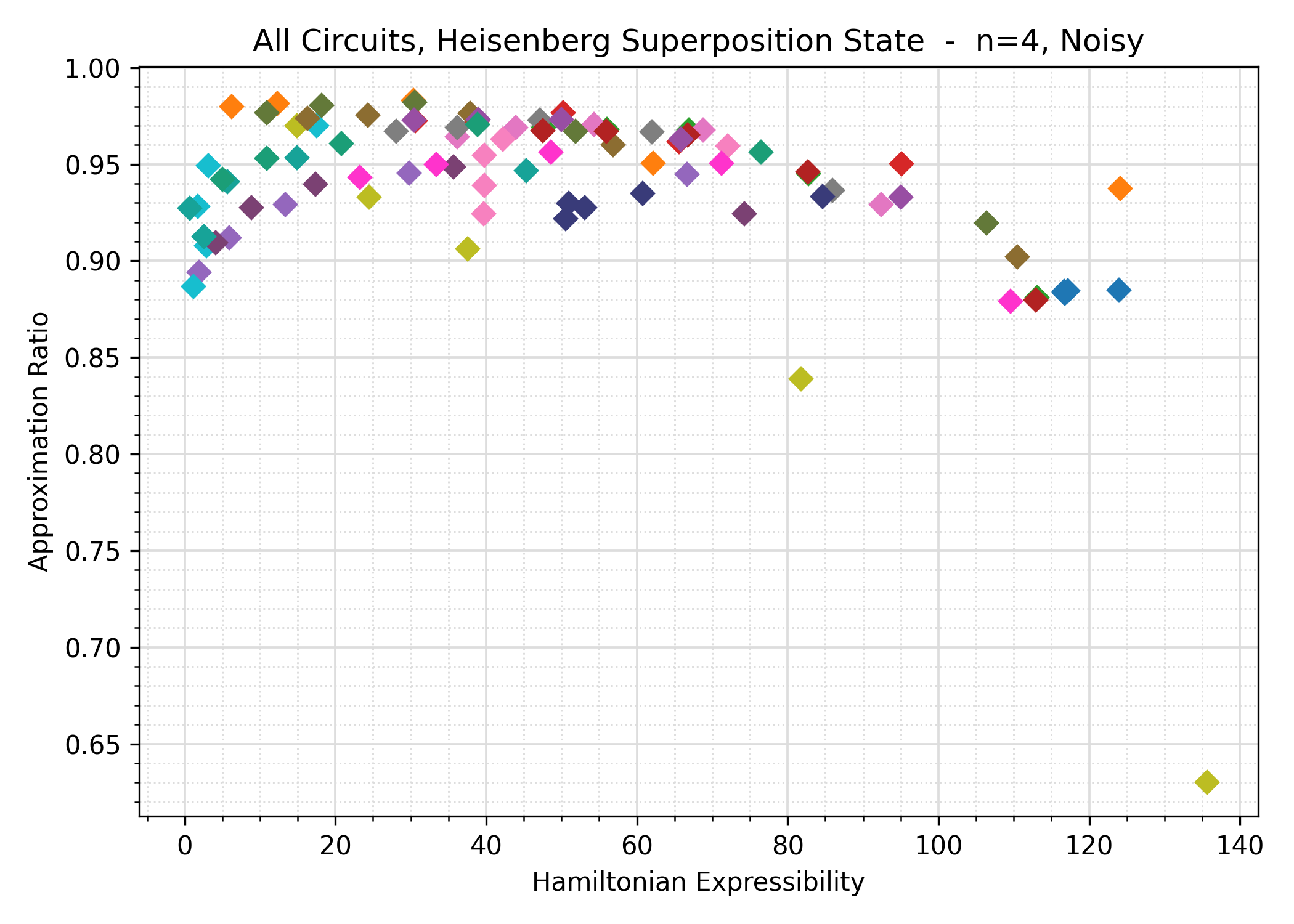}\hfill
 \caption{Average normalized approximation ratio as a function of Hamiltonian expressibility of all 4-qubit circuits over a specific instance of the Heisenberg Superposition State class, with an intermediate level of noise. Colours indicate the base circuit pattern from which each ansatz is derived. Ansätze built by varying the number of layers from the same pattern share the same colour. The average error rate is approximately $1.3 \times 10 ^{-3}$, given by noise parameters $T_1 = T_2 = 400 \mu s$, $err_1 = 4 \times 10^{-5}$, and $err_2 = 1 \times 10^{-3}$. }
 \label{fig:Heis_Interm_Noise}
\end{figure}

This hypothesis is supported by Figure~\ref{fig:Heis_Interm_Noise}, which shows the same relationship for all circuits applied to a Heisenberg Superposition State instance under intermediate noise conditions. In this case, the average error rate is approximately $1.5 \times 10^{-3}$, lower than that used in the previously analysed noisy setting. Under these conditions, the bell-shaped curve remains visible, but is less pronounced.

Based on these results, we conclude that the findings presented in Section~\ref{subsec:noiseless_results}, which support the use of highly Hamiltonian-expressive circuits for problems with non-diagonal Hamiltonians or superposition state solutions, remain valid in scenarios with very low (though not necessarily ideal) noise levels. Specifically, Figure~\ref{fig:Error_Tuning} shows that when the average error rate falls below $0.5 \times 10^{-3}$, the correlation coefficients already closely resemble those observed in the ideal case. In practice, our simulations achieve a comparable average error rate using the following noise parameters: $T_1 = 1500 \mu s$, $T_2 = 1500 \mu s$, $err_1 = 8 \times 10^{-6}$, and $err_2 = 2 \times 10^{-4}$.

\paragraph{Mutual Information}

\begin{figure}[!ht]
\begin{center}
 
 \includegraphics[width=0.85\textwidth]{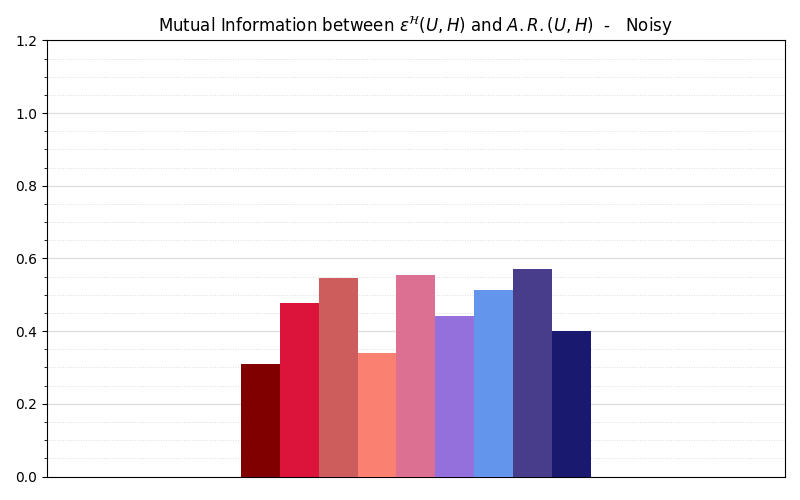}\hfill
 \includegraphics[width=\textwidth]{ultimate_legend.png}
 \caption{Average mutual information between Hamiltonian expressibility and average normalized approximation ratio, when adopting a noisy setting for the VQE. The values are obtained for circuits and problems of 4 qubits. Colouring in shades of red indicates problem classes with diagonal Hamiltonians or with a basis state solution, while blue colours represent problem classes based on non-diagonal Hamiltonians with a superposition state solution.}
 \label{fig:Noisy_Mutual_Info}
\end{center}
\end{figure}

We now present the results of our analysis of the mutual information between Hamiltonian expressibility and the average normalized approximation ratio under noisy conditions. Figure~\ref{fig:Noisy_Mutual_Info} shows, for each problem class, the mutual information values in the 4-qubit setting.

When comparing these results to the ideal (noise-free) scenario, we first observe a notable uniformity across all problem classes. This suggests that the informational power of Hamiltonian expressibility remains consistent regardless of problem type. Indeed, recalling the trends illustrated in Figure~\ref{fig:MinVertexNoisy} and \ref{fig:NoisyHeis_Is}, we note that although the relationship between Hamiltonian expressibility and normalized approximation ratio is different (some curves exhibit a monotonic increase while others show a decline), they appear to encode comparable levels of information content.

Furthermore, we observe a notable reduction in informational power for those problem classes whose solutions lie in superposition states, with mutual information values now ranging between $0.3$ and $0.6$. This finding aligns with our earlier conclusions regarding the increased difficulty of extracting useful insights from Hamiltonian expressibility metrics in noisy conditions for this problem category.
Nevertheless, the fact that mutual information does not vanish entirely, combined with the trend observations discussed in previous sections, leaves open potential avenues for future research into the precise nature of the relationship between expressibility and approximation ratio.

\section{Conclusions}
\label{sec:conclusions}
In this work, we analysed the role of Hamiltonian expressibility and Hamiltonian expressibility ratio in selecting the most suitable parametric quantum circuit within the context of Variational Quantum Algorithms. To achieve this, we developed an experimental protocol to estimate such metrics for a given circuit on a given problem described by a Hamiltonian matrix. We then picked a set of representative circuits of 4 and 8 qubits, designed a set of notable problem Hamiltonians, and computed the expressibility metrics for all circuits across all problems.

Upon analysing the results, we found that Hamiltonian expressibility improves when the depth of the circuit increases, eventually saturating either before or shortly after reaching a maximal expressive threshold, which corresponds to the sampling from the Haar uniform distribution. Furthermore, we discovered that Hamiltonian expressibility is actually a problem-oriented metric and can distinguish the ability of different circuits to explore the energy spaces of various types of problems. Notably, we observed how the energy spaces of problems encapsulated by a diagonal Hamiltonian are easier to explore in a manner similar to the Haar distribution. In these cases, circuits consistently achieve higher levels of expressibility compared to problems with non-diagonal Hamiltonians.

Most importantly, we trained each circuit to solve all problem instances using the VQE algorithm, under both ideal and noisy conditions. We measured the solution accuracy by averaging the normalized approximation ratios over multiple runs and investigated its relationship with Hamiltonian expressibility-based metrics. To this end, we computed the Pearson, Spearman, and Kendall Tau correlation coefficients, as well as the mutual information coefficient, between the accuracy metric and each expressibility measure.
From this correlation analysis, we derived our main contributions, which can be summarized as follows. 

In ideal or minimally disturbed settings:
\begin{itemize}
 \item For small-scale problems with diagonal Hamiltonians or basis state solutions, low Hamiltonian expressibility is preferable. Conversely, high expressibility proves advantageous for problems characterized by non-diagonal Hamiltonians or superposition state solutions.
 \item The correlation between Hamiltonian expressibility and VQE effectiveness tends to weaken as the number of qubits increases. This observation aligns with theoretical results linking high expressibility to the occurrence of barren plateaus and associated trainability issues.
\end{itemize}

This behaviour could be intuitively motivated by the structure of the underlying solution spaces. In particular, diagonal Hamiltonians or problems with basis state solutions correspond to simpler and more constrained landscapes, where increased exploration capability is largely unnecessary and may even be detrimental due to increased circuit complexity. In contrast, non-diagonal Hamiltonians with superposition ground states define richer and more complex energy landscapes, where higher expressibility becomes beneficial as it enables more effective exploration of the solution space.

Furthermore, in a noisy and small-scale scenario:
\begin{itemize}
 \item For problems with diagonal Hamiltonians or basis state solutions, low Hamiltonian expressibility becomes even more advantageous than in the noiseless case.
 \item For problems with non-diagonal Hamiltonians and superposition state solutions, circuits with low Hamiltonian expressibility perform better. However, for certain superposition state problems, such as instances of the Heisenberg XXZ model where the ground state is a superposition, it is preferable to select circuits with intermediate levels of Hamiltonian expressibility. In such cases, circuits with low expressibility may fail to sufficiently explore the energy landscape, while highly expressive circuits, typically deeper and more complex, are more prone to noise-induced errors.
\end{itemize}

Our results suggest that the Hamiltonian expressibility metric can serve as a useful tool for guiding ansatz selection, albeit primarily for small-scale problems, in both ideal and noisy settings.
In particular, problems with more complex solutions tend to benefit from circuit selection strategies informed by expressibility metrics. 

Moreover, there is considerable potential for further investigation into the relationship between Hamiltonian expressibility and effectiveness within the VQE algorithm, especially in noisy environments.
In particular, a dedicated study could be conducted to better understand the underlying factors behind the emergence of optimal expressibility ranges in specific settings, and to determine where such ranges lie for different problem instances or classes.
Such insights could enable more targeted and effective ansatz selection strategies.

Finally, novel metrics based on Hamiltonian expressibility, accounting more directly for the problem structure and the impact of noise, could be developed and may prove even more effective in guiding the choice of quantum circuits for VQAs. 
These metrics could also be integrated into circuit design frameworks based on adaptive strategies or machine learning techniques, further enhancing their effectiveness.

\backmatter

\section*{Acknowledgements}

We acknowledge the financial support from ICSC - ``National Research Centre in High Performance Computing, Big Data and Quantum Computing'', funded by the European Union - Next Generation EU.
We also acknowledge the support and computational resources provided by E4 Computer Engineering S.p.A.
Finally, we acknowledge the financial support from the PNRR MUR project PE0000023-NQSTI and from the MUR Progetti di Ricerca di Rilevante Interesse Nazionale (PRIN) Bando 2022 - project n. 20227HSE83 – ThAI-MIA funded by the European Union - Next Generation EU.

\begin{appendices}

\section{Circuit Specifics}
\label{app:circuit_specifics}
In this appendix, we present the structures of all circuits used in our experiments. Figure~\ref{fig:Circuits} shows the 19 circuit templates on 4 qubits. As shown, all circuits are composed of one- and two-qubit gates. Single-qubit gates are either non-parametric (Hadamard gates) or parametric (rotations $R_X$, $R_Y$, and $R_Z$). Two-qubit gates can also be non-parametric (controlled-X and controlled-Z gates) or parametric (controlled rotations $CR_X$, $CR_Y$, and $CR_Z$). In each circuit, a specific set of gates constitutes one layer, which is repeated up to five times in our experiments.

\begin{figure}[p]
    \centering

    \begin{subfigure}[t]{0.30\textwidth}
        \centering
        \begin{tikzpicture}
            \node[scale=0.55]{
                \begin{quantikz}
                    \ket{0} & \gate{R_X} & \gate{R_Z} & \\
                    \ket{0} & \gate{R_X} & \gate{R_Z} & \\
                    \ket{0} & \gate{R_X} & \gate{R_Z} & \\
                    \ket{0} & \gate{R_X} & \gate{R_Z} & \\
                \end{quantikz}
            };
        \end{tikzpicture}
        \caption{Circuit 1}
    \end{subfigure}\hfill
    \begin{subfigure}[t]{0.30\textwidth}
        \centering
        \begin{tikzpicture}
            \node[scale=0.55]{
                \begin{quantikz}
                    \ket{0} & \gate{R_X} & \gate{R_Z} &    &       & \targ{}    & \\
                    \ket{0} & \gate{R_X} & \gate{R_Z} &    & \targ{} & \ctrl{-1} & \\
                    \ket{0} & \gate{R_X} & \gate{R_Z} & \targ{} & \ctrl{-1} &         & \\
                    \ket{0} & \gate{R_X} & \gate{R_Z} & \ctrl{-1} &     &          & \\
                \end{quantikz}
            };
        \end{tikzpicture}
        \caption{Circuit 2}
    \end{subfigure}\hfill
    \begin{subfigure}[t]{0.30\textwidth}
        \centering
        \begin{tikzpicture}
            \node[scale=0.55]{
                \begin{quantikz}
                    \ket{0} & \gate{R_X} & \gate{R_Z} &    &       & \gate{R_Z}  & \\
                    \ket{0} & \gate{R_X} & \gate{R_Z} &    & \gate{R_Z} & \ctrl{-1} & \\
                    \ket{0} & \gate{R_X} & \gate{R_Z} & \gate{R_Z} & \ctrl{-1} &      & \\
                    \ket{0} & \gate{R_X} & \gate{R_Z} & \ctrl{-1} &     &          & \\
                \end{quantikz}
            };
        \end{tikzpicture}
        \caption{Circuit 3}
    \end{subfigure}

    \vspace{1em}

    \begin{subfigure}[t]{0.30\textwidth}
        \centering
        \begin{tikzpicture}
            \node[scale=0.55]{
                \begin{quantikz}
                    \ket{0} & \gate{R_Z} & \gate{R_X} &    &       & \gate{R_X}  & \\
                    \ket{0} & \gate{R_Z} & \gate{R_X} &    & \gate{R_X} & \ctrl{-1} & \\
                    \ket{0} & \gate{R_Z} & \gate{R_X} & \gate{R_X} & \ctrl{-1} &      & \\
                    \ket{0} & \gate{R_Z} & \gate{R_X} & \ctrl{-1} &     &          & \\
                \end{quantikz}
            };
        \end{tikzpicture}
        \caption{Circuit 4}
    \end{subfigure}
    \begin{subfigure}[t]{0.50\textwidth}
        \centering
        \begin{tikzpicture}
            \node[scale=0.55]{
                \begin{quantikz}
                    \ket{0} & \gate{R_X} & \gate{R_Z} & \gate{R_Z} & \gate{R_X} & \gate{R_Z} & & \\
                    \ket{0} & \gate{R_X} & \gate{R_Z} & \ctrl{-1} & \gate{R_X} & \gate{R_Z} & \gate{R_Z} & \\
                    \ket{0} & \gate{R_X} & \gate{R_Z} & \gate{R_Z} & \gate{R_X} & \gate{R_Z} & \ctrl{-1} & \\
                    \ket{0} & \gate{R_X} & \gate{R_Z} & \ctrl{-1} & \gate{R_X} & \gate{R_Z} & & \\
                \end{quantikz}
            };
        \end{tikzpicture}
        \caption{Circuit 7}
    \end{subfigure}
    
    \vspace{1em}

    \begin{subfigure}[t]{1.0\textwidth}
        \centering
        \begin{tikzpicture}
            \node[scale=0.55]{
                \begin{quantikz}
                    \ket{0} & \gate{R_X} & \gate{R_Z} &  & & \gate{R_Z} & & & \gate{R_Z} & & & \gate{R_Z} & \ctrl{3} & \ctrl{2} & \ctrl{1} & \gate{R_X} & \gate{R_Z} & \\
                    \ket{0} & \gate{R_X} & \gate{R_Z} &  & \gate{R_Z} & & & \gate{R_Z} & & \ctrl{2} & \ctrl{1} & \ctrl{-1} & & & \gate{R_Z} & \gate{R_X} & \gate{R_Z} & \\
                    \ket{0} & \gate{R_X} & \gate{R_Z} & \gate{R_Z} & & & \ctrl{1} & \ctrl{-1} & \ctrl{-2} & & \gate{R_Z} & & & \gate{R_Z} & & \gate{R_X} & \gate{R_Z} & \\
                    \ket{0} & \gate{R_X} & \gate{R_Z} & \ctrl{-1} & \ctrl{-2} & \ctrl{-3} & \gate{R_Z} & & & \gate{R_Z} & & & \gate{R_Z} & & & \gate{R_X} & \gate{R_Z} & \\
                \end{quantikz}
            };
        \end{tikzpicture}
        \caption{Circuit 5}
    \end{subfigure}\hfill

    \vspace{1em}

    \begin{subfigure}[t]{1.0\textwidth}
        \centering
        \begin{tikzpicture}
            \node[scale=0.55]{
                \begin{quantikz}
                    \ket{0} & \gate{R_X} & \gate{R_Z} &  & & \gate{R_X} & & & \gate{R_X} & & & \gate{R_X} & \ctrl{3} & \ctrl{2} & \ctrl{1} & \gate{R_X} & \gate{R_Z} & \\
                    \ket{0} & \gate{R_X} & \gate{R_Z} &  & \gate{R_X} & & & \gate{R_X} & & \ctrl{2} & \ctrl{1} & \ctrl{-1} & & & \gate{R_X} & \gate{R_X} & \gate{R_Z} & \\
                    \ket{0} & \gate{R_X} & \gate{R_Z} & \gate{R_X} & & & \ctrl{1} & \ctrl{-1} & \ctrl{-2} & & \gate{R_X} & & & \gate{R_X} & & \gate{R_X} & \gate{R_Z} & \\
                    \ket{0} & \gate{R_X} & \gate{R_Z} & \ctrl{-1} & \ctrl{-2} & \ctrl{-3} & \gate{R_X} & & & \gate{R_X} & & & \gate{R_X} & & & \gate{R_X} & \gate{R_Z} & \\
                \end{quantikz}
            };
        \end{tikzpicture}
        \caption{Circuit 6}
    \end{subfigure}\hfill

    \caption{The employed circuits. Rotation gates are parametric. Each circuit, from start to finish, consists of one layer that can be repeated. For Circuit 10, the layer starts from the red dashed line.}
    \label{fig:Circuits}
\end{figure}

\begin{figure}[p]
    \ContinuedFloat
    \centering
    
    \begin{subfigure}[t]{0.45\textwidth}
        \centering
        \begin{tikzpicture}
            \node[scale=0.55]{
                \begin{quantikz}
                    \ket{0} & \gate{R_X} & \gate{R_Z} & \gate{R_X} & \gate{R_X} & \gate{R_Z} & & \\
                    \ket{0} & \gate{R_X} & \gate{R_Z} & \ctrl{-1} & \gate{R_X} & \gate{R_Z} & \gate{R_X} & \\
                    \ket{0} & \gate{R_X} & \gate{R_Z} & \gate{R_X} & \gate{R_X} & \gate{R_Z} & \ctrl{-1} & \\
                    \ket{0} & \gate{R_X} & \gate{R_Z} & \ctrl{-1} & \gate{R_X} & \gate{R_Z} & & \\
                \end{quantikz}
            };
        \end{tikzpicture}
        \caption{Circuit 8}
    \end{subfigure}
    \begin{subfigure}[t]{0.25\textwidth}
        \centering
        \begin{tikzpicture}
            \node[scale=0.55]{
                \begin{quantikz}
                    \ket{0} & \gate{H} &    &       & \control{} & \gate{R_X} & \\
                    \ket{0} & \gate{H} &    & \control{} & \ctrl{-1} & \gate{R_X} & \\
                    \ket{0} & \gate{H} & \control{} & \ctrl{-1} &      & \gate{R_X} & \\
                    \ket{0} & \gate{H} & \ctrl{-1} &     &          & \gate{R_X} & \\
                \end{quantikz}
            };
        \end{tikzpicture}
        \caption{Circuit 9}
    \end{subfigure}\hfill
    \begin{subfigure}[t]{0.30\textwidth}
        \centering
        \begin{tikzpicture}
            \node[scale=0.55]{
                \begin{quantikz}
                    \ket{0} & \gate{R_X} & \gate{R_Z} & \gate{R_Z} & & \\
                    \ket{0} & \gate{R_X} & \gate{R_Z} & \ctrl{-1} & \gate{R_Z} & \\
                    \ket{0} & \gate{R_X} & \gate{R_Z} & \gate{R_Z} & \ctrl{-1} & \\
                    \ket{0} & \gate{R_X} & \gate{R_Z} & \ctrl{-1} & & \\
                \end{quantikz}
            };
        \end{tikzpicture}
        \caption{Circuit 16}
    \end{subfigure}\hfill
    \begin{subfigure}[t]{0.30\textwidth}
        \centering
        \begin{tikzpicture}
            \node[scale=0.55] {  
                \begin{quantikz}
                    \ket{0}  &  \gate{R_Y}  \slice{} &[1cm]      &       &            \control{} & \control{}  & \gate{R_X} & \\
                    \ket{0}  & \gate{R_Y}  &    &          \control{}   &  \ctrl{-1} & & \gate{R_X} &\\
                    \ket{0} & \gate{R_Y} &  \control{}    &  \ctrl{-1} &     &      & \gate{R_X} &\\
                    \ket{0} & \gate{R_Y}  & \ctrl{-1}  &     &       & \ctrl{-3 }        & \gate{R_X} & 
                \end{quantikz}
            };
        \end{tikzpicture}
        \caption{Circuit 10}
    \end{subfigure}\hfill
    \begin{subfigure}[t]{0.40\textwidth}
        \centering
        \begin{tikzpicture}
            \node[scale=0.55]{
                \begin{quantikz}
                    \ket{0} & \gate{R_Y} & \gate{R_Z} & \targ{} & & &  & \\
                    \ket{0} & \gate{R_Y} & \gate{R_Z} & \ctrl{-1} & \gate{R_Y} & \gate{R_Z} & \targ{} & \\
                    \ket{0} & \gate{R_Y} & \gate{R_Z} & \targ{} & \gate{R_Y} & \gate{R_Z} & \ctrl{-1} & \\
                    \ket{0} & \gate{R_Y} & \gate{R_Z} & \ctrl{-1} & & &  & \\
                \end{quantikz}
            };
        \end{tikzpicture}
        \caption{Circuit 11}
    \end{subfigure}\hfill
    \begin{subfigure}[t]{0.30\textwidth}
        \centering
        \begin{tikzpicture}
            \node[scale=0.55]{
                \begin{quantikz}
                    \ket{0} & \gate{R_Y} & \gate{R_Z} & \control{} & & &  & \\
                    \ket{0} & \gate{R_Y} & \gate{R_Z} & \ctrl{-1} & \gate{R_Y} & \gate{R_Z} & \control{} & \\
                    \ket{0} & \gate{R_Y} & \gate{R_Z} & \control{} & \gate{R_Y} & \gate{R_Z} & \ctrl{-1} & \\
                    \ket{0} & \gate{R_Y} & \gate{R_Z} & \ctrl{-1} & & &  & \\
                \end{quantikz}
            };
        \end{tikzpicture}
        \caption{Circuit 12}
    \end{subfigure}\hfill
    \begin{subfigure}[t]{0.60\textwidth}
        \centering
        \begin{tikzpicture}
            \node[scale=0.55]{
                \begin{quantikz}
                    \ket{0} & \gate{R_Y} & \gate{R_Z} &&& \ctrl{1} & \gate{R_Y} && \ctrl{3} & \gate{R_Z} && \\
                    \ket{0} & \gate{R_Y} &&& \ctrl{1} & \gate{R_Z} & \gate{R_Y} &&& \ctrl{-1} & \gate{R_Z} & \\
                    \ket{0} & \gate{R_Y} && \ctrl{1} & \gate{R_Z} && \gate{R_Y} & \gate{R_Z} &&& \ctrl{-1} & \\
                    \ket{0} & \gate{R_Y} & \ctrl{-3} & \gate{R_Z} &&& \gate{R_Y} & \ctrl{-1} & \gate{R_Z} &&& \\
                \end{quantikz}
            };
        \end{tikzpicture}
        \caption{Circuit 13}
    \end{subfigure}\hfill
    \begin{subfigure}[t]{0.60\textwidth}
        \centering
        \begin{tikzpicture}
            \node[scale=0.55]{
                \begin{quantikz}
                    \ket{0} & \gate{R_Y} & \gate{R_X} &&& \ctrl{1} & \gate{R_Y} && \ctrl{3} & \gate{R_X} && \\
                    \ket{0} & \gate{R_Y} &&& \ctrl{1} & \gate{R_X} & \gate{R_Y} &&& \ctrl{-1} & \gate{R_X} & \\
                    \ket{0} & \gate{R_Y} && \ctrl{1} & \gate{R_X} && \gate{R_Y} & \gate{R_X} &&& \ctrl{-1} & \\
                    \ket{0} & \gate{R_Y} & \ctrl{-3} & \gate{R_X} &&& \gate{R_Y} & \ctrl{-1} & \gate{R_X} &&& \\
                \end{quantikz}
            };
        \end{tikzpicture}
        \caption{Circuit 14}
    \end{subfigure}\hfill
    \begin{subfigure}[t]{0.31\textwidth}
        \centering
        \begin{tikzpicture}
            \node[scale=0.55]{
                \begin{quantikz}
                    \ket{0} & \gate{R_Y} & \targ{} &&& \ctrl{1} & \gate{R_Y} && \ctrl{3} & \targ{} && \\
                    \ket{0} & \gate{R_Y} &&& \ctrl{1} & \targ{} & \gate{R_Y} &&& \ctrl{-1} & \targ{} & \\
                    \ket{0} & \gate{R_Y} && \ctrl{1} & \targ{} && \gate{R_Y} & \targ{} &&& \ctrl{-1} & \\
                    \ket{0} & \gate{R_Y} & \ctrl{-3} & \targ{} &&& \gate{R_Y} & \ctrl{-1} & \targ{} &&& \\
                \end{quantikz}
            };
        \end{tikzpicture}
        \caption{Circuit 15}
    \end{subfigure}\hfill
    \begin{subfigure}[t]{0.25\textwidth}
        \centering
        \begin{tikzpicture}
            \node[scale=0.55]{
                \begin{quantikz}
                    \ket{0} & \gate{R_X} & \gate{R_Z} & \gate{R_X} & & \\
                    \ket{0} & \gate{R_X} & \gate{R_Z} & \ctrl{-1} & \gate{R_X} & \\
                    \ket{0} & \gate{R_X} & \gate{R_Z} & \gate{R_X} & \ctrl{-1} & \\
                    \ket{0} & \gate{R_X} & \gate{R_Z} & \ctrl{-1} & & \\
                \end{quantikz}
            };
        \end{tikzpicture}
        \caption{Circuit 17}
    \end{subfigure}\hfill
    \begin{subfigure}[t]{0.35\textwidth}
        \centering
        \begin{tikzpicture}
            \node[scale=0.55]{
                \begin{quantikz}
                    \ket{0} & \gate{R_X} & \gate{R_Z} & \gate{R_Z} &&& \ctrl{1} & \\
                    \ket{0} & \gate{R_X} & \gate{R_Z} &&& \ctrl{1} & \gate{R_Z} & \\
                    \ket{0} & \gate{R_X} & \gate{R_Z} && \ctrl{1} & \gate{R_Z} && \\
                    \ket{0} & \gate{R_X} & \gate{R_Z} & \ctrl{-3} & \gate{R_Z} &&& \\
                \end{quantikz}
            };
        \end{tikzpicture}
        \caption{Circuit 18}
    \end{subfigure}\hfill
    \begin{subfigure}[t]{0.31\textwidth}
        \centering
        \begin{tikzpicture}
            \node[scale=0.55]{
                \begin{quantikz}
                    \ket{0} & \gate{R_X} & \gate{R_Z} & \gate{R_X} &&& \ctrl{1} & \\
                    \ket{0} & \gate{R_X} & \gate{R_Z} &&& \ctrl{1} & \gate{R_X} & \\
                    \ket{0} & \gate{R_X} & \gate{R_Z} && \ctrl{1} & \gate{R_X} && \\
                    \ket{0} & \gate{R_X} & \gate{R_Z} & \ctrl{-3} & \gate{R_X} &&& \\
                \end{quantikz}
            };
        \end{tikzpicture}
        \caption{Circuit 19}
    \end{subfigure}

    \caption[]{The employed circuits (continued).}
\end{figure}

This circuit set was originally presented by \cite{sim_2019} and consists of notable circuit patterns from past studies. For instance, Circuits 5 and 6 were developed as programmable quantum circuits \cite{sousa_2006} and were applied to train the quantum autoencoder \cite{romero_2017}. Circuits 7 and 8 were used for the QVECTOR algorithm \cite{johnson_2017}. Circuit 9 was considered in \cite{wilson_2019}, and Circuit 10 followed the hardware-efficient circuit architecture from \cite{kandala_2017}. Circuits 11 and 12 were Josephson sampler circuits from \cite{geller_2018}. Circuits 13-15 and 18-19 were used for data classification \cite{schuld_2020}.

When adapting this set to the 8-qubit case, we preserve the original structural design while extending it to a larger number of qubits. For instance, Circuit 1 on 8 qubits consists of a first layer of $R_X$ gates applied to all qubits, followed by a layer of $R_Z$ gates applied to all qubits. As a more complex example, Circuits 5 and 6 include a layer of controlled rotations in which each qubit acts as a control for the others. The same structural pattern is extended to the 8-qubit case.

\FloatBarrier
    
\section{Formulation of Notable Optimization Problems}
\label{app:notable_problems}

In this appendix, we detail the optimization problems used in our experiments, providing their mathematical formulation and the corresponding Hamiltonian representation based on their mathematical definitions. We start by introducing the three QUBO problems analysed in our work, namely Maximum Cut, Maximum Clique, and Minimum Vertex Cover, followed by the Heisenberg XXZ problem, the Transverse Field Ising model, and the Adiabatic Optimization problem.

\subsection{Quadratic Unconstrained Binary Optimization Problems}
\label{app:combinat_problems}

In this section, we present the QUBO problems Maximum Cut, Maximum Clique, and Minimum Vertex Cover. 

An important preliminary note is that the Hamiltonians corresponding to these problems are diagonal matrices. As a result, their ground states always correspond to computational basis states. This behaviour is expected, as the cost functions of these problems are defined over binary variables, typically representing a node in a graph. Consequently, the solution space consists of the finite set of all possible bit-strings of length $n$, and the optimal solution is one of such bit-strings. This bit-string can then be directly mapped to a computational basis state.

\subsubsection{Maximum Cut Problem}

Given an undirected graph $G = (V, E)$, where $V$ is the set of $n$ vertices and $E$ is the set of edges, the Maximum Cut problem seeks a partition of the vertices into two disjoint subsets that maximizes the number of edges connecting vertices across the partition.

The Hamiltonian formulation of this problem for a graph with $n$ vertices is given by:
\begin{align}
H = \sum_{(i,j) \in E} \frac{1}{2}(Z_i Z_j - Id),
\label{eq:MaxCut_def}
\end{align}
\mbox{}\\
where $Z_i$ denotes the unitary matrix corresponding to the Pauli $Z$ operator applied to qubit $i$ in a register of $n$ qubits.

\subsubsection{Minimum Vertex Cover Problem}

Given an undirected graph $G = (V, E)$, the Minimum Vertex Cover problem aims to find the smallest subset $S \subseteq V$ such that every edge in $E$ has at least one endpoint in $S$.

The Hamiltonian formulation for this problem for a graph with $n$ vertices is:
\begin{align}
H = \sum_{i=1}^{n} \frac{1}{2}(Id - Z_i) + \sum_{(i,j) \in E} \frac{p}{4}(Id + Z_i + Z_j + Z_i Z_j),
\label{eq:MinVertex_def}
\end{align}
\mbox{}\\
where $p$ is a penalty hyper-parameter, and $Z_i$ is the Pauli $Z$ operator acting on qubit $i$.

\subsubsection{Maximum Clique Problem}

Given an undirected graph $G = (V, E)$, the Maximum Clique problem aims to identify the largest clique in the graph. A clique is a subset of fully connected vertices. More formally, a subset $S \subseteq V$ is a clique if the subgraph $G' = (S, E')$, where $E' = \lbrace (i, j) \in E \mid i,j \in S \rbrace$, is a fully connected graph.

The Hamiltonian formulation for this problem for a graph with $n$ vertices is:
\begin{align}
H = \sum_{i=1}^{n} -\frac{1}{2}(Id - Z_i) + \sum_{(i,j) \in \overline{E}} \frac{1}{2}(Id - Z_i - Z_j + Z_i Z_j),
\label{eq:MaxClique_def}
\end{align}
where $\overline{E}$ denotes the set of non-edges (\idest pairs of vertices not connected in $G$), and $Z_i$ is the Pauli $Z$ operator applied to qubit $i$.

\subsection{Heisenberg XXZ Problem}

The Heisenberg XXZ model~\citep{gopalakrishnan_2019} is a quantum spin chain model that describes interactions between spins on a one-dimensional lattice in the presence of a transverse magnetic field. It is a generalization of the standard Heisenberg model and is widely studied in condensed matter physics, particularly in the contexts of quantum magnetism, spin liquids, and quantum phase transitions.

Its Hamiltonian for $n$ qubits is given by:
\begin{align}
 H = \sum_{i=1}^n \left( X_i X_{i+1} + Y_i Y_{i+1} + \Delta Z_i Z_{i+1} \right) + g \sum_{i=1}^n Z_i,
 \label{eq:Heisengerg_def}
\end{align}
where $\Delta$ is the anisotropy parameter, $g$ is the relative strength of the external magnetic field, and $X_i$, $Y_i$, and $Z_i$ are the $2^n$-dimensional unitary matrices corresponding to the application of the Pauli $X$, Pauli $Y$, and Pauli $Z$ operators, respectively, to qubit $i$. We assume periodic boundary conditions, such that $n + 1 \equiv 1$.

\subsection{Transverse Field Ising Problem}

The Transverse Field Ising model~\citep{pfeuty_1970} is a quantum spin system widely used in condensed matter physics to investigate quantum phase transitions and critical phenomena. It extends the classical Ising model, which describes spins interacting on a lattice, by introducing a transverse magnetic field that induces quantum fluctuations.

Its Hamiltonian formulation for $n$ qubits is the following:
\begin{align}
 H = - J \sum_{i=1}^n X_iX_{i+1} - g \sum_{i=1}^n Z_i,
 \label{eq:Ising_def}
\end{align}
where $J$ indicates the interaction strength, $g$ is the magnitude of the transverse magnetic field, and $X_i$, $Z_i$ are the $2^n$-dimensional unitary matrices corresponding to the application of the Pauli $X$ and Pauli $Z$ gates, respectively, to qubit $i$.
Again, periodic boundary conditions are assumed, and therefore $n+1 \equiv 1$.

\subsection{Adiabatic Problem}

The adiabatic problem in quantum computing refers to the use of adiabatic quantum computation~\citep{albash_2018}, a computational paradigm that exploits quantum mechanical principles to solve optimization problems by gradually evolving a quantum system from an initial easy-to-prepare ground state to a final state that encodes the solution to the problem.

Essentially, the adiabatic problem involves finding the ground state of a Hamiltonian that encodes a specific optimization problem. If the system begins in the ground state of an initial Hamiltonian and is evolved slowly enough, the adiabatic theorem ensures that it will remain in the ground state, a process known as adiabatic evolution.

The general Hamiltonian formulation for an adiabatic problem is:
\begin{align}
 H(s) = -\frac{A(s)}{2} \sum_{i=1}^n X_i + \frac{B(s)}{2} H_p,
 \label{eq:Adiabatic_def}
\end{align}
where $A(s)$ and $B(s)$ are schedule functions that depend on the normalized time parameter $s \in [0, 1]$, $H_p$ is the problem Hamiltonian encoding the objective function, and $X_i$ denotes the Pauli $X$ operator acting on the qubit $i$.

\section{Sample Size}
\label{app:sample_size}

\begin{figure}[!ht]
 \centering
 \includegraphics[width=\textwidth{}]{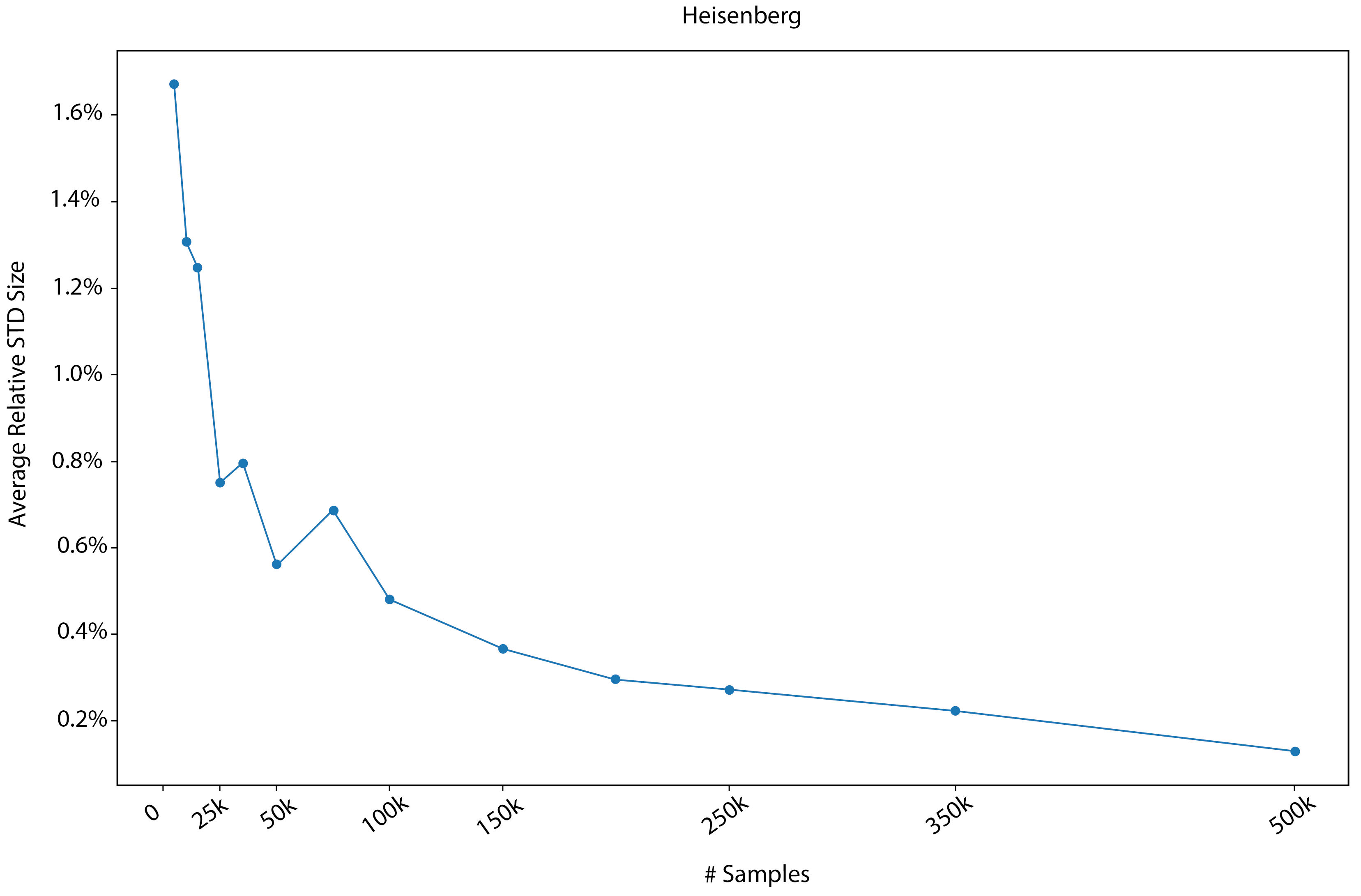}
 \caption{Average relative standard deviation for five estimates of $\mathcal{F}_{\text{Haar}}^{\mathcal{H}}(H)$ across all instances of the Heisenberg problem class, with respect to the sample size $k$.}
 \label{fig:sample_size}
\end{figure}

We now provide a justification for the choice of the sample size $k$ used in the Monte Carlo estimation of Hamiltonian expressibility and Hamiltonian expressibility ratio.
To ensure a relative accuracy of approximately $0.5\%$, we set the sample size to $k = 250,\!000$. This value was determined by evaluating and analysing the relative standard deviation across five independent estimates of $\mathcal{F}_{\text{Haar}}^{\mathcal{H}}(H)$ for different problem instances, when sampling unitaries directly from the Haar distribution.
The analysis revealed that the relative standard deviation decreases rapidly as the sample size $k$ increases, eventually plateauing for sufficiently large sample sizes. In all tested cases, convergence to the desired accuracy occurred for $k \leq 250,\!000$, consistently achieving a relative error below $0.5\%$. This behaviour is illustrated in Figure~\ref{fig:sample_size}, which shows a representative example based on the Heisenberg problem class.

\section{Thresholds Computation}
\label{app:thresholds}

In this section, we describe the implementation details and rationale behind the concept of \textit{maximum expressibility thresholds} for the Hamiltonian expressibility and Hamiltonian expressibility ratio metrics.

As discussed in Section~\ref{subsec:expressibility_estimation}, the use of finite sample sizes in the Monte Carlo procedure introduces a systematic bias in the estimation of frame potentials. This bias affects the computed values of Hamiltonian expressibility and expressibility ratio. This issue is particularly significant for two reasons. 

Firstly, it becomes essential to establish a benchmark for the maximum Hamiltonian expressibility that accounts for this limitation. Specifically, circuits that perfectly replicate the Haar unitary distribution and thus satisfy the unitary $2$-design condition, should achieve a Hamiltonian expressibility of $0$ and a Hamiltonian expressibility ratio of $1$. However, in practice, these ideal values are unattainable due to the inherent limitations of finite sampling.

Secondly, as shown in (\ref{eq:ham_expr_final_estimate}) and (\ref{eq:ham_expr_ratio_final_estimate}) in Section~\ref{subsec:expressibility_estimation}, if $\tilde{\mathcal{F}}^{\mathcal{H}}(U, H) < \mathcal{F}^{\mathcal{H}}_{\text{Haar}}(H)$ due to this statistical noise, the resulting expressibility may become ill-defined, yielding a negative value under the square root (for Hamiltonian expressibility) or dropping below the theoretical lower bound of $1$ (for Hamiltonian expressibility ratio).

It is therefore necessary to define empirical thresholds for determining when a circuit can be considered to replicate the Haar distribution and to help mitigate precision-related errors, ensuring that the definition of expressibility remains well defined under all sampling conditions. 

To achieve this, for a given Hamiltonian $H$, we employ the same experimental protocol used to estimate Hamiltonian expressibility, but, instead of using circuit-generated unitaries, we sample unitaries directly from the Haar distribution. These thresholds, denoted by $\tilde{\varepsilon}^{\mathcal{H}}_{\text{Haar}}(H)$ and $\tilde{\gamma}^{\mathcal{H}}_{\text{Haar}}(H)$, are defined as follows:
\begin{align*}
 \tilde{\varepsilon}^{\mathcal{H}}_{\text{Haar}}(H) & \coloneq \sqrt{ \left| \tilde{\mathcal{F}}^{\mathcal{H}}_{\text{Haar}}(H) - \mathcal{F}^{\mathcal{H}}_{\text{Haar}}(H) \right| }, \\
 \tilde{\gamma}^{\mathcal{H}}_{\text{Haar}}(H) & \coloneq 1 + \frac{ \left| \tilde{\mathcal{F}}^{\mathcal{H}}_{\text{Haar}}(H) - \mathcal{F}^{\mathcal{H}}_{\text{Haar}}(H) \right| }{ \mathcal{F}^{\mathcal{H}}_{\text{Haar}}(H) }.
\end{align*}
Here, $\tilde{\mathcal{F}}^{\mathcal{H}}_{\text{Haar}}(H)$ denotes the estimated ansatz-Hamiltonian frame potential computed using unitaries sampled from the Haar distribution. The use of the absolute value ensures that both thresholds are strictly greater than $0$ and $1$, respectively, regardless of sampling variability.

Now we can refine the estimates of Hamiltonian expressibility and Hamiltonian expressibility ratio by incorporating the thresholds $\tilde{\varepsilon}^{\mathcal{H}}_{\text{Haar}}(H)$ and $\tilde{\gamma}^{\mathcal{H}}_{\text{Haar}}(H)$, thereby ensuring the validity of the computed values.
The modified estimators are defined as follows:
{\normalsize
\begin{equation}
\label{eq:ham_expr_w_threshold}
\tilde{\varepsilon}^{\mathcal{H}}(U, H) \coloneq
\begin{cases}
 \sqrt{ \tilde{\mathcal{F}}(U, H) - \mathcal{F}^{\mathcal{H}}_{\text{Haar}}(H)} & \text{if } \tilde{\mathcal{F}}(U, H) \geq \mathcal{F}^{\mathcal{H}}_{\text{Haar}}(H), \\
 \min\left\{ \sqrt{ \left| \tilde{\mathcal{F}}(U, H) - \mathcal{F}^{\mathcal{H}}_{\text{Haar}}(H) \right| }, \tilde{\varepsilon}^{\mathcal{H}}_{\text{Haar}}(H) \right\} & \text{otherwise}.
\end{cases}
\end{equation}

\vspace{0.5em}

\begin{equation}
\label{eq:ham_ratio_w_threshold}
\tilde{\gamma}^{\mathcal{H}}(U, H) \coloneq
\begin{cases}
 \dfrac{\tilde{\mathcal{F}}(U, H)}{\mathcal{F}^{\mathcal{H}}_{\text{Haar}}(H)} & \text{if } \tilde{\mathcal{F}}(U, H) \geq \mathcal{F}^{\mathcal{H}}_{\text{Haar}}(H), \\
 \min\left\{ 1 + \dfrac{ \left| \tilde{\mathcal{F}}(U, H) - \mathcal{F}^{\mathcal{H}}_{\text{Haar}}(H) \right| }{ \mathcal{F}^{\mathcal{H}}_{\text{Haar}}(H) }, \tilde{\gamma}^{\mathcal{H}}_{\text{Haar}}(H) \right\} & \text{otherwise}.
\end{cases}
\end{equation}
}

These definitions ensure that the expressibility metrics remain well-defined and physically meaningful, even in cases where the estimated ansatz-Hamiltonian frame potential, $\tilde{\mathcal{F}}(U, H)$ falls below the Haar-Hamiltonian frame potential $\mathcal{F}^{\mathcal{H}}_{\text{Haar}}(H)$, a scenario that is theoretically inadmissible.
By taking the absolute value of the difference of frame potentials, we guarantee that (\ref{eq:ham_expr_w_threshold}) always yields a non-negative value and (\ref{eq:ham_ratio_w_threshold}) remains strictly greater than $1$, consistent with the theoretical bounds on these metrics.

Furthermore, the incorporation of the $\min$ function ensures that expressibility estimates do not exceed their respective maximum expressibility thresholds. This refinement is especially pertinent when $\tilde{\mathcal{F}}(U, H) < \mathcal{F}^{\mathcal{H}}_{\text{Haar}}(H)$, and the left-hand terms in the minima of (\ref{eq:ham_expr_w_threshold}) and (\ref{eq:ham_ratio_w_threshold}) surpass the corresponding maximal expressibility thresholds. In such instances, the expressibility is effectively clipped at the threshold, indicating that the circuit is maximally expressive.

\section{Pearson and Kendall Tau Correlation Results}

\label{app:Pearson_KT}
In this appendix, we report the Pearson and Kendall Tau correlation coefficients between Hamiltonian expressibility and the average normalized approximation ratio for the 4-qubit and 8-qubit ideal settings and for the 4-qubit noisy setting.

\begin{figure}
\begin{center}
 \includegraphics[width=0.7\textwidth]{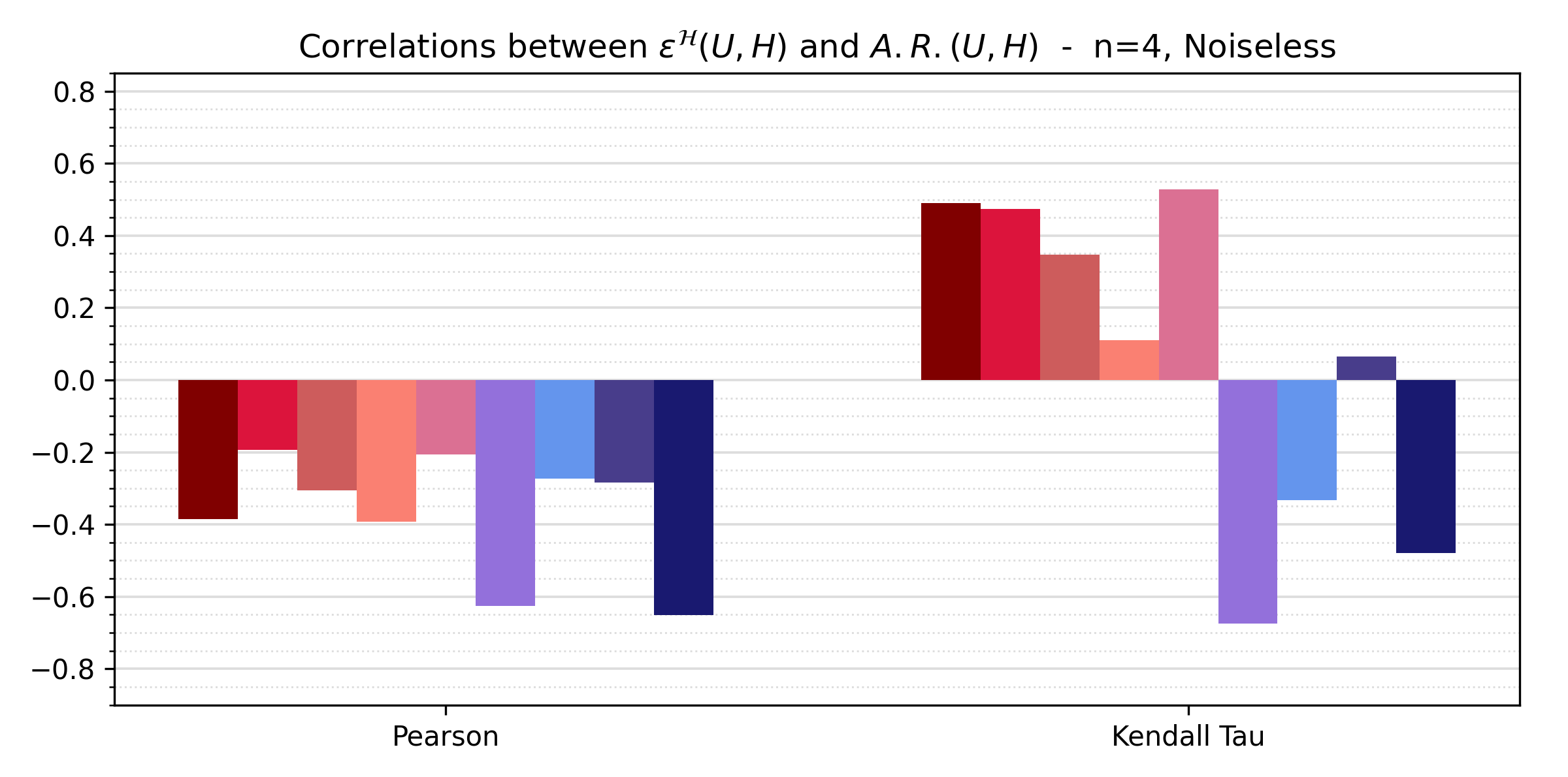} \hfill
 \includegraphics[width=0.7\textwidth]{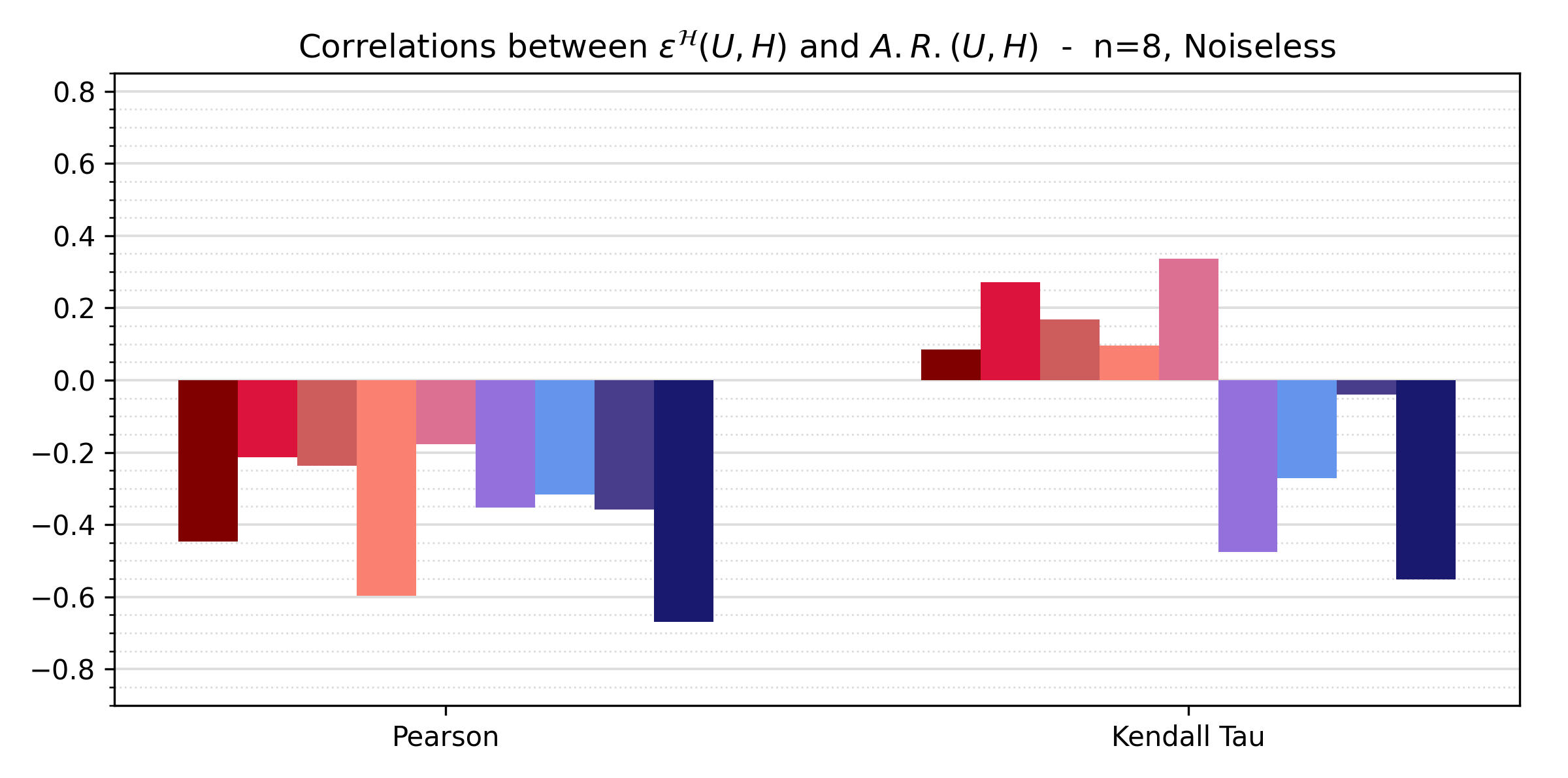}\hfill
 \includegraphics[width=0.7\textwidth]{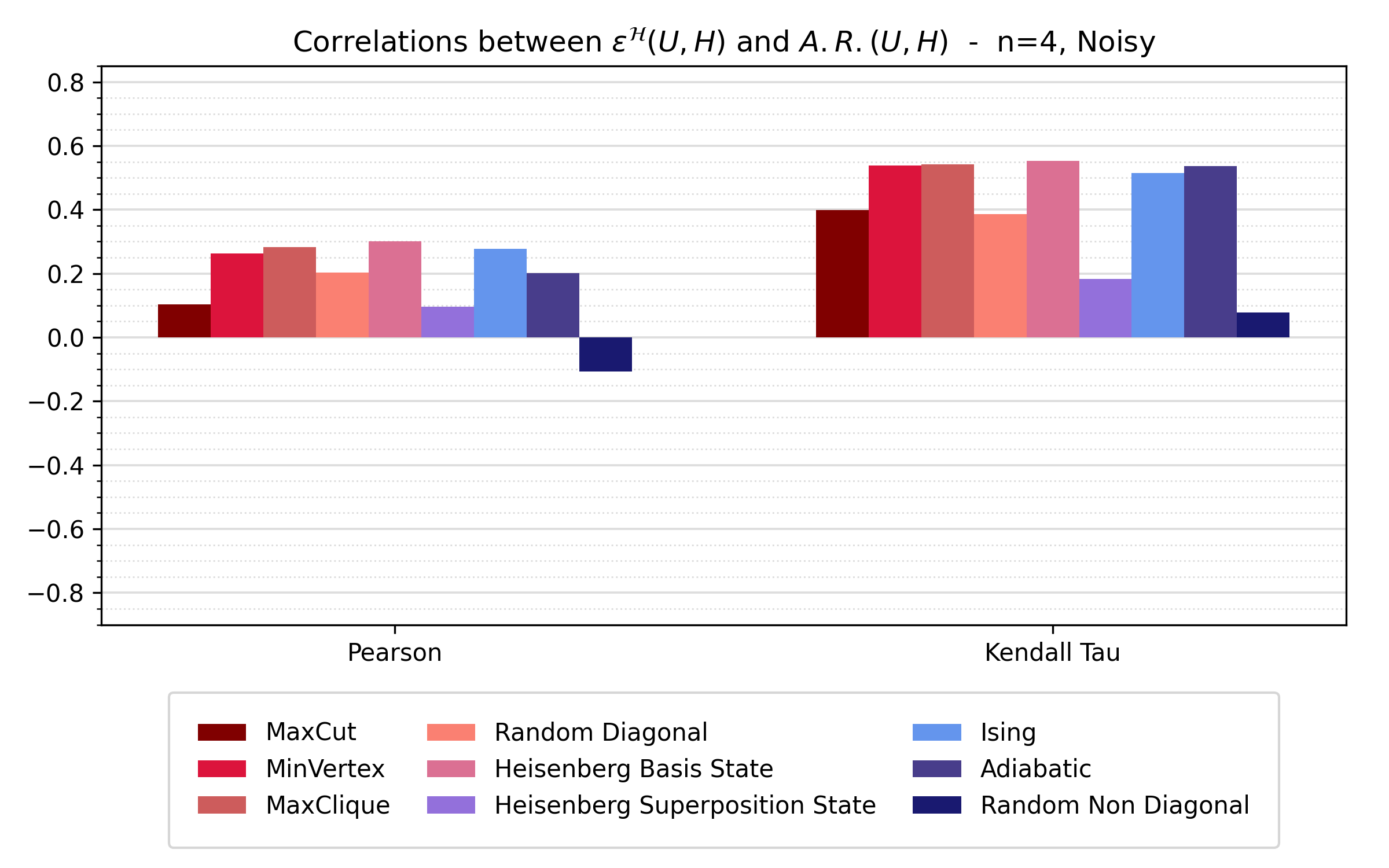}\hfill   
\end{center}
 \caption{Average Pearson and Kendall Tau correlation coefficients between Hamiltonian expressibility and the average normalized approximation ratio in a 4-qubit noiseless setting (top panel), an 8-qubit noiseless setting (middle panel), and a 4-qubit noisy setting (bottom panel). Colouring in shades of red indicates problem classes with diagonal Hamiltonians or with basis state solutions, while blue colours represent problem classes based on non-diagonal Hamiltonians with superposition state solutions.}
 \label{fig:Correlations_HamExpr_Pears_KT}
\end{figure}


For the 4-qubit ideal case (Figure~\ref{fig:Correlations_HamExpr_Pears_KT}, top panel), we observe relatively low Pearson correlation coefficient values for the MinVertex, MaxClique, Heisenberg Basis State, Ising, and Adiabatic classes, which suggests a weak linear correlation for these classes of problems. The MaxCut and Random Diagonal classes exhibit moderately negative Pearson correlation coefficients, suggesting the possible presence of a mild negative linear correlation between the considered quantities. Furthermore, the Heisenberg Superposition State and Random Non-Diagonal classes show stronger negative correlations, with coefficients as low as $-0.65$, providing evidence of a stronger decreasing linear relationship between the variables under consideration.

As for the Kendall Tau correlation coefficient in the 4-qubit ideal setting, we observe a similar behaviour to the one exhibited by the Spearman coefficient (Figure~\ref{fig:Correlations_4_Classes} in Section~\ref{subsec:noiseless_results}): moderately or highly negative coefficients for non-diagonal problem classes such as Adiabatic, Heisenberg Superposition State and Random Non-Diagonal (up to $-0.7$), and low to moderately positive coefficients for classes with a diagonal Hamiltonian or a basis state solution (up to $0.55$). Indeed, also Kendall Tau is designed to capture potential monotonic relationships, and so the same conclusions can be made: picking highly Hamiltonian-expressive circuits can be beneficial for problems with solutions in a superposition state, and detrimental for problems with solutions in computational basis states.

In the 8-qubit ideal case (Figure~\ref{fig:Correlations_HamExpr_Pears_KT}, middle panel), the Pearson coefficient ranges from slightly to moderately negative. However, many of these values are affected by outliers, like the one shown in Figure~\ref{fig:HamExpr_vs_ar_MaxCut_4} for the 4-qubit case. Once again, the Kendall Tau results follow the same pattern as the Spearman coefficient: a weakening in the contrast between problem classes with basis state solutions and those with superposition state solutions, possibly due to the onset of barren plateaus or similar trainability issues in complex and highly expressive circuits.

Finally, for the 4-qubit noisy case (Figure~\ref{fig:Correlations_HamExpr_Pears_KT}, bottom panel), we observe consistently low Pearson correlation coefficients across all problem classes: this indicates a lack of linear correlation between these metrics and the average normalized approximation ratio in the noisy setting. On the other hand, the Kendall Tau coefficient undergoes the same changes as the Spearman coefficient. The distinction observed in the ideal case between diagonal classes (or those with ground-state solutions, shown in shades of red) and classes with superposition state solutions (shown in shades of blue and violet) seems to disappear under noisy conditions. Most importantly, for non-diagonal classes, the Kendall Tau coefficients, previously moderately or strongly negative under ideal conditions, now exhibit low or even moderately positive values. This can be interpreted in the same way as for the Spearman coefficient, by analysing the trends between Hamiltonian expressibility and average normalized approximation ratio in Figure~\ref{fig:NoisyHeis_Is}, where in some cases highly expressive circuits are heavily penalized by noise, while in others an intermediate level of expressibility can be beneficial.

\section{Hamiltonian Expressibility Ratio Results}
\label{app:Ham_Ratio}

In this appendix, we report all correlation coefficients and the mutual information between Hamiltonian expressibility ratio and the average normalized approximation ratio for the 4-qubit and 8-qubit ideal settings and for the 4-qubit noisy setting.

\begin{figure}
\begin{center}
 \includegraphics[width=0.75\textwidth]{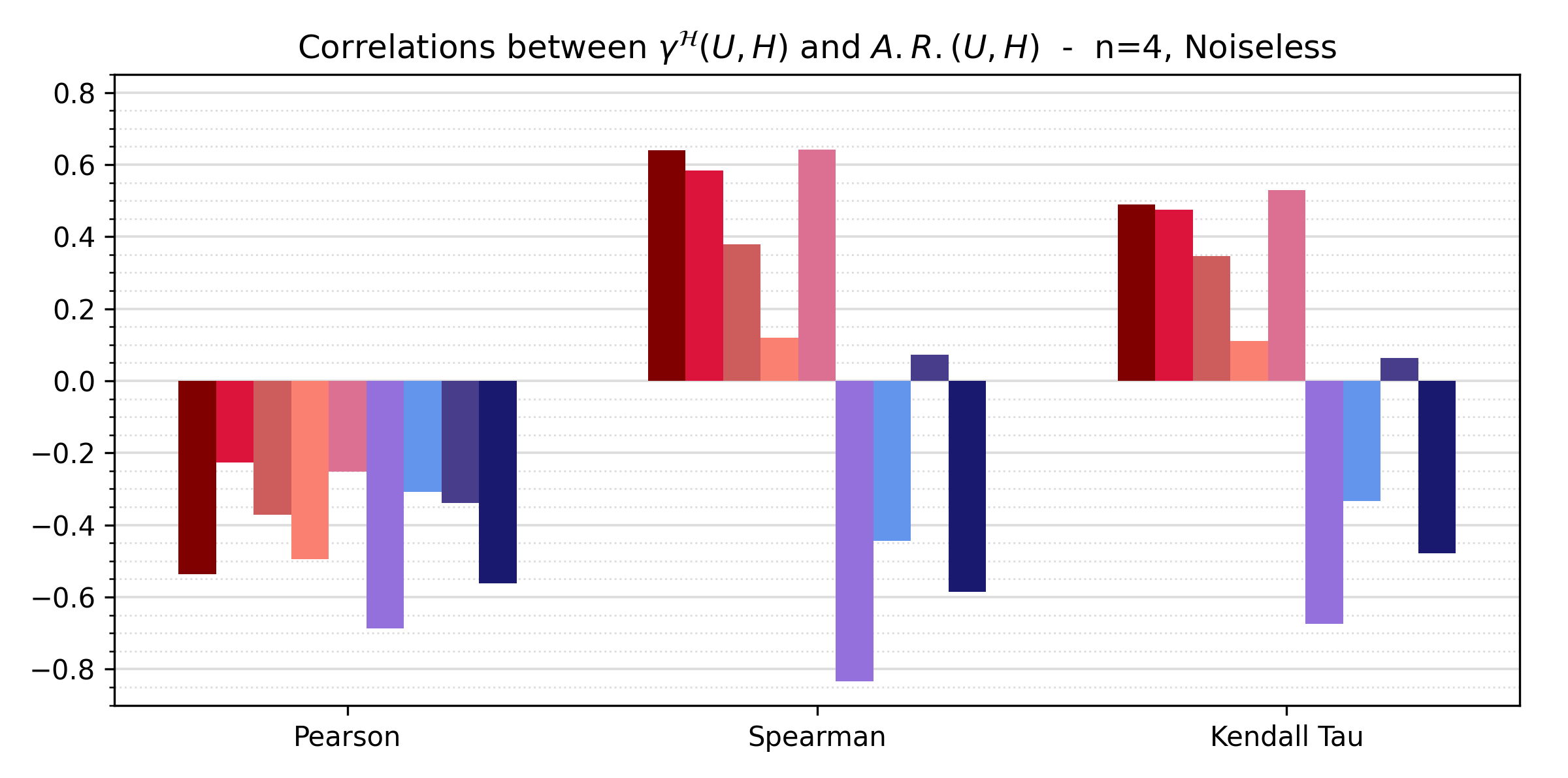} \hfill
 \includegraphics[width=0.75\textwidth]{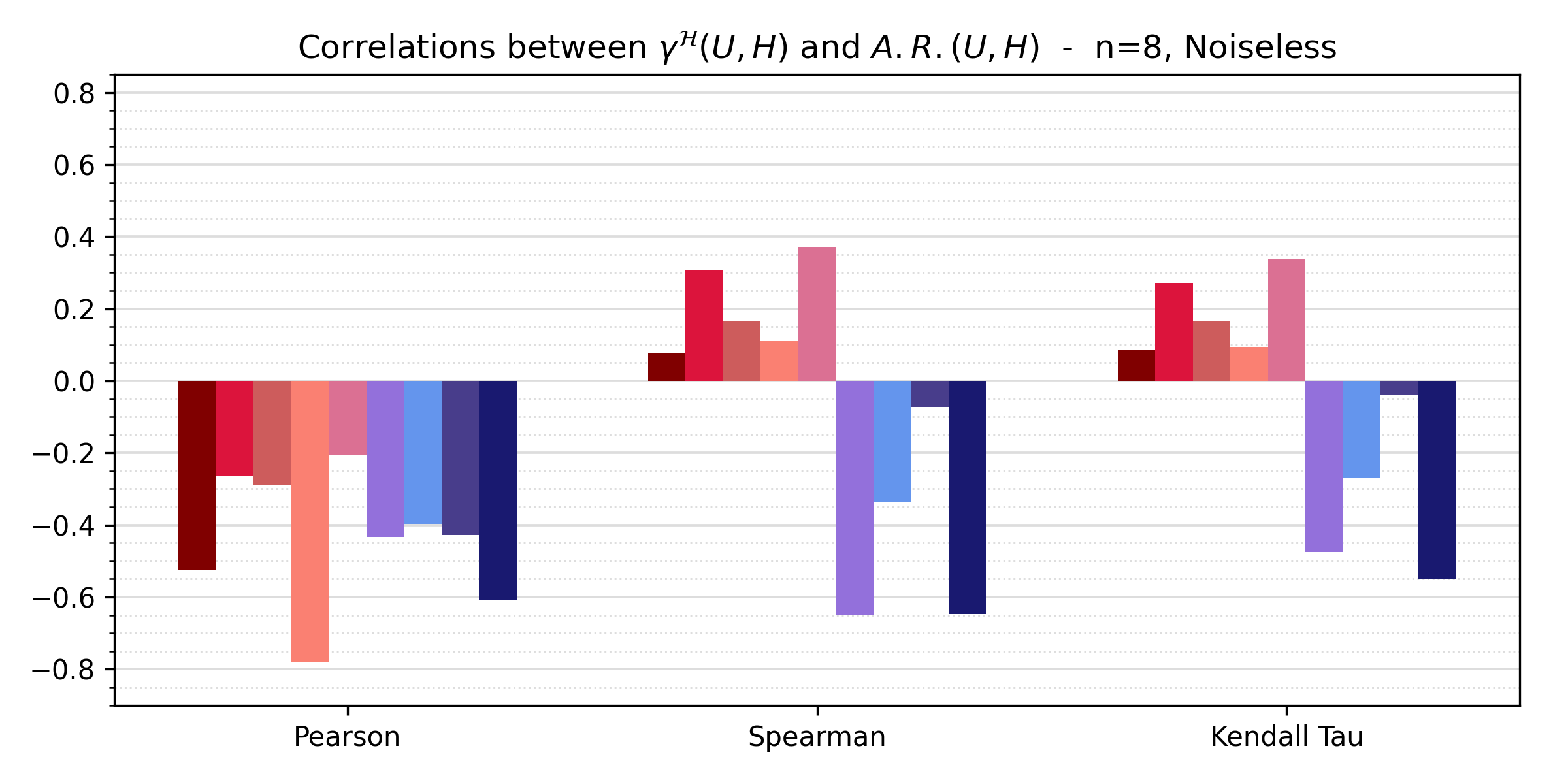}\hfill
 \includegraphics[width=0.75\textwidth]{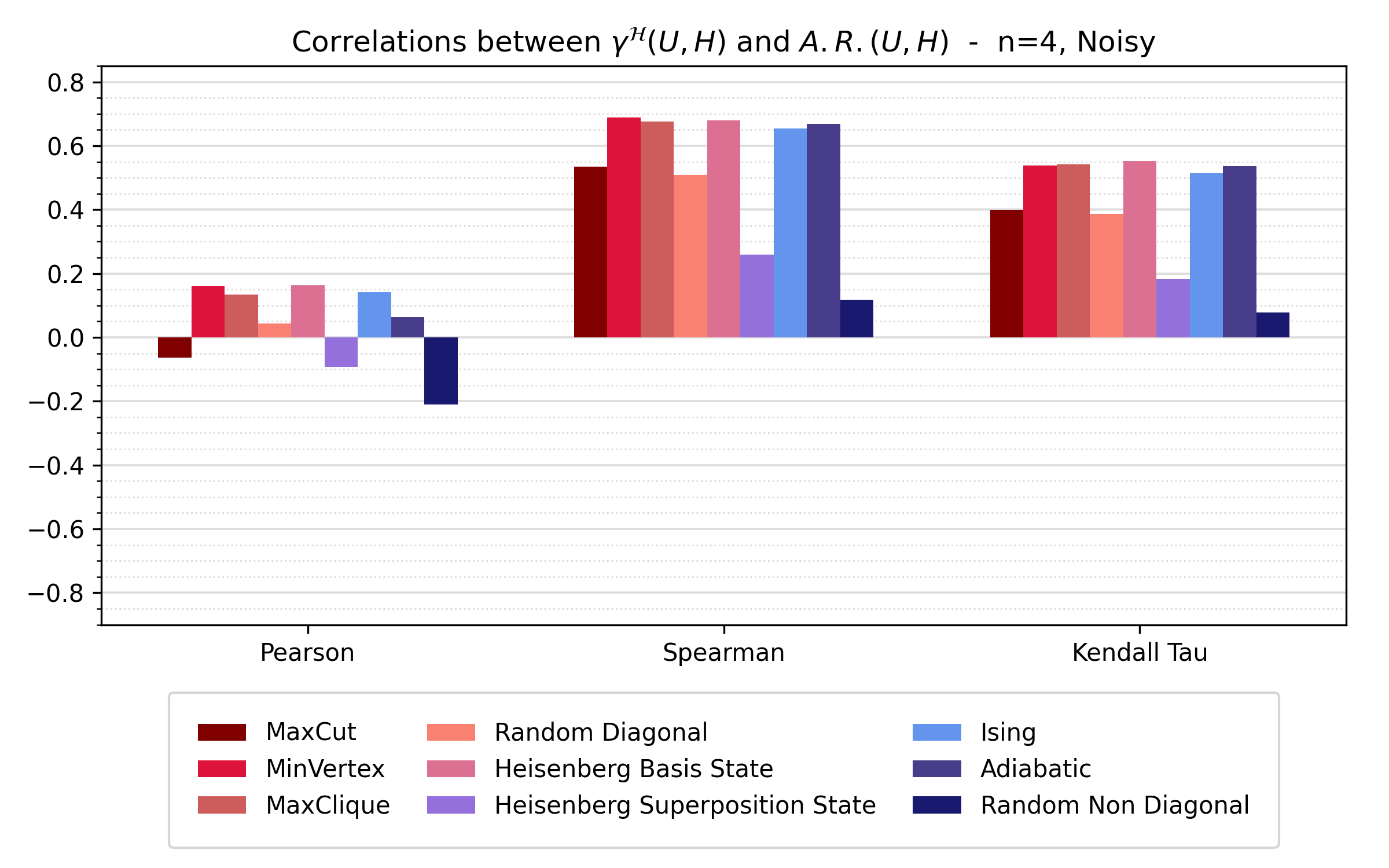}\hfill   
  \end{center}
 \caption{Average Pearson, Spearman, and Kendall Tau correlation coefficients between Hamiltonian expressibility ratio and the average normalized approximation ratio in a 4-qubit noiseless setting (top panel), an 8-qubit noiseless setting (middle panel), and a 4-qubit noisy setting (bottom panel). Colouring in shades of red indicates problem classes with diagonal Hamiltonians or with a basis state solution, while blue colours represent problem classes based on non-diagonal Hamiltonians with a superposition state solution.}
 \label{fig:Correlations_HamRatio}
\end{figure}

\begin{figure}[!ht]

\begin{center}
 
 \includegraphics[width=0.6\textwidth]{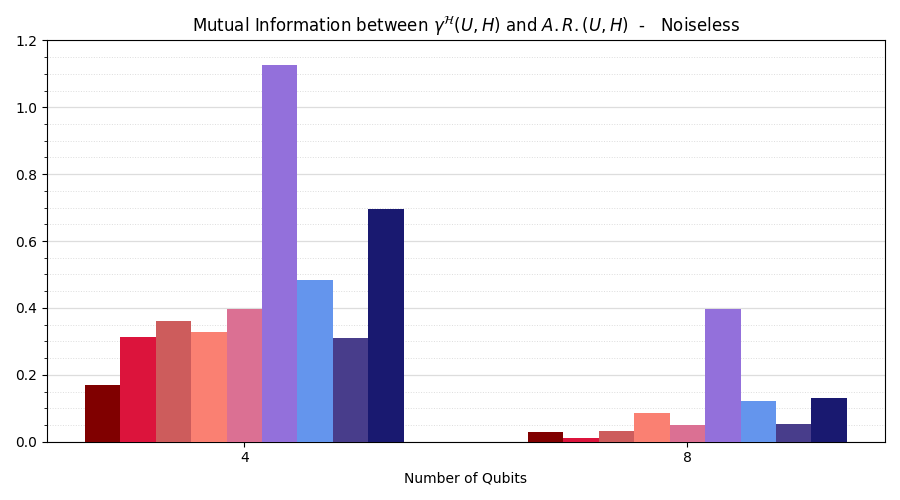}\hfill
 \includegraphics[width=0.4\textwidth]{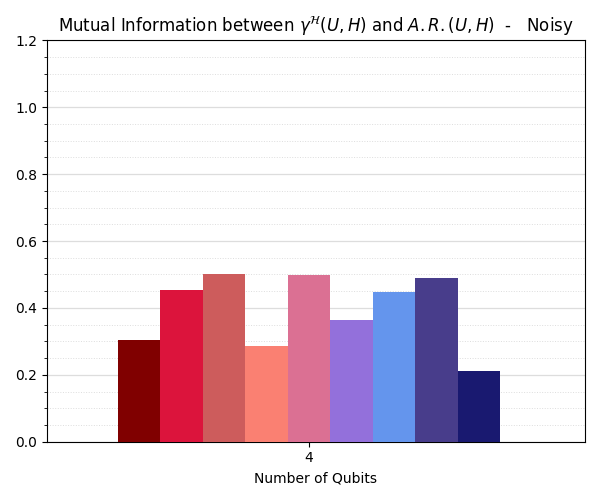}\hfill \\
 \includegraphics[width=\textwidth]{ultimate_legend.png}
 \caption{Average mutual information between Hamiltonian expressibility ratio and the average normalized approximation ratio for the ideal (left panel) and noisy (right panel) VQE settings. The values are obtained for circuits and problems with 4 and 8 qubits. Colouring in shades of red indicates problem classes with diagonal Hamiltonians or with a basis state solution, while blue colours represent problem classes based on non-diagonal Hamiltonians with a superposition state solution.}
  \label{fig:Mutual_Info_HamRatio}
\end{center}

\end{figure}

Figure~\ref{fig:Correlations_HamRatio} reports the obtained values of all correlation coefficients in all the considered settings. As shown, the results closely resemble those obtained with Hamiltonian expressibility (see Section~\ref{sec:results}, and Appendix~\ref{app:Pearson_KT}), and so the same considerations apply.
Similarly, Figure~\ref{fig:Mutual_Info_HamRatio} reports the mutual information values in all considered settings. Again, the results closely resemble those obtained with Hamiltonian expressibility (see Section~\ref{sec:results}), so the same considerations apply.

\section{Correlation with Number of Parameters and Parameter Dimension}
\label{app:N_Params_Param_Dim}
In this appendix, we compare the correlation results obtained using the Hamiltonian expressibility metric with those obtained from two additional quantities: the number of parameters of the ansatz $U(\theta)$, denoted by $P(U)$, and its Parameter Dimension, denoted by $D(U)$.

Let $\ket{\varphi(\theta)}$ be the output state of the parametrized ansatz $U(\theta)$. The Parameter Dimension \citep{Haug_2021} is a metric that quantifies the total number of independent parameters that the quantum state can effectively express. In order to define it, we first introduce the Quantum Fisher Information (QFI) matrix \citep{yamamoto_2019,Stokes_2020}, which naturally arises in gradient-based VQE optimization algorithms such as the Quantum Natural Gradient (QNG) method \citep{yamamoto_2019,Stokes_2020}. In the QNG framework, the parameters $\theta$ are updated according to:
\begin{align}
\theta_{k+1} = \theta_k - \eta \mathcal{F}^{-1}(\theta) \frac{\partial E(\theta)}{\partial \theta},
\label{eq:QNG}
\end{align}

where $\eta$ is a small learning rate, $\partial E(\theta)/\partial \theta$ is the gradient of the cost function, and $\mathcal{F}(\theta)$ is the QFI matrix, defined as:
\begin{align}
    \mathcal{F}_{ij}(\theta) := \mathrm{Re}(\braket{\partial_i\varphi| \partial_j \varphi} - \braket{\partial_i\varphi|\varphi} \braket{\varphi |\partial_i\varphi}),
    \label{eq:QFI}
\end{align}

where $\ket{\partial_i \varphi} := \partial \ket{\varphi(\theta)} / \partial \theta_i$ denotes the partial derivative of  $\ket{\varphi(\theta)}$.

The QFI matrix is a local, positive semi-definite matrix that captures the sensitivity of the state with respect to parameter variations. Its dimension is given by the number of parameters of the circuit $P(U)$, and its entries can be computed via second-order derivatives of the fidelity. In practice, the QFI can be efficiently estimated using the parameter-shift rule \citep{Wierichs_2022}:

\begin{align}
\mathcal{F}_{ij}(\theta) := -\frac{1}{8} \Big[ 
&|\braket{\varphi(\theta) | \varphi(\theta + (e_i + e_j)\pi/2)}|^2 
-|\braket{\varphi(\theta) | \varphi(\theta + (e_i - e_j)\pi/2)}|^2 \nonumber + \\ 
&- |\braket{\varphi(\theta) | \varphi(\theta + (-e_i + e_j)\pi/2)}|^2 
+ |\braket{\varphi(\theta) | \varphi(\theta - (e_i + e_j)\pi/2)}|^2
\Big],
\label{eq:QFI_Shift_Rule}
\end{align}
where $e_i$ denotes the unit vector along the $i$-th parameter direction.

We then define the Effective Quantum Dimension $Q(U(\theta))$ \citep{Haug_2021} of a parametrized circuit $U(\vartheta)$ as:
\begin{align}
Q(U(\theta)) := \mathrm{rank}(\mathcal{F}(\theta))
\label{eq:Quantum_Dimension}
\end{align}
The Parameter Dimension $D(U)$ is then defined by \citep{Haug_2021}, for sufficiently random parameter configurations $\theta_{\mathrm{random}}$, as: 
\begin{align}
D(U) \simeq Q(U(\theta_{random})),
\label{eq:Parameter_Dimension}
\end{align}

We evaluate both the number of parameters $P(U)$ and the Parameter Dimension $D(U)$ for all considered circuits (see Section~\ref{subsec:circuits_and_hamiltonians} and Appendix~\ref{app:circuit_specifics}), and compute their correlation coefficients and mutual information with the average normalized approximation ratio. The results are summarized in Tables~\ref{tab:Spearman_comparison}, and \ref{tab:MI_Comparison}.

\begin{table}
    \def\arraystretch{1.3}
    \begin{tabular}{|>{\centering\arraybackslash}m{1.9cm}|>{\centering\arraybackslash}m{3.6cm}|>{\centering\arraybackslash}m{3cm}|>{\centering\arraybackslash}m{3cm}|}
    \hline
        Problem class&Hamiltonian Expressibility $\varepsilon^{\mathcal{H}}(U,H)$&Number of parameters $P(U)$&Parameter Dimension $D(U)$\\
        \hline
        \hline
       MaxCut& $0.64 \pm 0.12 $ &  $-0.75 \pm 0.11 $ &$\bm{-0.76 \pm 0.10 }$ \\
       \hline
       MinVertex&  $\bm{0.58 \pm 0.28}$ & $-0.58 \pm 0.22$ & $-0.58 \pm 0.22$\\
       \hline
        MaxClique& $0.38 \pm 0.44 $ &  $\bm{-0.43 \pm 0.36} $ &$0.42 \pm 0.38$ \\
       \hline
       Random Diagonal&  $0.12 \pm 0.42$ & $\bm{-0.20 \pm 0.36}$ & $-0.19 \pm 0.36$\\
       \hline
       Heisenberg Basis State& $\bm{0.64 \pm 0.16 } $ &  $-0.59 \pm 0.17 $ &$-0.60 \pm 0.15$ \\
       \hline
       Heisenberg Superposition State&  $\bm{-0.83 \pm 0.05}$ & $0.70 \pm 0.03$ & $0.72 \pm 0.01$\\
       \hline
        Ising&  $\bm{-0.44 \pm 0.43}$ & $0.39 \pm 0.37$ & $0.39 \pm 0.40$\\
       \hline
       Adiabatic&  $0.07 \pm 0.42$ & $-0.08 \pm 0.43$ & $\bm{-0.09 \pm 0.44}$\\
       \hline
       Random Non Diagonal&  $\bm{-0.59 \pm 0.16}$ & $0.51 \pm 0.16$ & $0.54 \pm 0.16$\\
       \hline

    \end{tabular}
    \caption{Average Spearman correlation coefficient between Hamiltonian expressibility (first column), number of parameters (second column), and Parameter Dimension (third column), with respect to the average normalized approximation ratio. The values are obtained for circuits and problems of 4 qubits, when adopting a noiseless setting for the VQE. Each value is reported along with its standard deviation. Bold text indicates the strongest correlation of each class.}
    \label{tab:Spearman_comparison}
\end{table}

Table~\ref{tab:Spearman_comparison} reports the average Spearman correlation coefficients between Hamiltonian expressibility, number of parameters, and Parameter Dimension, with respect to the average normalized approximation ratio, across all considered problem classes.

We observe that, for most problem classes, Hamiltonian expressibility exhibits a stronger correlation than both $P(U)$ and $D(U)$, particularly for non-diagonal problem classes and problems whose solutions lie in superposition states. This further supports our conclusion regarding the usefulness of Hamiltonian expressibility in ideal settings for such a class of problems. The Adiabatic class represents an exception; however, none of the considered metrics shows a significant correlation for such a class, a behaviour which can be attributed to the hybrid nature of this class, as discussed in Section~\ref{subsec:noiseless_results}.

Additionally, we note that the correlations associated with $P(U)$ and $D(U)$ consistently exhibit the opposite sign compared to those of $\varepsilon^{\mathcal{H}}(U,H)$. This can be explained by the “lower-is-better” nature of the Hamiltonian expressibility metric and its relationship with circuit depth and complexity. As discussed in Section~\ref{subsubsec:relationship_with_depth}, Hamiltonian expressibility approaches zero as the number of layers, and hence the number of parameters, increases. Consequently, more expressive circuits (with lower numerical values of Hamiltonian expressibility) typically correspond to higher values of $P(U)$ and $D(U)$, leading to correlations of opposite sign.

Regarding the mutual information results (Table~\ref{tab:MI_Comparison}), Hamiltonian expressibility again appears to be the most informative metric for the Heisenberg superposition state and Random Non-Diagonal classes, which can be regarded as representative of problems with superposition state solutions. In these cases, it achieves the highest absolute mutual information values $1.12$ and $0.81$, respectively. On the other hand, Parameter Dimension proves to be more informative for problem classes with solutions in the computational basis, albeit with relatively modest mutual information values, and also for the Ising and Adiabatic classes.

Overall, these results confirm that Hamiltonian expressibility is particularly well-suited for problems whose solutions lie in superposition states. Moreover, its problem-dependent nature, arising from its explicit dependence on the structure of the Hamiltonian, provides a stronger justification for its use and for the observed variation in its behaviour across different problem classes, compared to $P(U)$ and $D(U)$, which are entirely problem-agnostic.

\begin{table}
    \def\arraystretch{1.3}
    \begin{tabular}{|>{\centering\arraybackslash}m{1.9cm}|>{\centering\arraybackslash}m{3.6cm}|>{\centering\arraybackslash}m{3cm}|>{\centering\arraybackslash}m{3cm}|}
    \hline
            Problem class&Hamiltonian Expressibility $\varepsilon^{\mathcal{H}}(U,H)$&Number of parameters $P(U)$&Parameter Dimension $D(U)$\\
        \hline
        \hline
       MaxCut  & $0.10 \pm 0.077 $ &  $0.50 \pm 0.097 $ &$\bm{0.53 \pm 0.06 }$ \\
       \hline
       MinVertex  &  $0.28 \pm 0.09$ & $0.33 \pm 0.16$ & $\bm{0.41 \pm 0.11}$\\
       \hline
        MaxClique  & $0.31 \pm 0.12 $ &  $0.41 \pm 0.09 $ &$\bm{0.50 \pm 0.09}$ \\
       \hline
       Random Diagonal  &  $0.33 \pm 0.15$ & $0.25 \pm 0.13$ & $\bm{0.37 \pm 0.13}$\\
       \hline
       Heisenberg Basis State & $0.38 \pm 0.06 $ &  $0.34 \pm 0.08 $ &$\bm{0.45 \pm 0.07}$ \\
       \hline
       Heisenberg Superposition State  &  $\bm{1.12 \pm 0.14}$ & $0.49 \pm 0.09$ & $0.94 \pm 0.09$\\
       \hline
        Ising  &  $0.49 \pm 0.16$ & $0.40 \pm 0.14$ & $\bm{0.72 \pm 0.20}$\\
       \hline
       Adiabatic  &  $0.29 \pm 0.12$ & $0.28 \pm 0.10$ & $\bm{0.50 \pm 0.12}$\\
       \hline
       Random Non Diagonal  &  $\bm{0.81 \pm 0.24}$ & $0.37 \pm 0.09$ & $ \bm{0.81 \pm 0.15}$\\
       \hline

    \end{tabular}
    \caption{Average mutual information between Hamiltonian expressibility (first column), number of parameters (second column), and Parameter Dimension (third column), with respect to the average normalized approximation ratio. The values are obtained for circuits and problems of 4 qubits, when adopting a noiseless setting for the VQE. Each value is reported along with its standard deviation. Bold text indicates the highest value of each class.}
    \label{tab:MI_Comparison}
\end{table}

\end{appendices}

\bibliography{sn-bibliography}

\end{document}